\documentclass[a4paper,12pt,twoside]{book}

\usepackage{amsmath,amssymb,enumerate,epsfig,subfigure,caption2
}
\usepackage{overpic}
\usepackage{subfigure}
\usepackage[english]{babel}
\usepackage{amsmath}
\usepackage{amssymb}
\usepackage{epsfig}
\usepackage{graphics,psfrag,rotating}
\usepackage{graphicx}
\usepackage{dcolumn}
\usepackage{bm}
\title{Some cosmological and astrophysical aspects of modified gravity theories}
\author{
\\
{\bf \'Alvaro de la Cruz Dombriz}\footnote{dombriz@fis.ucm.es}\\
\\
PhD thesis
}
\date{under the supervision of
\\
\vskip0.65cm
Dr.\ Antonio Dobado Gonz\'alez and Dr.\ Antonio  L\'opez Maroto \\
\vskip3cm
\epsfig{file=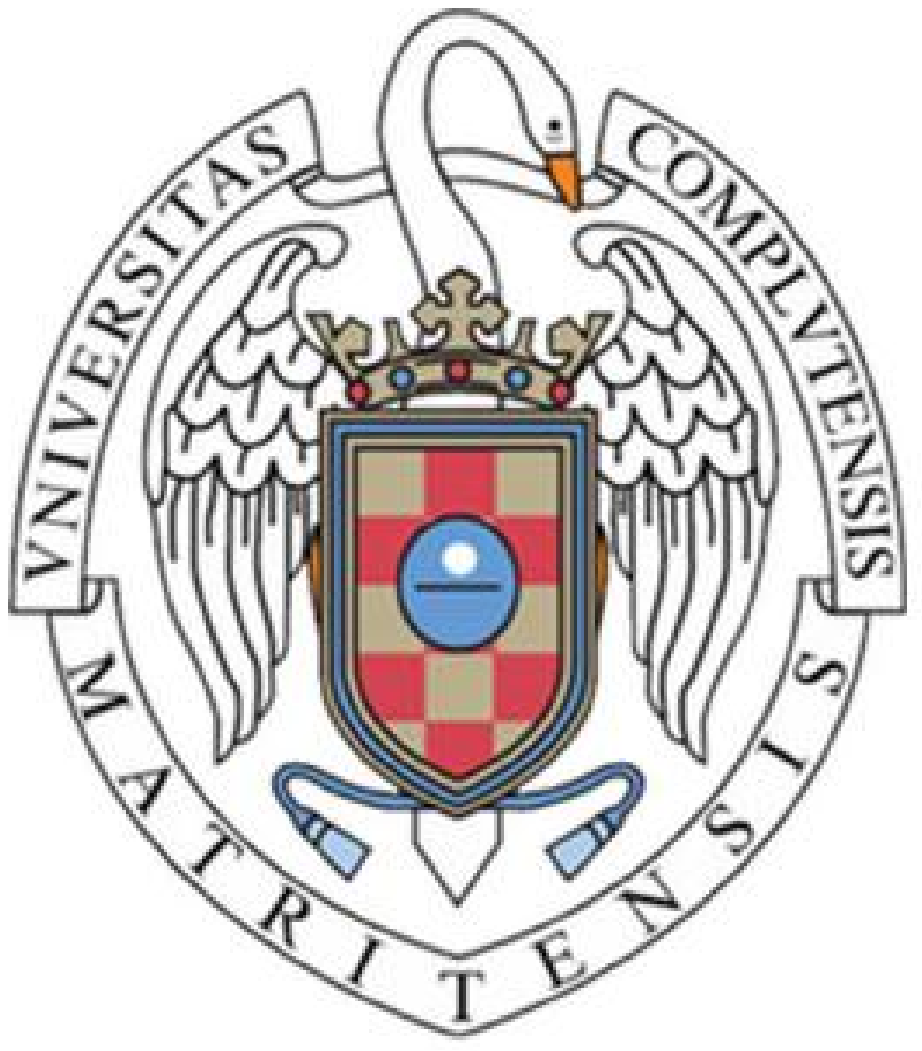,height=3cm}
\vskip 0.6 cm Universidad Complutense de Madrid\\
2010
}

\newlength\encuadernacion \newlength\longB 
\renewenvironment{itemize}
   {\begin{list}
      {}%
      {\setlength{\topsep}{0pt}%
       \setlength{\partopsep}{0pt}%
       \setlength{\itemsep}{-4pt}%
       \setlength{\labelsep}{5pt}%
       \setlength{\itemindent}{0pt}%
      }%
   }%
   {\end{list}}%

\makeatletter
 \renewcommand{\maketitle}{\begin{titlepage}%
  \let\footnotesize\small
  \let\footnoterule\relax
  \let \footnote \thanks
  \null\vfil
  \vskip 60\p@
  \begin{center}%
    {\LARGE\bf \@title \par}%
    \vskip 3em%
    {\large
     \lineskip .75em%
      \begin{tabular}[t]{c}%
        \@author
      \end{tabular}\par}%
      \vskip 1.5em%
    {\large \@date \par}
  \end{center}\par
  \@thanks
  \vfil\null
  \end{titlepage}%
  \setcounter{footnote}{0}%
  \global\let\thanks\relax
  \global\let\maketitle\relax
  \global\let\@thanks\@empty
  \global\let\@author\@empty
  \global\let\@date\@empty
  \global\let\@title\@empty
  \global\let\title\relax
  \global\let\author\relax
  \global\let\date\relax
  \global\let\and\relax
}
\makeatother

\makeatletter
 \newcommand\figcaption{\def\@captype{figure}\caption}
\makeatother

\def\a{\alpha}
\def\b{\beta}

\def\t{\tau}

\def\id3x{\int\!\! d^3\!\vec{x}}
\def\idx{\int\!\! d^4\!x}

\def\rig>{\right>}

\newcommand{\bea}{\begin{eqnarray}}
\newcommand{\eea}{\end{eqnarray}}
\newcommand{\beann}{\begin{eqnarray*}}
\newcommand{\eeann}{\end{eqnarray*}}
\newcommand{\ba}{\begin{array}}
\newcommand{\ea}{\end{array}}

\def\g5{\gamma_{5}}

\def\idx3{\int\! d^{3}\!\vec{x}\,}
\def\idx{\int\! d^{4}\!x\,}

 \def\g {\gamma}

 \def\a {\alpha}
\def\b {\beta}

\begin{document}
\frontmatter
\maketitle
\newpage
\
\newpage
\begin{flushright}
\ \\ \par
\vspace{7cm}
{\it A Carmen y Julio Dombriz},\\
\vskip 0.2cm
cuyos ojos\\
no alcanzaron este d\'ia,\\
pero cuyas palabras\\
me ense\~naron a hablar.
\end{flushright}

\begin{centering}
\ \\ \par
\vspace{5.5cm}
\begin{minipage}{0.2\textwidth}
\
\end{minipage}
\begin{minipage}{0.6\textwidth}
{\bf Acknowledgements}\par\ \par
It is my wish to thank my advisors, Dr.\ Antonio Dobado Gonz\'alez and Dr.\ Antonio L\'opez Maroto, for introducing me to
research in Theoretical Physics. It is always a pleasure to work with them.
I also want to thank Dr.\ Jos\'e Alberto Ruiz Cembranos, an {\it effective} supervisor,
for fruitful conversations and his continuous both personal and professional encouragement during these years.
I am also grateful to the Departamento de F\'{\i}sica Te\'orica I at the Universidad Complutense de Madrid, the head of department
and staff, for providing the conditions allowing me to carry out this work. Part of it was done during a short stay at the
Department of Physics and Astronomy at University of California Irvine under the advice of Dr.\ Jonathan Feng; I am greatly indebted to him.
This work would not have been possible without the financial support of the Spanish Ministry of Education.
\\

I would like to thank in a profound and sincere manner my colleagues - and friends nevertheless - at
the office: Jos\'e Beltr\'an, Alejandro Berm\'udez and Juan M. Torres.
A particular dedicatory must be offered to Javier Almeida, Rub\'en
Garc\'ia, Sophie Hirel, Gil Jannes, Carlos Tamarit, Ignacio Vleming and Marc Wouts
for their friendship and loyalty during these years, particularly
in difficult moments. A special thought is for the two muses of my present life, Beatriz
Seoane and Lourdes Tabares. {\it Verae amicitiae sempiternae sunt}.
\\

Lastly, but not the least, I want to thank my parents, for the opportunities they gave to me that
they never had, and my family, who has been supporting me, even when they did not understand very much
what I was doing. {\it Verba volant, sed scripta manent}.

\end{minipage}
\end{centering}
\newpage
\begin{flushright}
\ \\ \par
\vspace{7cm}
\begin{minipage}{0.5\textwidth}
\emph{Beauty is truth, truth beauty,\\that is all
    Ye know on earth,\\ and all ye need to know.  }\par\ \par
John Keats,\\ \emph{Ode on a Grecian Urn}
\end{minipage}
\end{flushright}
%
%
\newcommand{\Od}{{\cal O}}
\newpage
\
\makeatletter
\renewcommand \thesection {\thechapter.\@arabic\c@section}
\renewcommand{\section}{\@startsection {section}{1}{\z@}%
                                   {-3.5ex \@plus -1ex \@minus -.2ex}%
                                   {2.3ex \@plus.2ex}%
                                   {\normalfont\Large\bfseries}}
\renewcommand\theequation
  {\ifnum \c@chapter>\z@ \thechapter.\fi \@arabic\c@equation}
\makeatother
\setcounter{tocdepth}{2}
\tableofcontents
\newpage
\mainmatter
%
\chapter*{Preface\markboth{\MakeUppercase{Preface}}{}}
\addcontentsline{toc}{chapter}{Preface}
\label{chap:Preface}
The twentieth century witnessed the development of both
gravitation and cosmology as modern scientific disciplines
subjected to observations. These observations have been performed
through terrestrial particle detection devices, telescopes and
satellites that allow to
verify theoretical predictions and to
rule out proposed theoretical models. With the turning of the new
century, called to be the century of precision cosmology, new
perspectives have been unveiled with recent
experiments such as WMAP, PLANCK or SDSS. These last experiments are
able to determine with higher and higher accuracy the features of the Cosmic Microwave Background (CMB), the
distribution of large scale structures and
the fundamental cosmological parameters which describe our universe on the largest scales.
Despite the improvements in the observational side, a fundamental
gravitational theory, which is renormalizable
from a quantum field theory point of view and applicable
to arbitrary scales,
from micro-gravitational tests, passing through solar system tests, to cosmological scales, is still lacking.

General relativity, in spite of being the most successful
gravitational theory in the last one hundred years, has left some
of these
problems without satisfactory answer. Although within the string theory
paradigm it would be possible to find a consistent quantum theory of gravity, this
is not the case of general relativity which turns out to be nonrenormalizable
as a perturbative field theory.
Moreover, if this theory is used to construct the
standard cosmological model, where the fluid content is given by
standard matter and radiation, it cannot account for the observed
accelerated expansion of the universe on sufficiently large scales. In
fact, it needs to be supplemented by some dark energy contribution
to accommodate this accelerated regime. On the other hand, general
relativity with gravitating luminous matter cannot account either
for the observed rotation curves of galaxies. A
dark matter contribution needs to be introduced to reconcile data with theoretical
predictions within this paradigm.

Instead of adding new elements
in the cosmological content, which try to accommodate observations
with general relativity, those problems
might show that the theoretical framework in cosmology should be enlarged
by alternative gravity theories.
This thesis will try to contribute to the understanding of those still open
issues by considering two recently proposed alternative and complementary
theories to general relativity. We shall consider some relevant aspects of
those models related to recent experimental results.

%
%

The present work is organized
in the way that follows: First, we will briefly introduce in Chapter \ref{chap:Introduction}
some modified gravity theories and their corresponding formalisms.
In this chapter
special attention will be paid to $f(R)$ gravities by summarizing the main features of this
paradigm 
in the metric formalism.
%
Then some geometrical results for $f(R)$ theories and both cosmological and gravitational
constraints usually imposed over such functions will be provided. Other alternative modified gravities, the
brane worlds, will then be
reviewed. Here
we shall introduce both the notion of brane excitations, the branons, and some
topologically nontrivial solutions, the brane-skyrmions. We shall finish
the chapter by providing some insight about the possibility of mini black holes
detection in the Large Hadron Collider (LHC) as a signature for the validity of these
modified gravity theories.
%

The second chapter will deal with $f(R)$ theories
which try to
provide a cosmological acceleration mechanism with no need for introducing
any extra dark energy contribution in the cosmological components. To do so, we shall use some
reconstruction procedures which start either from a given solution of the cosmological scale
factor for an homogeneous and isotropic metric or
from an effective equation of state. In particular, those $f(R)$ functions able
to mimic Einstein-Hilbert plus cosmological constant solutions will be obtained. In this realm, $f(R)$ theories
will be shown to be able to mimic the cosmological evolution generated by any
perfect fluid with constant equation of state.
%
%

Then the third chapter will be devoted to the computation of cosmological perturbations
for $f(R)$ theories. Since in Chapter \ref{chap:$f(R)$ theories} the modified Einstein equations will
have been studied as background equations, it is quite natural when modifying general relativity by $f(R)$
models, to ask about
the first order perturbed equations for these theories and what consequences
in the growing of these perturbations may appear. This is the {\it leitmotiv} in this
chapter. Throughout it, special attention will be paid to the possibility of obtaining a completely general
differential equation for the evolution of perturbations and its particularization for the so-called
sub-Hubble scales will be explicitly shown. The mentioned differential equation in those scales will
be shown to be very useful to understand the regime validity of some approximations widely accepted
in the literature 
and to rule out that some proposed $f(R)$ models could be cosmologically viable.

%
%
The introduction of modified gravity theories, with or without extra dimensions, may lead
to the existence of new solutions with respect to those of general relativity. In that sense, the
research about spherically symmetric solutions is of particular interest. For instance, it may shed some light on the number of extra dimensions, the fundamental scale of gravity or the required restrictions to be imposed over the parameters of those theories. The possible detection of mini black holes at the LHC in the coming years will be
a turning point to discover
certain properties of the underlying gravity theory. For
this reason, chapters 4 and 5 will be devoted to the study of spherical solutions
in extra dimensions theories. In particular, spherically symmetric and static black-hole solutions coming from $f(R)$ theories in an arbitrary number of dimensions will be studied in Chapter 4 whereas Chapter 5 will be focused on studying
a particular topologically nontrivial solution with spherical symmetry appearing in brane-world models -- different
from the well-known black-hole solutions -- the so-called brane-skyrmions.

Hence, in the fourth chapter we shall focus on the
study of black holes in $f(R)$ gravity theories in
an arbitrary number of dimensions.
We shall concentrate on the existence of black-hole solutions and
%
%
we shall study which will be their inherited or
different features with respect to those in general relativity.
With this purpose we shall study constant
curvature solutions for $f(R)$ theories as well as perturbative
solutions around
the standard Schwarzschild-anti-de Sitter
geometry.
An important part of this chapter will be then devoted to the
thermodynamics of Schwarzschild-anti-de Sitter black holes in $f(R)$ theories.
This research will prove that for $f(R)$ gravities, there
exists a thermodynamical viability condition
which is related to one of the conditions which ensure
gravitational viability for $f(R)$ models.

In the fifth chapter we will thoroughly study other kind of spherically
symmetric solutions in brane-world theories that are not black holes. These
solutions, the brane-skyrmions, are topologically nontrivial configurations arising in the
presence of these extra dimensions theories. In this context, the recent claim of detection
of an unexpected feature in the CMB, referred to as the cold spot, will be explained as a topological
defect on the brane. After performing some calculations, it will be shown
that results obtained are in complete agreement with those in the literature that tried to
explain that cold spot as a texture of a non-linear sigma model. The physical interpretation of
these results and future prospects will finish this chapter.
%
%
%

At the end of each chapter, we shall
include the corresponding conclusions. These conclusions are summarized
all together in the sixth chapter, which is followed by an appendix
where more detailed formulae for the calculations performed in the third chapter are shown.
\chapter{Introduction to modified gravity theories}
\label{chap:Introduction}
%
%
\section{Motivation}
\label{sec:Int:Motivation}
From its very beginning, it was questioned whether general relativity (GR) was the unique correct theory
among other theories for gravitation. Thus
for instance Weyl in \cite{Weyl} and Eddington in \cite{Eddington}
included higher order invariants in the gravitational action.
Those attempts were neither experimentally nor theoretically
motivated, but it was soon proved that
the Einstein-Hilbert (EH) action
was not renormalizable and therefore could not be conventionally
quantized. In fact, this action
needs to be supplemented by higher order terms in order
for the resultant theory
to be one-loop renormalizable \cite{Utiyama&DeWitt,
tHooft&Veltman}.
More recent research has shown that when
quantum loop corrections in field theory or
higher order corrections in the low
energy string dynamics are considered, the
effective low energy gravitational action includes higher order
curvature invariants \cite{Birrell&Davies, Buchbinder,
Vilkovisky}.

Such results encouraged the interest in higher order gravity theories, i.e., modifications
of the EH gravitational action which include higher order curvature invariants. 
Nonetheless, those new added contributions were thought to be relevant only
in very strong gravity regimes, such as at scales close to the Planck scale
and therefore in the early universe or near black hole singularities. 
However, these corrections were not expected to affect gravitational phenomenology
neither at low curvature nor at low energy regimes, and therefore they were assumed to be
negligible at large scales such as those involved in the late universe evolution.

Very recent evidence coming from both astrophysics and cosmology
have revealed the unexpected accelerated expansion 
of the universe. Different
data from type Ia supernovae (SNIa) surveys \cite{typeIa1, typeIa2,
typeIa3}, large structure formation and delicate measurements of
the 
CMB anisotropies, particularly
those from the Wilkinson Microwave Anisotropy Probe (WMAP)
\cite{WMAP}, have concluded that our universe is expanding at an
increasing rate. This fact sets the very urgent problem of finding
the cause for this speed-up since standard GR with ordinary matter
and radiation is not able to do so. Usual explanations for this
fact have been categorized to belong to one of the following three
classes:
\begin{enumerate}
\item The first type of explanations reconciles this acceleration with GR by invoking a strange cosmic fluid, dark energy (DE) (see \cite{review} and references therein), with a state equation relating its pressure and energy density in the following way
\begin{eqnarray}
P_{\text{DE}} =\omega_{\text{DE}}\,\rho_{\text{DE}}
\label{DE_eqn_state}
\end{eqnarray}
where $\omega_{\text{DE}}<-1/3$ is required to provide acceleration in the usual Einstein equations as is described in Section \ref{sect:fR:Standard Einstein's eqns for FLRW}. This state equation shows that the DE fluid has a large negative pressure. For the particular case $\omega_{\text{DE}}=-1$, this fluid behaves just as a cosmological constant $\Lambda$.
Within this approach of DE in the form of a cosmological constant, recent data obtained by WMAP \cite{WMAP} provide the following cosmological content distribution: $4.6\%$ corresponds to ordinary baryonic matter, $22.7\%$ to cold dark matter and $72.9 \%$ to DE.
%
This is the so-called concordance or $\Lambda$-Cold Dark Matter model ($\Lambda\text{CDM}$) which is supplemented with some inflation mechanism usually through some scalar field, the inflaton.
The main problem of this kind of description is that the fitted $\Lambda$
value seems to be about 55 orders of magnitude smaller than
the expected vacuum energy of matter fields, this is the so-called
{\it cosmological constant problem}. From a more philosophical
point of view, the DE description
also presents the so-called {\it coincidence problem}. This problem 
wonders why the DE and matter densities are so close in order of
magnitudes precisely in these days, i.e. in the present
cosmological era, even though for both the cosmological past and
future that is not the case. This kind of problems comes to claim
that the $\Lambda\text{CDM}$ model could be regarded as an
empirical fit to data with a poorly motivated gravitational
theory behind
and therefore, it should be considered as a
phenomenological approach of the underlying correct cosmological
theory.
\\
\item The second type of explanations consider a dynamical DE by introducing
a new scalar field. They are the so-called quintessence theories. The theories which introduce
an extra scalar field in the gravitational sector of the action are usually referred
to as scalar-tensor theories. Some very interesting subcases
of such theories are the so-called Brans-Dicke theories 
which are going to be explained in detail in the Section \ref{sec:Int:fR:Brans-Dicke theories}.
\\
\item Finally the third one consists of trying to explain the cosmic acceleration as a
consequence of new gravitational physics \cite{Carroll1, otras}. For instance, modifications
to the EH gravitational action have been widely considered in the literature \cite{S,B,M,Carroll,DGP,Cembranos2006}. More
recently, vector-tensor theories of gravity and the electromagnetic field itself have also been proposed
as compelling DE candidates \cite{Jose}.

Some of those theories add higher or lower powers of the
scalar curvature, the Riemann and Ricci tensors or their
derivatives \cite{Maroto&Dobado:1993}. Lovelock theories and
$f(R)$ gravity theories are some examples of these attempts. In recent
years, some $f(R)$ proposals have even tried to reconcile
dark matter through a gravitational sector modification
\cite{Cembranos:2008} or to explain both the current cosmic
speed-up and early inflation simultaneously \cite{Nojiri:2003}. The core of
Chapter 2 will thoroughly deal with some attempts of $f(R)$
theories to circumvent the necessity of introducing DE to explain
the cosmic acceleration.
\end{enumerate}
%

On the other hand, other open issues in the Standard Model (SM) of
elementary particles, namely the {\it hierarchy problem}, could
also be related to the fundamental gravity theory.
Thus, this problem
appears in the renormalization procedure in theories containing
scalar fields. In such theories
the renormalized scalar masses
are expected to be given by the cut-off of the theory, i.e., the Planck
scale. Therefore an extreme fine tuning is required in order
to get the expected mass for scalars, in
particular the Higgs mass.
If on the contrary the fundamental scale of gravitation is
close to the electroweak scale, the corresponding
cut-off would be of the same order as the expected Higgs mass
and an extreme fine tuning would not be required.

With the aim of solving this problem, large extra dimensions theories
have recently been considered. Unlike ancient Kaluza-Klein theories, with
compactified Planck scale size extra dimensions, recent brane-world
models may contain much larger extra dimensions. In order to avoid
the presence of Kaluza-Klein towers of copies of SM particles with similar
masses, these models restrict SM particles to propagate on the brane, whereas
only gravity can propagate in the whole bulk space. In this way, the
fundamental gravity scale can be reduced to the electroweak scale and
the gauge {\it hierarchy problem}
is avoided.

Brane-world (BW) theories may also explain the observed
accelerated expansion of the universe \cite{DGP} and as will be
shown in Section \ref{sec:Int:BW:Excitations in BW:Branons}, they
present excitations which can produce weakly interacting massive
particles (WIMPs), which are natural candidates for the observed
dark matter \cite{DM2}. Let us finally remark that such modified extra
dimensions gravity theories, as will be explained in Section
1.10, may give rise to the production of black holes (BHs) of
little size at the LHC whose eventual detection may give valuable
information about the dimensionality of space-time.

In the following sections of this chapter we shall deal with
different aspects of the already mentioned modified gravity
theories, both $f(R)$ theories and brane-world theories: in
Section \ref{sec:Int:fR:Action and field equations} we shall
present some generalities about the formalism which will be used
throughout the thesis. Thus in Section \ref{sec:Int:fR:Modified
Einstein's equations}, the modified Einstein equations derived
from $f(R)$ theories in the metric formalism will be presented. Then
in Section \ref{sec:Int:fR:Brans-Dicke theories}, the equivalence
of such theories with Brans-Dicke theories will be briefly
sketched. Some geometrical results for $f(R)$ gravities which were
originally published in \cite{BH_Dombriz} will be presented in
Section \ref{sec:Int:fR:Geometrical results}. They deal with 
constant curvature solutions and analytical conditions to reproduce
Einstein's equations for the EH action with or without cosmological
constant.
Concerning BW theories, the main concepts for those models are
analyzed in Section \ref{sec:Int:BW:BW theory}, while a
study of BW excitations, called branons, is presented in Section
\ref{sec:Int:BW:Excitations in BW:Branons}. Topologically
nontrivial configurations, called skyrmions are introduced in
Section \ref{sec:Int:BW:Skyrmions} and some gravitational
consequences of those theories will be summarized in the final Section
\ref{sec:Int:BW:BH}.
\section{Generalities}
\label{sec:Int:fR:Action and field equations}
The gravitational action for GR in an arbitrary number of dimensions $D$ is given by the so-called EH action
\begin{eqnarray}
S_{\,\text{EH}}\,=\,\frac{1}{2\kappa}\int \text{d}^{D}x\sqrt{\mid g\mid}\,R\,.
\label{action_EH}
\end{eqnarray}
Here, $\kappa\equiv8\pi G_D$ where $G_D\equiv M_D^{2-D}$ holds for the $D$-dimensional gravitational
constant, with $M_D$ the gravitational fundamental scale, $g$ is the metric determinant and $R$ is the Ricci scalar
defined from the metric tensor.

With the aim of modifying the EH action, gravitational action for $f(R)$ theories, considered as
generalizations of GR, may be written as
\\
\begin{eqnarray}
S_{G}\,=\,\frac{1}{2\kappa}\int \text{d}^{D}x\sqrt{\mid g\mid}\,(R+f(R))\,.
\label{action}
\end{eqnarray}
\\
From either actions given in equations \eqref{action_EH} or \eqref{action}, the field equations, giving rise to the
so-called standard and modified Einstein equations respectively, can be derived by using different variational
principles. Two such variational principles have been mainly considered in the literature: on the one hand, the
standard metric formalism considers that the connection 
is metric dependent and therefore the only present fields in the gravitational sector are those coming from the metric tensor. On the other hand, there exists the so-called Palatini variational principle where metric and connection are assumed to be
independent fields. In this case the action is varied with respect to both of them. Whereas for an action linear in $R$ such as
that in expression \eqref{action_EH} both formalisms lead to the same field equations, this is no longer true for
nonlinear gravity theories (see \cite{Sotiriou} for an exhaustive review on nonmetric formalisms). In this thesis, we
shall restrict ourselves to the metric formalism. For that purpose, we shall assume that the connection is the usual Levi-Civita connection given by
\begin{eqnarray}
\Gamma^{\alpha}_{\;\mu\nu}\,\equiv\,\frac{1}{2}g^{\alpha\gamma}\left(\frac{\partial g_{\gamma\nu}}{\partial x^{\mu}}+\frac{\partial g_{\mu\gamma}}{\partial x^{\nu}}-\frac{\partial g_{\mu\nu}}{\partial x^{\gamma}}\right)
\label{Levi-Civita connection}
\end{eqnarray}
where, as in the rest of the work, Einstein's convention for implicit summation is assumed.

At this stage, let us point out that the convention to be used for the metric signature will be
$(+,\,-,\,...,\,-)$, i.e., positive sign for temporal coordinate
whereas negative sign for spatial ones. With respect to the
Riemann tensor definition, our conventions will be
\begin{eqnarray}
R^\mu_{\;\;\nu\alpha\beta}\,\equiv\,\frac{\partial\,\Gamma^{\mu}_{\nu\alpha}}{\partial x^{\beta}}
-\frac{\partial\,\Gamma^{\mu}_{\nu\beta}}{\partial x^{\alpha}} +\Gamma^{\mu}_{\sigma\beta}\Gamma^{\sigma}_{\nu\alpha}
-\Gamma^{\mu}_{\sigma\alpha}\Gamma^{\sigma}_{\nu\beta}.
\label{Riemann}
\end{eqnarray}
From expression \eqref{Riemann}, the corresponding Ricci tensor and scalar curvature are obtained straightforwardly and they read respectively as follows
\begin{eqnarray}
R_{\mu\nu}\,\equiv\,R^{\alpha}_{\;\;\mu\alpha\nu}\,\,\,;\,\,\,R\,\equiv\,R^{\alpha}_{\,\,\alpha}.
\label{Ricci&R}
\end{eqnarray}


In addition to the already explained gravitational sector, the
energy content may be introduced in the cosmological content
through energy-momentum tensors, which will describe the different
components such as dust matter, radiation, dark matter, etc. which
are present in the cosmological content of the universe. For
each different type of fluid content ($\alpha$), assumed
from now on to behave as a
perfect fluid, the corresponding energy-momentum tensor is given
by
\begin{eqnarray}
T^{(\alpha)}_{\mu\nu}=(P_{\alpha}+\rho_{\alpha})u_\mu^{(\alpha)}\,u_\nu^{(\alpha)}-P_{\alpha}\,g_{\mu\nu}
\label{perfect_fluid}
\end{eqnarray}
where $P_{\alpha}$, $\rho_{\alpha}$ and $u^{\mu\,\,(\alpha)}$ are the pressure, energy density and 4-velocity of the $\alpha$ component respectively. 
Therefore the total energy-momentum tensor will be nothing but
\begin{eqnarray}
T_{\mu\nu}\,\equiv\,\sum_{\alpha}T^{(\alpha)}_{\mu\nu}
\label{total_EM_tensor}
\end{eqnarray}
for all possible fluid contributions. The most usual approach is to consider barotropic fluids where $P_{\alpha}=P_{\alpha}(\rho_{\alpha})$ and very often the relation between these two quantities is linear through
an equation of state
\begin{eqnarray}
P_{\alpha}\,=\,\omega_{\alpha}\rho_{\alpha}
\label{eq_state}
\end{eqnarray}
where for instance $\omega_{\alpha}=-1,0,1/3$ if cosmological constant, dust matter or radiation are the considered fluids respectively.
DE fluids with constant equation of state are given by the condition $\omega_{\text{DE}}<-1/3$ whereas phantom candidates for DE obey $\omega_{\text{DE}}<-1$.
In our approach to modify GR, to be rigorously implemented in
Chapter 2, DE will appear as a modification of the gravitational
sector itself so no DE component will be explicitly included in
the content expressed by the summation \eqref{total_EM_tensor}. In
this case the cosmic acceleration will be a consequence of the
modification of the gravitational action by the presence of a 
$f(R)$ term.
Let us finish this section by mentioning that each fluid component is assumed to be conserved separately since no interaction among fluids is considered. This fact also implies the conservation of the total energy-momentum tensor straightforwardly.

\section{Modified Einstein equations}
\label{sec:Int:fR:Modified Einstein's equations}
Now that the previous generalities have been presented, the
modified Einstein equations in the metric formalism for $f(R)$ gravity
theories may be found by performing variations of the
gravitational action \eqref{action}
with respect to the metric and equaling the result to minus the
energy-momentum tensor times $\kappa$
providing the following equations:

\begin{equation}
(1+f_R)R_{\mu\nu}-\frac{1}{2}(R+f(R))g_{\mu\nu}+{\cal D}_{\mu\nu}f_R\,=\,-\kappa\,T_{\mu\nu}
\label{fieldtensorialequation}
\end{equation}
where $f_R\,\equiv\,\text{d}f(R)/\text{d}R$ and
\begin{eqnarray}
{\cal D}_{\mu\nu}\equiv \nabla_{\mu}\nabla_{\nu}-g_{\mu\nu}\square
\end{eqnarray}
with $\square\,\equiv\,\nabla_{\alpha}\nabla^{\alpha}$ and $\nabla$ is the usual
covariant derivative.

Taking the trace of the equation \eqref{fieldtensorialequation} we get:
\begin{eqnarray}
R(1+f_R)-\frac{D}{2}(R+f(R))+(1-D)\square f_R\,=\,-\kappa\,T
\label{trace_eqn}
\end{eqnarray}
which provides a differential relation between $R$ and $T$ unlike GR where this relation
is just algebraic. An interesting point to stress at this stage is that in general, vacuum
solutions, i.e. $T_{\mu\nu}\equiv 0$, do not imply straightforwardly $R=0$ solutions.

By computing the covariant derivative of \eqref{fieldtensorialequation}, it is found that the l.h.s.\ of those
equations vanishes identically, so the covariant derivative for the r.h.s.\ of equations
\eqref{fieldtensorialequation} must obey the conservation equations 
\begin{eqnarray}
\nabla_{\mu}T^{\mu}_{\,\,\,\nu}\,=\,0
\label{motion_eqns}
\end{eqnarray}
where this identity does not depend explicitly on $f(R)$ but only
on the energy-momentum tensor components and metric tensor
elements.

Two particular simple choices for $f(R)$ may be considered in the equations \eqref{fieldtensorialequation}:
\begin{enumerate}
\item  $f(R)\equiv0$, which allows to recover the standard Einstein equations without cosmological constant, i.e.,
\begin{eqnarray}
G_{\mu\nu}\,\equiv\,R_{\mu\nu}-\frac{1}{2}R\,g_{\mu\nu}\,=\,-\kappa\,T_{\mu\nu}
\label{EH_fieldequations}
\end{eqnarray}
where the conservation equations \eqref{motion_eqns} still hold.

\item A second simple choice would be $f(R)\equiv-(D-2)\,\Lambda_{D}$. This choice allows to recover the standard Einstein equations in $D$ dimensions with nonvanishing cosmological constant $\Lambda_D$, i.e.,
\begin{eqnarray}
R_{\mu\nu}-\frac{1}{2}R\,g_{\mu\nu}+\frac{D-2}{2}\Lambda_{D}g_{\mu\nu}\,=\,-\kappa\,T_{\mu\nu}
\label{LCDM_fieldequations}
\end{eqnarray}
where the particular choice of the $\Lambda_D$ normalization will be explained below.
Let us note that the equations \eqref{motion_eqns} again hold.
Notice that in this case the new piece in the previous equation \eqref{LCDM_fieldequations} proportional to $\Lambda_D$ can be moved to the r.h.s. and then an energy-momentum tensor $(T_{\Lambda_D})_{\mu\nu}$ can be defined as follows
\begin{eqnarray}
(T_{\Lambda_{D}})_{\mu\nu}\,\equiv\,\frac{D-2}{2}\frac{\Lambda_{D}}{\kappa}g_{\mu\nu}.
%
%
%
\label{DE_cte_tensor}
\end{eqnarray}
In this case, both density and pressure from the cosmological constant contribution may be written
for any number of dimensions as:
\begin{eqnarray}
\rho_{\Lambda_{D}}\,\equiv\,\frac{D-2}{2}\frac{\Lambda_{D}}{\kappa}\,\,\,;\,\,\,P_{\Lambda_{D}}\,\equiv\,-\frac{D-2}{2}\frac{\Lambda_{D}}{\kappa}
\label{rho&P_Lambda}
\end{eqnarray}
since $P_{\Lambda_{D}}\,=\,-\rho_{\Lambda_{D}}$ is the state equation for a cosmological constant.
\end{enumerate}

Finally let us point out that the equations \eqref{fieldtensorialequation} may be expressed {\it \`{a} la Einstein} by writing all extra terms due to the $f(R)$ presence on the r.h.s. One can try to recover the standard form of the Einstein equations as follows
\\
\begin{eqnarray}
G_{\mu\nu}\,&\equiv&\,R_{\mu\nu}-\frac{1}{2}g_{\mu\nu}R\,=\,\frac{-\kappa}{1+f_R}\left(T_{\mu\nu}+T^{eff}_{\mu\nu}\right)
%
%
\label{fieldtensorialequation_Einstein_manner}
\end{eqnarray}
\\
where an effective energy-momentum tensor has been defined as
\begin{eqnarray}
T^{eff}_{\mu\nu}\,\equiv\,\frac{1}{\kappa}\left[{\cal D}_{\mu\nu}f_{R}-\frac{1}{2}(f(R)-R\,f_R)g_{\mu\nu}\right].
\label{eff_energy_momentum_tensor}
\end{eqnarray}
This energy-momentum tensor does not necessarily obey the strong
energy condition which holds in ordinary fluids (dust matter, radiation, etc.)
do.
\newpage
\section{Equivalence with Brans-Dicke theories}
\label{sec:Int:fR:Brans-Dicke theories}
From a classical field theory perspective, it is always possible to redefine the fields of a given theory in order to express the field equations in a more attractive way which would be easier either to handle or to solve. The price to pay is to introduce new auxiliary fields and even to perform either renormalizations or conformal transformations.

It is widely assumed that two theories are dynamically equivalent if, under a suitable redefinition of either gravitational or matter fields, one can make the field equations to coincide. Nevertheless, some controversy has appeared in recent times especially when conformal transformations are used to redefine fields (see for instance  \cite{Magnano:1993} and \cite{Faraoni:1999} and references therein).

As mentioned in Section \ref{sec:Int:Motivation}
a possibility to construct
alternative theories of gravity are the scalar-tensor theories
which are based upon the introduction of
an extra scalar field which modifies the gravitational sector. Those theories are still metric theories in the sense that the newly introduced fields do not couple
to the fluid contributions.

The gravitational action for a general scalar-tensor theory in $D$ dimensions is
\begin{eqnarray}
S_{ST}\,=\,
\int \text{d}^{D}x\sqrt{\mid g\mid}\,\left[\frac{y(\phi)}{2}R-\frac{\omega(\phi)}{2}(\partial_{\mu}\phi\,\partial^{\mu}\phi)-U(\phi)\right].
\label{action_ST}
\end{eqnarray}
By choosing $y(\phi)=\phi/\kappa$, $\omega(\phi)=\omega_0/(\kappa\phi)$ and $U(\phi)=V(\phi)/\kappa$, the action
\begin{eqnarray}
S_{BD}\,=\,\frac{1}{2\kappa}\int \text{d}^{D}x\sqrt{\mid g\mid}\,\left[\phi\,R-\frac{\omega_0}{\phi}(\partial_{\mu}\phi\partial^{\mu}\phi)-V(\phi)\right]
\label{action_BD_1}
\end{eqnarray}
is obtained from \eqref{action_ST}. This is the action for the Brans-Dicke theories which is obviously a particular case of scalar-tensor theories.

It can be shown that $f(R)$ gravities within the metric formalism are nothing but a Brans-Dicke theory with Brans-Dicke parameter $\omega_0=0$. This fact is easily proven as follows: a new field $\chi$ is introduced and for the sake of simplicity let us define
\begin{eqnarray}
F(R)\,\equiv\,R+f(R).
\end{eqnarray}

Thus the action \eqref{action} can be seen to be equivalent to the action
\begin{eqnarray}
S_{\chi}\,=\,\frac{1}{2\kappa}\int \text{d}^{D}x\sqrt{\mid g\mid}\,\left[F(\chi)+\frac{\text{d}F(\chi)}{\text{d}\chi}(R-\chi)\right]
\label{action_BD}
\end{eqnarray}
since if a variation of \eqref{action_BD} with respect to $\chi$ is performed, the equation which is found reads:
\begin{eqnarray}
\frac{\text{d}^{2}f(\chi)}{\text{d}\chi^{2}}(R-\chi)\,=\,0
\label{chi_variation}
\end{eqnarray}
and thus $\chi=R$ provided $\text{d}^{2}f(\chi)/\text{d}\chi^2\neq0$. Therefore the original action \eqref{action} is recovered. Defining now the scalar field $\phi$ as $\phi\equiv\text{d}F(\chi)/\text{d}\chi$ and introducing a potential $V(\phi)$ as follows
\begin{eqnarray}
V(\phi)\equiv\chi(\phi)\phi-F(\chi(\phi))
\label{V_phi}
\end{eqnarray}
the action \eqref{action_BD} takes the form
\begin{eqnarray}
S_{\phi}\,=\,\frac{1}{2\kappa}\int \text{d}^{D}x\sqrt{\mid g\mid}\,\left(\phi R-V(\phi)\right)
\label{action_BD_2}
\end{eqnarray}
\\
which is exactly the same as \eqref{action_BD_1} if $\omega_0=0$ is imposed.

By including the corresponding fluid sector given by an energy-momentum tensor $T_{\mu\nu}$, the
field equations derived from \eqref{action_BD_2} are
\begin{eqnarray}
G_{\mu\nu}\,=\,-\frac{\kappa}{\phi}\,T_{\mu\nu}-\frac{1}{2\phi}g_{\mu\nu}V(\phi)+\frac{1}{\phi}{\cal D}_{\mu\nu}\phi
\label{eqns_BD}
\end{eqnarray}
\begin{eqnarray}
R\,=\,\frac{\text{d}V(\phi)}{\text{d}\phi}
\end{eqnarray}
where the trace of \eqref{eqns_BD}
\begin{eqnarray}
(D-1)\square\phi+\frac{D}{2}V(\phi)+\frac{2-D}{2}\phi\frac{\text{d}V}{\text{d}\phi}\,=\,-\kappa\,T
\label{trace_BD}
\end{eqnarray}
gives the dynamics of $\phi$ in terms of the matter content.

Let us finally note that if $f_{RR}\,\equiv\,\text{d}^{2}f(R)/\text{d}R^{2}$
vanishes, the equivalence between the two theories cannot be guaranteed
as can be seen from equation \eqref{chi_variation}.
On the other hand, the resulting Brans-Dicke equivalent theory
makes clear that $f(R)$ gravity theories have just one more extra
degree of freedom than standard EH gravity. The apparent absence of kinetic
term in the action \eqref{action_BD_2} must not be thought of as the absence of
dynamics in $\phi$ since this
scalar is dynamically related to the matter fields, as can be seen from
expression \eqref{trace_BD}. Thus $\phi$, or equivalently $f(R)$, is
indeed a dynamical degree of freedom.
\newpage
\section{Geometrical results}
\label{sec:Int:fR:Geometrical results}
In this section we present different geometrical results obtained from the 
modified Einstein equations which were obtained in Section 1.3. Particular interest will be devoted in Subsection \ref{subsect:fR:Vacuum} to vacuum solutions. Then, the possibility of mimicking the usual GR results using $f(R)$ functions will be addressed in Subsection \ref{subsect:fR:EH_solutions_reproduced}. These results were originally presented in \cite{BH_Dombriz}.

\subsection{Vacuum solutions}
\label{subsect:fR:Vacuum}
Let us consider the EH action \eqref{action_EH} in $D$ dimensions with nonvanishing cosmological constant. In this case the equations \eqref{LCDM_fieldequations} can be studied in vacuum, i.e. $T_{\mu\nu}$ vanishes for all its components and therefore
\begin{eqnarray}
 R_{\mu\nu}-\frac{1}{2}R\,g_{\mu\nu}+\frac{D-2}{2}\Lambda_D\,g_{\mu\nu}=0
\label{LCDM_eqn}
\end{eqnarray}
whose solutions satisfy
\begin{eqnarray}
R_{\mu\nu}=\Lambda_D g_{\mu\nu}\,\,\,;\,\,\,R= D \Lambda_D
\label{R_vacuum}
\end{eqnarray}
which motivated our choice for $\Lambda_D$ normalization in Section 1.3. Equations
\eqref{R_vacuum} provide the conditions to be accomplished by a metric $g_{\mu\nu}$
to allow vacuum solution in this case. If now one considers
the $f(R)$ general case provided by the equations \eqref{fieldtensorialequation},
one may wonder about the condition for the existence of constant curvature solutions, $R_0$ from now on, in a vacuum scenario. Thus, the equations \eqref{fieldtensorialequation} may be simplified to become
\begin{eqnarray}
 R_{\mu\nu}\,(1+f_R)-\frac{1}{2}\,g_{\mu\nu}\,(R+f(R))\,=\,0.
\end{eqnarray}

Note that the term involving ${\cal D}_{\mu\nu}f_R$ in \eqref{fieldtensorialequation} has disappeared since it vanishes when constant curvature is assumed. Taking the trace in the previous equation we get
\begin{equation}
2R\,(1+f_R)-D\,(R+f(R))\,=\,0.
\label{dif}
\end{equation}
If $R_0$ is a root of the previous equation,
%
an effective cosmological constant may be defined as $\Lambda_D^{eff}\equiv R_{0}/D$. Provided
the condition
$1+f'(R_0)\neq 0$ is satisfied, $R_0$ fulfills:
\begin{eqnarray}
 R_{\mu\nu}\,=\,\frac{R_{0}+f(R_0)}{2(1+f_{R}(R_0))}\,g_{\mu\nu}.
\end{eqnarray}
\\
Let us illustrate this procedure considering a simple model:
\begin{eqnarray}
f(R)=\frac{g_1}{R}+g_0
\end{eqnarray}
which has been widely studied in the literature
(see for instance \cite{Carroll_D70_2004} where $D=4$ and $g_0=0$). Then the constant curvature solutions -- for an arbitrary number of dimensions $D$ -- are
\begin{eqnarray}
R_0\,=\,\frac{-D g_0\pm\sqrt{D^2(g_0^2-4g_1)+16 g_1}}{2(D-2)}
\end{eqnarray}
which reduce for $D=4$ to the expression
\begin{equation}
R_0=-g_0\pm\sqrt{g_0^2-3g_1}\,.
\end{equation}

For the EH case in $D=4$ with cosmological constant $\Lambda_{}\equiv\Lambda_{4}$, i.e. $g_1=0$ and $g_0=-2\Lambda_{}$, the
constant curvature solutions are both $R_0=4\Lambda_{}$ and $R_0=0$ and for the vanishing cosmological constant case, i.e. $g_0=0$,  $R_0=\pm\sqrt{-3g_1}$ is obtained.

As a different approach, one can consider equation \eqref{dif} as a differential equation for
the $f(R)$ function so that the corresponding
solution would admit any curvature $R$ value. The
solution of \eqref{dif} is just:
\begin{equation}
f(R)\,=\,\alpha R^{D/2}-R
\end{equation}
where $\alpha$ is an arbitrary constant. Thus the gravitational action \eqref{action} becomes
\begin{equation}
S_G\,=\,\frac{\alpha}{2\kappa}\int \text{d}^Dx\sqrt{\mid g\mid}\,R^{D/2}
\label{RD2}
\end{equation}
which has solutions of constant curvature for arbitrary $R$. The
reason is that this action is scale invariant since the ratio $\alpha/\kappa$ is a dimensionless constant.

\subsection{Some EH solutions reproduced by $f(R)$ theories}
\label{subsect:fR:EH_solutions_reproduced}
Now we shall address the issue of finding some general criteria to mimic, by using
general $f(R)$ gravities, some solutions of the EH action not necessarily of constant
scalar curvature and either with or without a cosmological constant term.

Let the metric tensor $g_{\mu\nu}$ be a solution of EH gravity with cosmological constant, i.e. such that the equations \eqref{LCDM_fieldequations} are fulfilled. Then the same metric tensor $g_{\mu\nu}$ will be a solution for \eqref{fieldtensorialequation}
provided the following compatibility equation
\begin{eqnarray}
f_{R}\,R_{\mu\nu}-\frac{1}{2}\,g_{\mu\nu}\left[f(R)+(D-2)\Lambda_D\right]+{\cal D}_{\mu\nu}f_{R}\,=\,0
\label{comp}
\end{eqnarray}
is fulfilled. Note that the fluid content comprised in $T_{\mu\nu}$ has been considered to be strictly the same as in \eqref{LCDM_fieldequations}. This allowed us to cancel this term out in order to obtain the compatibility equation \eqref{comp}. In  Section \ref{sect:fR:Reconstruction}, a slight deviation of fluid contents between EH and $f(R)$ approaches will be permitted.

Some particularly interesting cases in which to apply this approach are the following:
\begin{enumerate}
\item The simplest case is obviously vacuum, i.e. $T_{\mu\nu}\equiv 0$, with vanishing cosmological constant $\Lambda_{D}=0$.
Then the equations \eqref{LCDM_fieldequations} become:
\begin{equation}
R_{\mu\nu}=\frac{1}{2}R g_{\mu\nu}
\end{equation}
which imply $R_0=0$ and $R_{\mu\nu}=0$. Consequently $g_{\mu\nu}$ is also a solution of any $f(R)$ gravity provided the following condition
\begin{eqnarray}
f(0)\,=\,0
\end{eqnarray}
is accomplished as seen from \eqref{comp}. This is for instance the case
if $f(R)$ is analytical around $R=0$ and it can be written as follows:
\begin {equation}
f(R)\,=\,\sum_{n=1}^{\infty}f_{n}\,R^{n}.
\end{equation}
 \item If the cosmological constant is different from zero ($\Lambda_{D}\neq 0$), but still 
$T_{\mu\nu}\equiv0$, the constant curvature results given in \eqref{R_vacuum} are again obtained. Then the compatibility equation \eqref{comp} reduces to \eqref{dif} with $R_0=D\Lambda_D$. In other words, $g_{\mu\nu}$ is also a solution of the $f(R)$ case provided
\begin{eqnarray}
f(D\Lambda_D)\,=\,\Lambda_D(2-D+2\,f_{R}(D\Lambda_D)).
\label{cond_LambdaD_vs_f}
\end{eqnarray}

Notice also that in this situation, i.e. nonvanishing $\Lambda_D$ and vacuum, according to the result in \eqref{RD2} there would also be a solution for any $R_0$ in the particular case  $f(R)\,=\,\alpha\,R^{D/2}-R$.
\\
\\
\item If the considered case is $\Lambda_D\,=\,0$ and conformal matter ($T\,\equiv\,T^{\mu}_{\,\mu}\,=\,0$), then the equations \eqref{LCDM_fieldequations} would imply
\begin{eqnarray}
R_0\,=\,0\,\,\,;\,\,\, R_{\mu\nu}\,=\,-\kappa\,T_{\mu\nu}
\label{eq_conformal_matter}
\end{eqnarray}
which will have a metric tensor $g_{\mu\nu}$ as solution.
Therefore, provided
\begin{eqnarray}
 f(0)\,=\,0\,\,\,;\,\,\,f_{R}(0)\,=\,0,
\end{eqnarray}
the same $g_{\mu\nu}$ is also a solution of any $f(R)$ gravity. This result could have particular interest in
cosmological  calculations for ultrarelativistic matter (i.e. conformal) dominated universes.
\\
\item Again in the conformal matter case with nonvanishing $\Lambda_D$, constant curvature
\begin{eqnarray}
R_0\,=\,D \Lambda_{D}
\end{eqnarray}
is a solution for \eqref{LCDM_fieldequations} for a given metric $g_{\mu\nu}$ which is also a solution of $f(R)$ provided again that the condition \eqref{cond_LambdaD_vs_f} is satisfied.
\\
\item Finally for the general case with no assumption about $\Lambda_D$ nor about $T_{\mu\nu}$, the metric tensor $g_{\mu\nu}$ will be a solution for any $f(R)$ gravity but for a modified energy momentum tensor $\overline{T}_{\mu\nu}$ given by:
\begin{eqnarray}
\overline{T}_{\mu\nu}\,&\equiv&\,T_{\mu\nu}-\frac{1}{\kappa}\left\{f_{R}\,R_{\mu\nu}
-\frac{1}{2}\left[f(R)+(D-2)
\Lambda_D\right]g_{\mu\nu}+{\cal D}_{\mu\nu}f_{R}\right\}.\nonumber\\
&&
\end{eqnarray}
\end{enumerate}
\section{Constraints on $f(R)$ theories to ensure viability}
\label{sec:Int:fR:Constraints}
$f(R)$ gravity models turn out to be severely constrained
in order to provide consistent theories of gravity. In this section
we review both cosmological and strictly gravitational
conditions presented in \cite{Silvestri_D77_2008}. Some
relevant bibliography will also be provided.

The usual four conditions that are required for a viable $f(R)$ theory are:

{\bf 1}. $f_{RR}\geq0$ for high curvatures \cite{Faraoni:2006}.
This is the requirement for a classically stable high-curvature
regime and for the existence of a matter dominated phase in the
cosmological evolution. In the opposite case, an instability, referred to in the
literature as the 'Dolgov-Kawasaki'  or  'Ricci scalar'  or  'matter'
instability, would appear. Indeed, if $f_{RR}$ is smaller than zero,
then the extra degree of freedom of the theory would behave as a
ghost.
This stability condition  may also be recovered in studies of cosmological perturbations \cite{Hu&Sawicki_May_2007} and it can be given a simple physical interpretation as in \cite{Faraoni:2007} where if an effective $D$ dimensional gravitational constant is defined as
$G_{eff}\equiv G_{D}/(1+f_R)$ then
\begin{eqnarray}
\frac{\text{d}G_{eff}}{\text{d}R}\,=\,-\frac{f_{RR}}{(1+f_R)^2}G_{D}.
\label{G_eff_derivative}
\end{eqnarray}
It is easy to notice from the previous equation that if
$f_{RR}<0$, $G_{eff}$ would increase as $R$ grows since $R$
itself generates larger and larger curvature via equation
\eqref{trace_eqn}. Such a mechanism would act to
destabilize the theory with no stable ground state since if a
small curvature starts growing it will do so without limit and the
system would run away.
If on the contrary $f_{RR}\geq0$, a negative
feedback mechanism operates to compensate
the growth of $R$ and consequently the runaway
behaviour will not appear
\footnote{Note that in this analysis
$1+f_R$ has been supposed to be positive (i.e. $G_{eff}>0$) as will be
required from the second condition below to ensure viability.}.

{\bf 2}. $1+f_{R}>0$ for all Ricci scalar curvature values. This
condition ensures the effective Newton's constant to be positive
at all times as can be seen from equation
\eqref{fieldtensorialequation_Einstein_manner} and the graviton
energy to be positive. This condition will also be proven in
Chapter \ref{chap:Black holes} to be required to recover standard
thermodynamics of Schwarzschild-anti-de Sitter BHs in $f(R)$
theories.

{\bf 3}. $f_{R}<0$ ensures ordinary GR behaviour is recovered at
early times. Together with the condition $f_{RR}>0$, it implies that
$f_{R}$ should be negative and a monotonically growing function of $R$
in the range $-1<f_{R}<0$.

{\bf 4}. $\vert f_{R}\vert\ll 1$  at recent epochs. This is imposed
by local gravity tests \cite{Hu&Sawicki_May_2007}, although it is still not clear what is the actual
limit on this parameter and some controversy still remains about the required $|f_{R}|$ value
\cite{Cembranos2006, Olmo2005}. This condition also implies that the cosmological
evolution at late times resembles that of $\Lambda\text{CDM}$. In any case, this constraint
is not required if one is only interested in building models for cosmic acceleration.

Let us summarize this section by saying that viable $f(R)$ models
can be constructed to be compatible with local gravity tests and
other cosmological constraints \cite{varia}.
\section{Brane-world theories}
\label{sec:Int:BW:BW theory}
%
As mentioned in the Motivation section, many of the SM extensions try to
solve open issues in modern physics such
as the {\it hierarchy problem}. Some approaches try to answer those questions
by introducing extra spatial dimensions, where the number of dimensions of the total space (bulk space) $D=4+\delta$. Those attempts were
first proposed independently by Kaluza \cite{Kaluza_1921} and Klein \cite{Klein_1926} and many proposals
followed throughout the twentieth century \cite{R&S:1999:1&2,ADD}.


First proposals for large extra dimensions were provided in \cite{ADD}: in this model, the SM matter is confined
in a spatial 3-dimensional manifold and the brane itself is considered not to be a gravitational source. Hence
the background metric is assumed to be Minkowskian. Gravitational fields are the only fields able to propagate
through the whole bulk space. Therefore gravity also propagates in the extra $\delta$ dimensions which are for simplicity often compactified in a toroidal shape whereby all extra dimensions acquire a radius $R_B$.

One of the most important consequences of this hypothesis is the relation between the fundamental gravitational scale in $D$ dimensions $M_D$ and the Planck scale $M_{P}$ which is not a fundamental constant any more but the effective gravitational constant in the theory reduced to 4 dimensions. In fact one may write
\begin{eqnarray}
M_{P}^2\,\equiv\,V_{\delta}\,M_{D}^{2+\delta}
\label{M_F}
\end{eqnarray}
where $V_{\delta}$ is the compactified volume in $\delta$ dimensions \cite{ADD}, for instance in the toroidal case
$V_{\delta}=(2\pi R_B)^{\delta}$. The expression \eqref{M_F} allows to reduce the fundamental gravitational scale to
the electroweak scale, $M_{D}\sim\,$TeV, if extra dimensions are large enough. For instance, compactification scales
of $R_{B}^{-1}\sim 10^{-3}\,\text{eV}$ to 10 MeV provide this effect for extra dimensions $\delta\sim2$ and 7
respectively. By reducing the fundamental scale to $M_{D}$, gravitational effects may be detectable in experiments
involving energies of this order \cite{ADD} as will be explained in the Section 1.9.
\section{Excitations in brane worlds: branons}
\label{sec:Int:BW:Excitations in BW:Branons}
Since no relativistic object may be considered as rigid in relativistic theories, the
3-brane, when embedded in the total space-time,
may present fluctuations. These fluctuations were originally studied in \cite{Sundrum}. In the extra dimensions
of the BW models, such fluctuations are usually referred to as branons. They give rise to new states whose low energy
dynamics has been widely studied \cite{Bando:1999,Kugo:1999,Creminelli:2000}.
A vast bibliography can be found dealing with branons \cite{DM, Branons:Theory},
their predicted detection in future colliders experiments
\cite{Branons:Colliders}  and the explanation that they may provide for the origin of dark matter \cite{DM2, Branons:DM}.

Let us consider a $D$ dimensional bulk space $\mathcal{M}_D$ wherein the brane lies embedded and
that for simplicity we shall assume to be factorized in the form $\mathcal{M}_{D}=\mathcal{M}_{4}\times B$
where $\mathcal{M}_4$ is a 4-dimensional space-time and $B$ is a $\delta$-dimensional compactified manifold. The
brane is therefore assumed to lie on the $\mathcal{M}_4$ space-time manifold. As already mentioned, the gravitational
contribution of the brane itself will not be considered.

Let us denote the coordinates over the manifold $\mathcal{M}_{D}$ as $\{x^{\mu},\,y^{m}\}$
with $\mu=0,1,2,3$ and  $m=1,2,...,\delta$ and
the ansatz for the total space $\mathcal{M}_{D}$ bulk metric will be
\begin{eqnarray}
G_{MN}=\left(
\begin{array}{cc}
\tilde g_{\mu\nu}(x)&\\ &-\,\tilde g'_{mn}(y)
\end{array}\right)
\label{metric_brane}
\end{eqnarray}
with signature $\left(+,-,-,-\,;\,-,...,-\right)$.

In the absence of the 3-brane, this metric possesses an isometry group that is assumed to
be of the form $G(\mathcal{M}_D)=G(\mathcal{M}_4)\times G(B)$. The presence
of the brane spontaneously breaks the symmetry to some subgroup
$G(\mathcal{M}_4)\times H$ with $H\subset G(B)$ some subgroup of $G(B)$. Therefore the quotient space
$K\,=\,G(\mathcal{M}_{D})/(G(\mathcal{M}_4)\times H)\,=\,G(B)/H$ may be defined.

\begin{figure}
\begin{center}
{\epsfxsize=10.0cm\epsfbox{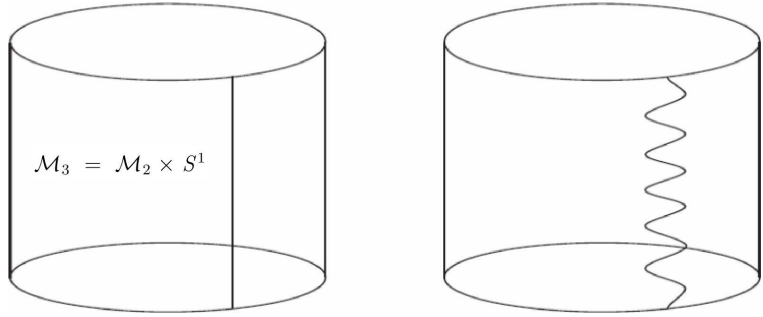}}
 \end{center}
\caption{Brane with trivial topology in
$\mathcal{M}_3\,=\,\mathcal{M}_2 \times {\it S}^{1}$ as originally
presented in reference \cite{CDM}. The fundamental brane state is
plotted on the left whereas on the right side an excited state is
presented.} \label{figura21_BW}
\end{figure}

The position of the brane can be parameterized as
$Y^{M}(x)\equiv\{x^\mu,Y^{m}(x)\}$ where the first four coordinates of the total
space have been chosen to be the space-time coordinates
corresponding to the brane $\{x^{\mu}\}$. Let us assume that the
brane is located at a point on $B$, i.e., $Y_{0}\equiv\{Y^{m}(x)\}$
corresponds to the fundamental state of the brane. In this case its
induced metric in the ground state is just $g_{\mu\nu}\equiv
\tilde g_{\mu\nu}\equiv G_{\mu\nu}$. However, when brane
excitations (branons) are present, the induced metric becomes
\begin{eqnarray}
g_{\mu\nu}\,=\,\partial_{\mu}Y^{M}\partial_{\nu}Y^{N}\,G_{MN}\,=\,
\tilde g_{\mu\nu}-\partial_\mu Y^m \partial_\nu Y^n \tilde g'_{mn}.
\label{induced_metric}
\end{eqnarray}
This situation may be illustrated by the simple Figure
\ref{figura21_BW} where a 1-brane (string) is represented within a
total space with two spatial coordinates
$\mathcal{M}_{3}\,=\,\mathcal{M}_{2}\times S^{1}$.

Since the brane creation mechanism is in principle unknown, or at least out of the scope of the present
section, let us assume that the brane dynamics is described by an effective action, so
we are allowed to consider for this action the most general expression which is invariant
under brane coordinates reparametrizations. Therefore, it is very common to perform
an expansion in derivatives of the induced metric given by
equation \eqref{induced_metric} to describe the brane dynamics.
Then, the first order of this effective action would describe the
brane dynamics at low energies and it is usually referred to as
the Dirac-Nambu-Goto ($NG$) action:
\begin{eqnarray}
S_{NG}\,=\,-f^4\int \mbox{d}^4x\sqrt{|g|}
\label{action_NG}
\end{eqnarray}
where a constant $f$ with energy units appears, which
may be identified with the brane tension $\tau\equiv f^4$ and $\text{d}^4 x \sqrt{|g|}$
is the brane volume element.
As was mentioned above, the presence of the brane will break any existing isometry of $B$
except those which leave the point $Y_{0}$ on $B$ invariant. In other words, the group $G(B)$
is spontaneously broken to $H(Y_{0})$ denoting the $Y_{0}$ isotropy group.

The brane excitations with respect to the broken Killing fields in $B$ correspond to the zero modes
and they are parameterized by the branon fields $\pi^{\alpha}(x)$,
$\alpha=1,...,k$ where $k\equiv\text{dim}(G(B))-\text{dim}(H)$.
These fields $\pi^{\alpha}(x)$ may be interpreted like the
corresponding coordinates in the quotient manifold $K=G(B)/H$.

In particular, for a fundamental state independent of the position $Y_{0}$ of the brane in the $B$ space, the
action of an element of $G(B)$ over $Y_{0}$ will take $Y_0$ to the point on $B$ with coordinates
\begin{eqnarray}
Y^{m}(x)\,\equiv\,Y^{m}(Y_{0}, \pi^{\alpha}(x))\,=\,Y_{0}^{m}+\frac{1}{\sqrt{2\kappa} f^2}\,\xi_{\alpha}^{m}(Y_0)\pi^{\alpha}(x)+\Od(\pi^2)
\label{Ym_transformed}
\end{eqnarray}
where
the branons fields normalization
is performed through  $\kappa=8\pi/M_{P}^2$. At this stage it is important to stress that coordinates for the transformed point given by \eqref{Ym_transformed} only depend on $\pi^{\alpha}(x)$, i.e., on the corresponding transformation parameters of the broken generators.

If $B$ is considered to be an homogeneous space, the isotropy
group does not depend on the particular chosen point where the
brane lies, i.e. $H(Y_0)=H$. In this case $B$ is homeomorphic
to the coset $K=G(B)/H$ which is the space of the Goldstone bosons
associated to the spontaneous isometry breaking - transverse
translations  - produced by the presence of the brane. Thus the
transverse translations of the brane - branons - can be considered
as Goldstone bosons on the coset $K$ and the branon fields can be
defined as coordinates  $\pi^\alpha$ on $K$, which are chosen to be proportional
to $B$ coordinates, since the number of Goldstone bosons is equal
to $\text{dim}(B)$, as:
\begin{eqnarray}
\pi^{\alpha}\,=\,\frac{v}{R_B}\delta^{\alpha}_{m}\,Y^{m}
\label{coordinates_pi_alpha}
\end{eqnarray}
where
\begin{eqnarray}
v=f^2 R_{B}
\label{v_f_R_B}
\end{eqnarray}
is the typical size of the coset $K$ and $R_{B}$ is the typical size, in length units, of the compactified space $B$.

Therefore, according to the previous assumption \eqref{coordinates_pi_alpha}, it is obvious that
\begin{eqnarray}
\partial_{\mu}Y^{m}(x)\,=\,\frac{\partial Y^{m}}{\partial\pi^{\alpha}}\,\partial_{\mu}\pi^{\alpha}\,=\,\frac{1}{\sqrt{2\kappa} f^2}\,\xi^{m}_{\alpha}(Y_0)\,\partial_{\mu}\pi^{\alpha}\,+\,\Od(\pi^2)
\label{Y_derivatives}
\end{eqnarray}
and the induced metric on the brane \eqref{induced_metric} is rewritten in terms of the branon fields $\pi$ as
\begin{eqnarray}
g_{\mu\nu} \,=\,\tilde g_{\mu\nu}-\frac{1}{f^4}\,h_{\alpha\beta}(\pi)\partial_{\mu}\pi^{\alpha}\,\partial_{\nu}\pi^{\beta}
\label{induced_metric_2}
\end{eqnarray}
where $h_{\alpha\beta}$ is the $K$ metric which is easily obtained from the $B$ metric
\begin{eqnarray}
h_{\a\b}(\pi)\,=\,f^4\,\tilde g'_{mn}(Y(\pi))\,\frac{\partial Y^m}{\partial \pi^{\a}}\frac{\partial Y^n}{\partial \pi^{\b}}
\label{h_definition}
\end{eqnarray}
as explained in \cite{DM}. In more complicated cases
in which translational isometries in the bulk space are
not only spontaneously but also explicitly broken, the
metric $g_{\mu\nu}$ could also be a function of the extra dimension
coordinates $\{y^{m}\}$.  Then it is possible to show that branons may become
massive. In fact in \cite{DM2} these massive branons were shown to
behave as WIMPs and thus form
natural candidates for dark matter in this kind of scenario.

Therefore for small brane excitations in a background metric $\tilde g_{\mu\nu}$, the effective action
\eqref{action_NG} can be expressed as a derivative expansion as follows:
\begin{eqnarray}
S_{eff}[\pi]\,=\,S_{eff}^{(0)}[\pi]+S_{eff}^{(2)}[\pi]+S_{eff}^{(4)}[\pi]+...
\label{S_eff}
\end{eqnarray}
where the corresponding zeroth order is
\begin{eqnarray}
S_{eff}^{(0)}[\pi]\,=\,-f^4\int_{\mathcal{M}_4}\text{d}^4 x\sqrt{|\tilde g|}.
\label{S_eff_0}
\end{eqnarray}
Note that $S_{eff}^{(2)}[\pi]$ and $S_{eff}^{(4)}[\pi]$ hold for
contributions to the effective action containing two and four
derivatives of the branon fields respectively. Let us finish this
digression by remarking that the term $S_{eff}^{(2)}[\pi]$, with
two field derivatives, is
nothing but the 
non-linear sigma model 
action associated to the coset space $K$.

\section{Brane-skyrmions}
\label{sec:Int:BW:Skyrmions}
Apart from branons, the brane may support other states due to the nontrivial homotopies of the coset space $K$ such as strings, monopoles or skyrmions. This fact
appears due to the possibility of wrapping around the extra dimension space $B$ giving rise to nontrivial topological configurations as
was studied in detail in the reference \cite{CDM}.

In fact, texture-like configurations, called brane-skyrmions, arise when the third homotopy group of $K$ is nontrivial. In particular, for
\begin{eqnarray}
\pi_3(B)=\pi_3(K)=\Bbb{Z},
\end{eqnarray}
the third homotopy group will be the minimal one supporting the
existence of those nontrivial configurations\footnote{Note that
$\pi_3(B)=\pi_3(K)$ if $B$ is an homogeneous space.}.
\begin{figure}
\begin{center}
{\epsfxsize=4.0cm\epsfbox{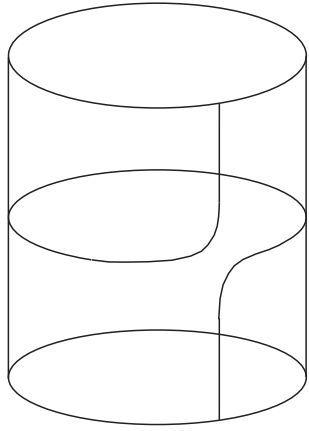}}
 \end{center}
\caption{Brane-skyrmion configuration with $n_{W}=1$ and
nonvanishing size in $\mathcal{M}_3\,=\,\mathcal{M}_2 \times {\it S}^{1}$
as originally presented in \cite{CDM}.}
\label{figura_BW_skyrmions}
\end{figure}
Those brane-skyrmions can be nicely understood in geometrical
terms as some kind of holes \cite{Shifman} in the brane which make it possible to
pass through them along the $B$ space. This is because in the core
of the topological defect the symmetry is reestablished. In
particular, in the case we are interested in, the broken symmetry
is basically the translational symmetry along the
extra-dimensions.

In order to simplify the calculations, we shall consider an homogeneous compactified manifold $B$
and the coset space $K$ homeomorphic to $SU(2)$ and equivalently to $S^3$. Then
\begin{eqnarray}
B\simeq K \simeq SU(2) \simeq S^3.
\end{eqnarray}
This fact allows to hold a third homotopy group $\Bbb{Z}$.

Furthermore, let us introduce spherical coordinates on both spaces, $\mathcal{M}_4$ and $K$ as follows: in
$\mathcal{M}_4$ we denote the coordinates $\{t,r,\theta,\varphi\}$ with $\phi
\in [0,2\pi)$, $\theta \in[0,\pi]$ and $r \in [0,\infty)$. On the
coset manifold $K$, the spherical coordinates are denoted
$\{\chi_K,\theta_K,\phi_K\}$ with $\phi_K \in [0,2\pi)$, $\theta_K
\in [0,\pi]$ and $\chi_K \in [0,\pi]$. Notice that
such coordinates cover the whole spherical manifolds and relate
to the physical branon fields (local normal geodesic coordinates on
$K$) by:
\begin{eqnarray}
\pi_1&=&v \sin\chi_K\sin\theta_K\cos\phi_K,\nonumber\\
\pi_2&=&v \sin\chi_K\sin\theta_K\sin\phi_K,\label{esfk}\\
\pi_3&=&v\sin\chi_K\cos\theta_K.\nonumber
\label{eqns_pi}
\end{eqnarray}

The coset metric in spherical coordinates is written as
\begin{eqnarray}
h_{\alpha\beta}=\left(
\begin{array}{ccc}
v^2&\\ &v^2\sin^2(\chi_K) &\\ & & v^2
\sin^2(\chi_K)\sin^2(\theta_K)
\end{array}\right).
\end{eqnarray}
\subsection{Static brane-skyrmions}
For static configurations, it can be proven that the mass for the brane-skyrmion may be obtained directly as:
\begin{eqnarray}
M[\pi]\,=\,-\int_{\mathcal{M}_3}\text{d}^3 x\,\mathcal{L}_{eff}\,=\,f^4 \int_{\mathcal{M}_3} \text{d}^3 x \sqrt{|g|}
\end{eqnarray}
where the effective Lagrangian comes from the expression \eqref{action_NG}. 
In general this expression is divergent due to the contribution of the zeroth order term when $\sqrt{|g|}$ is expanded in branon fields derivatives, reflecting
the fact that the brane is an infinite object
with finite tension. To prevent that, we substract this
term in order to get a brane-skyrmion finite mass, i.e.:
\begin{eqnarray}
M_{S}[\pi]\,\equiv\,M[\pi]-M[0]\,=\,f^4\int_{\mathcal{M}_3}\text{d}x^3\sqrt{|g|}-M[0].
\label{skyrmion_mass}
\end{eqnarray}
%
%
%
%
As was shown in the previous section, the  $\pi^{\a}$ fields are mappings from
the $\mathcal{M}_{4}$ manifold to the coset manifold $K$. For static, i.e. time independent, field configurations, these
could be understood as mappings from the corresponding spatial 3-dimensional hypersurface
($\mathcal{M}_3$) to the coset space ($S^3$ in the case we are studying). For finite energy
configurations, fields should vanish at the spatial infinity and $\mathcal{M}_{3}$ can be
compactified to $S^3$. Therefore one may write $\pi^{\alpha}: S^3 \rightarrow S^3$. Since the
third homotopy group of $S^3$ is $\Bbb{Z}$, the mappings can be classified by an integer number $n_W$. Thus,
branons can be identified with the topologically trivial configurations $n_W=0$, whereas
those configurations with $n_W\neq 0$ will be denoted as brane-skyrmions.

Consequently, for static skyrmions this mapping may be implemented in the following way:
\begin{eqnarray}
\phi_{K}\,=\,\phi\,\,\,\,;\,\,\,\,
\theta_{K}\,=\,\theta\,\,\,\;\,\,\,\,
\chi_{K}\,=\,F(r)
\end{eqnarray}
with the boundary conditions $F(0)-F(\infty)=n_{W}\pi$ for a winding number $n_{W} \neq 0$.

In this case, $M_{S}[\pi]$ may be written as a $F(r)$ functional
and the correct mass for this kind of skyrmions is obtained by
minimizing $M_{S}[F]$ in the space function with adequate boundary
conditions. 

From expression \eqref{skyrmion_mass}, it can be proven that the skyrmion is point-like, stable and its
mass becomes:
\begin{eqnarray}
M_{S}\,=\,2\pi^2 f^4 R_{B}^3
\end{eqnarray}
\newpage
\section{Gravitational signatures at the LHC}
\label{sec:Int:BW:BH}
As already mentioned, it can be
seen from the equation \eqref{M_F}, the fundamental gravitational
scale could be as low as the electroweak scale if extra dimensions are
large enough. Since the LHC is operating at a center of mass
energy of $\sqrt{s}=14\,\text{TeV}$, if the fundamental scale of
gravitation is $M_{D}\sim\text{TeV}$, both the production and
decay of Schwarzschild mini BHs at high ratio becomes possible
\cite{Dimopoulos}.

These BHs once produced would decay into SM particles with a clean
signature and a low background. Several features of such objects
could then be extracted from experimental data: for instance BH
masses $M_{\text{BH}}$ may be determined very precisely due to the
absence of missing energy and their temperature could be extracted
from the energy spectrum of the products. Thus the correlation
between these two quantities may provide relevant information able
to determine the number of extra dimensions, and therefore the
fundamental scale of gravity. On the other hand the Hawking
evaporation law could be tested experimentally.

The total cross section when two partons collide at the LHC with an impact
parameter less than the Schwarzschild radius $R_S$ is of order
\begin{eqnarray}
\sigma(M_{\text{BH}})\,\approx\,\pi R_{S}^2=\frac{1}{M_{D}^2}\left[\frac{M_{\text{BH}}}{M_{D}}\left(\frac{8\Gamma(\frac{D-1}{2})}{D-2}\right)\right]^{\frac{2}{D-3}}
\label{BH_cross_section}
\end{eqnarray}
and it does not contain small coupling constants. If
$M_{D}\sim\text{TeV}$ the cross section is of order
$\text{TeV}^{-2}\approx 400\,\text{pb}$ and therefore BHs will be
produced copiously. The total production cross section ranges from
0.5 nb for $M_D=2\,\text{TeV}$, $D=11$ to 120 fb for
$M_{D}=6\,\text{TeV}$ $D=7$. For $M_{D}\sim1\,\text{TeV}$, the LHC
-- with a peak of luminosity of $30\,\text{fb}^{-1}/\text{year}$
-- will produce $10^7$ BH/year.

Experimental signatures rely on two qualitative properties: on the
one hand, the absence of small couplings as seen from expression
\eqref{BH_cross_section} and on the other hand, the flavor
independence nature of BHs decays as will be explained in the
following paragraph. Note that when $M_{\text{BH}}$ approaches
$M_D$, some stringy corrections to the previous assumptions may arise
but semiclassical arguments remain valid as long as
$M_{\text{BH}}\gg M_{D}$.

Once the BHs have been produced they decay following a process
governed by their Hawking temperature $T_{H}\sim1/R_{S}$ with an
associated wavelength $\lambda=2\pi/T_{H}$ larger than the BH size
and therefore BHs would emit, in a first approximation, as point
radiators mostly in the s-waves. This indicates that BHs decay
equally to particles on the brane and in the bulk since the decay
is only sensitive to the radial coordinate. If the approximation
$M_{\text{BH}} \gg T_{H}$ is made, the average multiplicity of
particles $\langle N \rangle$
 produced in the BH evaporation is given by:
\begin{eqnarray}
\langle N \rangle\,=\,
\frac{2\sqrt{\pi}}{D-3}\left(\frac{M_{\text{BH}}}{M_{D}}\right)^{\frac{D-2}{D-3}}\left(\frac{8\Gamma\left(\frac{D-1}{2}\right)}{D-2}\right)^{\frac{1}{D-3}}.
\label{mean_N_BH_evaporation}
\end{eqnarray}
Since the decay is thermal, it does not discriminate between particle species
(of the same mass and spin) and therefore BHs decay, roughly speaking, with the
same probability to all SM particles. The signal of hard primary leptons and hard photons
is quite clean with a negligible background since the production of SM leptons or photons
occur at much smaller rate than BH production \cite{Dimopoulos}.

The way to determine $M_{\text{BH}}$ and $T_{H}$ deals with the
study of decay products and the fits of the energy spectrum of
those products to the Planck formula respectively. Once those
two quantities are determined, they could provide some evidence of
the Hawking radiation and of the fact that the observed events indeed come
from BH evaporation and not from any other mechanism.

The relation between those two quantities, $M_{\text{BH}}$ and
$T_{H}$ obtained independently, may shed light about the
dimensionality of the space since it can be proved that
\begin{eqnarray}
\text{log}(T_{H})\,=\,-\frac{1}{D-3}\,\text{log}(M_{\text{BH}})+\text{constant}
\end{eqnarray}
where the constant does not depend on $M_{\text{BH}}$. Therefore
the previous equation provides a direct method to determine the
dimensionality $D$ of the space as the slope of this relation.

The experimental signatures outlined above allow us to state that if the
fundamental scale of gravitation is of order TeV, as suggested
in BW scenarios, some important physical consequences may appear. In fact, colliders study of BHs
-- eventually produced at a high rate in accelerators such as LHC -- could help
revealing the main features of physics in the vicinity of the electroweak scale or even determining
the total number of dimensions of the space-time.
\chapter[Dark energy in {\it f(R)} theories]{Dark energy in {\it f(R)} theories}
\label{chap:$f(R)$ theories}
\section{Introduction}
\label{sect:fR:Introduction}
%

As was commented in Chapter 1, when
the modified Einstein equations were rewritten {\it \`{a} la Einstein}, the presence of a function $f(R)$ in the gravitational sector modifying the usual EH Lagrangian may be understood as the introduction of an effective fluid which is not restricted to hold the usual energy conditions. Therefore $f(R)$ functions may be used
to explain the present cosmological acceleration. Historically, some $f(R)$ models were proposed to modify GR at short scales, i.e., high energies trying to explain inflation, as for instance $f(R)\propto R^{2}$, but no interest was paid in those models to provide a mechanism to cause late time acceleration. First attempts to induce cosmological acceleration considered $f(R)\propto 1/R$ but those models turned out to be in conflict with solar system tests \cite{Chiba} and even to be unstable when matter is introduced \cite{Dolgov&Kawasaki}.

Before studying these issues, let us mention that $f(R)$ models,
apart from satisfying those gravitational and cosmological
conditions given in Section \ref{sec:Int:fR:Constraints}, should verify some extra conditions of cosmological
viability. For instance,
they have to include a background evolution with Big Bang
nucleosynthesis (BBN) and both radiation and matter dominated
cosmological eras. This fact will be explicitly studied in this
chapter in Section \ref{sect:fR:Reconstruction}. On the other
hand, they must provide cosmological perturbations compatible with
cosmological constraints from CMB and large scale structures
(LSS). This fact will be studied thoroughly in Chapter 3.

The present chapter is organized as follows: in Section
\ref{sect:fR:Standard Einstein's eqns for FLRW} we shall revise
the standard approach to describe the cosmological evolution in
the $\Lambda\text{CDM}$ model in a homogeneous, isotropic and
spatially flat metric. In the following Section
\ref{sect:fR:Modified Einstein's equations for FLRW universe} we
shall generalize the usual Einstein equations when $f(R)$ gravity
theories are present. Then we shall study in Section
\ref{sect:fR:Cosmological viability} the cosmological viability
conditions for $f(R)$ theories to hold a dust matter dominated era
followed by a late time acceleration and some interesting models
which have been proposed to be viable will be provided in Section
\ref{sect:fR:Some viable f(R) models}. Then, Section
\ref{sect:fR:Reconstruction} will be the core of the chapter and
it will be devoted to study if $f(R)$ theories are able to mimic
standard $\Lambda\text{CDM}$ evolution. These $f(R)$ models will
possess vacuum solutions with null scalar curvature what allows to
recover some GR solutions usually considered.

To finish this chapter, we shall study in Section
\ref{sect:fR:Arbitrary fluid} how the modification of the
gravitational sector by $f(R)$ models may mimic the influence of
perfect fluids (parameterized by a constant equation of state) in
the cosmological evolution without any presence of such a fluid in the
fluid content.
%
%
The chapter will finish with Section \ref{sect:fR:Conclusions} by drawing some attention over
the main obtained conclusions.
%
%

The results presented in this chapter were originally published in
\cite{Dobado&Dombriz}.
\section{Standard Einstein equations in a FLRW universe}
\label{sect:fR:Standard Einstein's eqns for FLRW}
Since the {\it leitmotiv} of this chapter is to study cosmological solutions, our universe, which is assumed to be isotropic and homogeneous at large enough scales for fundamental observers, may be represented with a $D=4$ dimensional Friedmann-Lema\^{i}tre-Robertson-Walker (FLRW) metric
\begin{equation}
\text{d}s^2\,=\,\text{d}t^2-a^2(t)\left(\frac{\text{d}r^2}{1-kr^2}+r^2\text{d}\Omega_{2}^2\right)
\label{metric_t}
\end{equation}
expressed in cosmic time $t$ and where $a(t)$ is usually referred
to as the scale factor. Alternatively, this metric may be
expressed in conformal time $\tau$, defined by the relation
$\text{d}t\equiv a(\tau)\text{d}\tau$ and thus this metric becomes
\begin{equation}
\text{d}s^2\,=\,a^2(\tau)\left(\text{d}\tau^2-\frac{\text{d}r^2}{1-kr^2}-r^2\text{d}\Omega_{2}^2\right)
\label{metric_tau}
\end{equation}

In this metric, the Hubble parameter may be defined in either
cosmic or conformal time as
\begin{eqnarray}
H(t)\equiv \frac{\text{d}a(t)/\text{d}t}{a(t)}\equiv\frac{\dot{a}}{a} \,\,\,;\,\,\, \mathcal{H}\equiv\frac{\text{d}a(\tau)/\text{d}\tau}{a(\tau)}\equiv \frac{a'(\tau)}{a(\tau)}
\end{eqnarray}
respectively and the identity $aH\equiv\mathcal{H}$ is straightforwardly inferred.

For the values of the parameter $k$ smaller, equal or bigger than
zero, the universe is spatially hyperbolic, flat or spherical
respectively. In the following calculations we shall be
considering $k=0$. This choice is justified according to WMAP
data \cite{WMAP}
where the results obtained for $\Lambda\text{CDM}$ model are:
$\Omega_k\,\equiv\,-k/H_{0}^2$, with $H_{0}\equiv H(t_{today})= 100 h\,\text{km}\,\text{s}^{-1}\text{Mpc}^{-1}$ and $h=0.699\pm0.018$ and $-0.0133<\Omega_{k}<0.0084$ ($95\%$ CL). Therefore
terms related with $k$ will be subdominant in either Friedmann's
or generalized Friedmann equations to be presented in following
the section.

Considering the previously introduced metric \eqref{metric_t} and perfect fluids given by \eqref{perfect_fluid} for the present fluids, the only two independent Einstein equations for $D=4$ are the Friedmann and the acceleration equations respectively, which may be written in cosmic time as:
\begin{eqnarray}
H^2\,\equiv\,\left(\frac{\dot{a}}{a}\right)^2\,=\,\frac{8\pi
G}{3}\,\sum_{\alpha}\,\rho_{\alpha}
\label{FLRW_GR_eqn00}
\end{eqnarray}
\begin{eqnarray}
\frac{\ddot{a}}{a}\,=\,-\frac{8\pi G}{6}\sum_{\alpha}(\rho_{\alpha}\,+\,3P_{\alpha})
\label{FLRW_GR_eqnii}
\end{eqnarray}
where $G\equiv G_{4}$ is the gravitational constant in four dimensions and the
summation over subindex $\alpha$ holds for the present fluids contributions (baryons, radiation, dark matter, DE, etc.). If positive
cosmological acceleration is required, i.e. $\ddot{a}>0$, the condition to be accomplished from expression \eqref{FLRW_GR_eqnii} would
be $\sum_{\alpha}(\rho_{\alpha}\,+\,3P_{\alpha})<0$.
This condition would require that if only a perfect fluid is present, its state equation would satisfy the condition
$\omega_{\alpha}<-1/3$
which is not the case for standard fluids, such as for instance dust matter and radiation.
On the contrary a cosmological constant does provide positive acceleration in equation \eqref{FLRW_GR_eqnii}
since its state equation satisfies $\omega_{\Lambda}=-1$.

For this metric, the energy-momentum conservation equations
\label{motion_eqn} lead, in cosmic and conformal time
respectively, to the following equations:
\begin{eqnarray}
&&\dot{\rho}_{\alpha}+3(1+\omega_{\alpha})H\,\rho_{\alpha}\,=\,0
\nonumber\\
&&\rho_{\alpha}'+3(1+\omega_{\alpha})\mathcal{H}\rho_{\alpha}\,=\,0
\label{eq_dif_densidad}
\end{eqnarray}
which hold separately for each fluid whose state equations are
assumed to be $P_{\alpha}=\omega_{\alpha}\rho_{\alpha}$. Previous
equation is integrated to give:
\begin{equation}
\rho_{\alpha}(t)\,=\,\rho_{\alpha}(t_{0})\left(\frac{a(t_{0})}{a(t)}\right)^{3(1+\omega_{\alpha})}
\label{matterdensity}
\end{equation}
where $t_0$ is an arbitrary time and the corresponding scale
factor is $a(t_0)$.

By using the definition given in \eqref{Riemann} and
\eqref{Ricci&R}, the Ricci scalar curvature for FLRW spatially
flat metric is written in terms of the scale factor and its
derivatives as follows
\begin{equation}
R\,=\,6\left[\left(\frac{\dot{a}}{a}\right)^2
+\frac{\ddot{a}}{a}\right]
\,=\,\frac{6}{a^2}(\mathcal{H}'+\mathcal{H}^2).
\label{R}
\end{equation}

Let us finish this section by rewriting the previous Friedmann
equation \eqref{FLRW_GR_eqn00}. To do so, let us divide that equation by
$H_{0}^2\,\equiv\,H^{2}(t_0)$ and consider as present fluids dust
matter ($\omega_{M}=0$), radiation ($\omega_{\text{Rad}}=1/3$) and cosmological constant $\Lambda$
($\omega_{\Lambda}=-1$). Thus, taking into account
\eqref{matterdensity} for each present fluid, we get:
\begin{eqnarray}
\frac{H^{2}(t)}{H_{0}^2}\,=\,\Omega_{M}\,a(t)^{-3}+\Omega_{\text{Rad}}\,a(t)^{-4}+\Omega_{\Lambda}
\label{FLRW_GR_eqn00_simp}
\end{eqnarray}
where we have used the notation:
\begin{eqnarray}
\Omega_{M}\equiv \frac{8\pi G_{}\rho_{M}(t_{0})}{3H_{0}^2(t_0)}
\,\,\,; \,\,\,
\Omega_{\text{Rad}}\equiv\frac{8\pi
G_{}\rho_{\text{Rad}}(t_{0})}{3H_{0}^2(t_0)}
\,\,\,; \,\,\,
\Omega_{\Lambda_{}}\equiv \frac{\Lambda_{}}{3H_{0}^2 (t_{0})}
\label{omegas}
\end{eqnarray}
and the normalization of the scale factor $a(t_0)=1$ has been
used. Note that if expression \eqref{FLRW_GR_eqn00_simp} is
evaluated at $t=t_0$, then
$\Omega_{M}+\Omega_{\text{Rad}}+\Omega_{\Lambda_{}}\,\equiv\,1$.
\section{Modified Einstein equations in a FLRW universe}
\label{sect:fR:Modified Einstein's equations for FLRW universe}
Inserting the metric \eqref{metric_t} for $D=4$ in the equations \eqref{fieldtensorialequation} and assuming also energy-momentum tensor as given in \eqref{perfect_fluid} for a fluid with energy density $\rho_0$ and pressure $P_0$, the only independent modified Einstein equations are
\begin{eqnarray}
3(1+f_R)\frac{\ddot{a}}{a}-\frac{1}{2}(R+f(R))-3\frac{\dot{a}}{a}\dot{R} f_{RR}\,=\,-8\pi G\,\rho_0
\label{00sin}
\end{eqnarray}
\begin{eqnarray}
(1+f_R)(\dot{H}+3H^{2})-\frac{1}{2}(R+f(R))-\frac{1}{a}\frac{\text{d}}{\text{d}t}\left(a^2 \dot{R} f_{RR}\right)\,=\,8\pi G\,P_0
\label{iisin}
\end{eqnarray}
and in conformal time $\tau$, these equations are given by
\begin{eqnarray}
\frac{3\mathcal{H}'}{a^2}(1+f_{R})-\frac{1}{2}(R+f(R))
-\frac{3\mathcal{H}}{a^2}f_{R}' \,=\,-8\pi G\,\rho_{0}
\label{00_bg_eqn}
\end{eqnarray}

\begin{eqnarray}
\frac{1}{a^2}(\mathcal{H}'&+&2\mathcal{H}^2)(1+f_{R})
-\frac{1}{2}(R+f(R))-\frac{1}{a^2}(\mathcal{H} f_{R}'+f_{R}'')
=\,8\pi G\,P_0.
\label{ii_bg_eqn}
\end{eqnarray}
Remind that dot denotes here derivative with respect to time $t$ whereas $\tau$ derivative was denoted with prime.
A very useful equation to use in the following calculations
is the combination \eqref{ii_bg_eqn} minus \eqref{00_bg_eqn} which becomes
\begin{eqnarray}
2(1+f_{R})(-\mathcal{H}'+\mathcal{H}^2)+2\mathcal{H}f_{R}'-f_{R}''\,=\,8\pi G (\rho_{0}+P_0)a^{2}.
\label{bg_density_plus_pression}
\end{eqnarray}

At this stage we should note that, for instance, according to
equations  \eqref{00sin}, \eqref{iisin} together with \eqref{R},
it is clear
that modified Einstein equations are not second order in
derivatives any more, but at least third order in the scale factor
derivatives provided $f_{RR}\neq0$.
%
\section{Cosmological viability for $f(R)$ dark energy models}
\label{sect:fR:Cosmological viability}
In this section we shall revise the conditions that a model for DE
given by a $f(R)$ theory must fulfill in order to be
cosmologically viable: i.e., any viable $f(R)$ model should have a
matter dominated phase long enough to provide the adequate
cosmological evolution prior to a late time acceleration phase. As
a matter of fact, equations \eqref{00sin} and \eqref{iisin} with
dust matter as the unique present fluid, can be rewritten in the
form of a system of autonomous equations \cite{Amendola&al:2007}.
In that reference two variables, $m$ and $r$ are introduced as
follows:
\begin{eqnarray}
m\,\equiv\,\frac{R\,f_{RR}}{1+f_R}\,\,\,;\,\,\,r\,\equiv\,-\frac{R(1+f_R)}{R+f(R)}.
\end{eqnarray}
Both the dynamics and stability of that autonomous system are determined by six critical points $P_{1,...,6}$ -- according to the notation in \cite{Amendola&al:2007} -- that appear in the system resolution.
\subsection{Critical points and stability}
According to the results presented in \cite{Amendola&al:2007}, both points $P_5$ and $P_6$ satisfy
\begin{eqnarray}
m(r)=-r-1.
\label{m-r}
\end{eqnarray}
If in the previous equation
$m$ is assumed to be constant, the condition \eqref{m-r} holds
straightforwardly from two other equations in the autonomous
system. In this case the points $P_{2,...6}$ always exist while
$P_1$ and $P_4$ are present for values $m=1$ and $m=-1$
respectively. The critical points $P_5$ and $P_6$ which give the
exact matter era evolution, i.e., $a(t)\propto t^{2/3}$, exist
only for $m=0$ ($P_5$) or for $m=-(5\pm\sqrt{73})/12$ ($P_6$) but
the latter corresponds to a vanishing matter density and obviously
it does not give a standard matter era.

If on the contrary $m$ is not assumed to be constant, the number
of solutions depends on the particular $f(R)$ choice, but only
$P_{1,5,6}$ can be accelerated and only $P_{5}$ might give rise to
matter era. This last situation would require $m\simeq0$ to
resemble the standard matter era evolution. Summarizing the result
in this case, only trajectories passing near $P_5$ with
$m\backsimeq0$ at $r\backsimeq-1$ and landing on an accelerated
attractor would give a viable cosmological evolution.

\subsection{Classification of $f(R)$ models}
By studying all possible trajectories in the $m$ and $r$ variables of 
the already mentioned autonomous system, it can be
shown \cite{Amendola&al:2007} that a classification of $f(R)$ models can be based entirely upon geometrical properties of the curve $m(r)$.
These two variables allow to classify the $f(R)$ models in four
different classes: {\bf I}, {\bf II}, {\bf III} and {\bf IV}
depending on the existence of a standard matter epoch and a final
accelerated expansion as follows: on the one hand, the viable
matter dominated epoch requires
\begin{eqnarray}
m\approx+0\,\,\,;\,\,\,\frac{\text{d}m}{\text{d}r}>-1
\label{viable_matter}
\end{eqnarray}
at $r\approx-1$.
On the other hand, the late time acceleration epoch requires to fulfill one of the following two conditions:
a de Sitter acceleration follows the matter epoch if and only if
\begin{eqnarray}
{\bf 1.}\,\,\,\,0\leq m(r)\leq1\,\,\, \text{at}\,\,\, r\,=\,-2
\label{acceleration_viable_2}
\end{eqnarray}
whereas a non-phantom accelerated attractor follows the matter
dominated epoch if and only if
\begin{eqnarray}
{\bf 2.}\,\,\,\,m\,=\,-r-1\,\,\,;\,\,\, \frac{\sqrt{3}-1}{2}<m\leq1\,\,\,;\,\,\, \frac{\text{d}m}{\text{d}r}<-1.
\label{acceleration_viable_1}
\end{eqnarray}

For instance, according to the previous requirements over $m$ and $r$ variables, models of the type $f(R)=\alpha R^{-n}-R$ and $f(R)=\alpha R^{-n}$ do not satisfy these conditions for any $n>0$ and $n<-1$ and are consequently cosmologically nonviable.

The main features of each class of models are:\\
\begin{itemize}
\item {\bf Class I}: this class covers all the cases for which the curve $m(r)$ does not connect the accelerated attractor with the standard matter point $(r,m)=(-1,0)$ either because $m(r)$ does not pass near that matter point, i.e., $m(r\rightarrow-1)\neq0$, or because the branch of $m(r)$ that accelerates is not connected with the standard matter point. Moreover, instead of having a standard matter phase given by a scale factor $a(t)\propto t^{2/3}$,
these $f(R)$ models possess a peculiar scale factor behaviour $a(t)\propto t^{1/2}$ before accelerating epoch and are therefore unsuitable models.
\\
\item {\bf Class II}:  for these $f(R)$ models the $m(r)$ curve
does connect the upper vicinity $(m>0)$ of $(r,m)=(-1,0)$ with a
critical point able to provide acceleration. Therefore models here
have a matter epoch and are asymptotically equivalent (hardly
distinguishable) to $\Lambda\text{CDM}$ model
($\omega_{eff}\equiv-1-2\dot{H}/3H^2 =-1$), i.e., they are
asymptotically de Sitter and observationally acceptable. These
models satisfy both equations \eqref{viable_matter} and
\eqref{acceleration_viable_2}.
\\
\item {\bf Class III}: these $f(R)$ models may possess an
approximated matter era but as a transient state followed by a
final and strongly phantom attractor at late-time. This is due to
the fact that the $m(r)$ curve intersects the critical line
$m(r)=-r-1$ at $-1/2<m<0$. The approximated matter era is a very
fast transitient phase and only a narrow range of initial
conditions may allow it. Since matter era is practically unstable,
these models are generally ruled out by the observations.
\\
\item {\bf Class IV}:  for models of this class the connection between the upper vicinity of the point $(r,m)=(-1,0)$ to the region located on the critical line $m(r)=-r-1$ is possible. Therefore these models are observationally acceptable: they possess an approximate standard matter epoch followed by a non-phantom acceleration with an effective equation of state $\omega_{eff}\equiv-1-2\dot{H}/3H^2 > -1$, thus these models posses a standard DE behaviour. These models satisfy both equations \eqref{viable_matter} and \eqref{acceleration_viable_1}.
\end{itemize}

{\bf Classes II} and {\bf IV}  have therefore some chance to be cosmologically viable but the basin of the attractor has to be determined to provide acceptable trajectories according to the already mentioned analysis fully performed in \cite{Amendola&al:2007}. In Table \ref{table_fR_DE_models} the previous analysis have been applied to some $f(R)$ models usually considered in the literature.
\begin{table}
\centering \small{
\begin{tabular}{||c|c|c|c|c||}
\hline $f(R)$ models & $m(r)$ & Class I & Class II & Class III \\
\hline
\hline
$-R+\alpha R^{-n}$ & $-1-n$ & $n>-0.713$ & $-$  & $-1\,<\,n\,<\,-0.713$ \\
\hline
$\alpha R^{-n}$ & $-\frac{n(1+r)}{r}$ & $n>0$ & $n\in(-1,0)$, $\alpha<0$  & $-$ \\
\hline
$-R+R^{p}\left[\text{log}\,\left(\alpha R\right)\right]^q$ & $\frac{(p+r)^2}{qr}-1-r$ & $p\neq1$ & $p=1$, $q>0$  & $-$ \\
\hline
$-R+R^{p} \text{exp}\,q R$ & $-r+\frac{p}{r}$ & $p\neq1$ & $-$  & $-$ \\
\hline
$-R+R^p \text{exp}(q/R)$ & $-\frac{p+r(2+r)}{r}$ & $p\neq1$ & $p=1$  & $-$ \\
\hline
%
%
%
\end{tabular}
} \caption{\footnotesize{Classification of some $f(R)$ DE models
presented in \cite{Amendola&al:2007}. None of these models belongs
to {\bf Class IV}. Models that belong to {\bf Class II} for the
provided parameter intervals, at least satisfy the conditions to
have a matter era followed by a de Sitter attractor.
}}
\label{table_fR_DE_models}
\end{table}
\section{Some cosmologically viable $f(R)$ models}
\label{sect:fR:Some viable f(R) models}
In this section we provide three $f(R)$ models already presented in the literature which claim to be cosmologically viable.

{\it a)}{\bf $f(R)=\lambda R_0\left[\left(1+\frac{R^2}{R_{0}^2}\right)^{-n}-1\right]$}

This model was originally considered in reference \cite{Starobinsky2} with $n,\lambda >0$ and $R_0$ of the order
of the presently observed effective cosmological constant. Then $f(0)=0$ and the cosmological constant is claimed
to disappear in flat space-time but $f_{RR}(0)$ is negative and therefore, according to condition {\bf 1} in Section
\ref{sec:Int:fR:Constraints}, flat space-time would be unstable.

For scalar curvatures $R\gg R_0$, $f(R)$ tends to $\lambda R_0$ and
the model would behave as the EH case with an effective
cosmological constant. On the other hand, de Sitter space-time
with curvature $R_1>0$ is also a vacuum solution provided
$R_1(f_{R}(R_1)-1)=2f(R_1)$ according to equation \eqref{dif} for
$D=4$ and $R=R_1$. Thus, this case would present an effective
cosmological constant $\Lambda(R_1)=R_1/4$.

For this model it can be proved that conditions {\bf 1} and {\bf
2} of Section \ref{sec:Int:fR:Constraints} are
satisfied in the curvature interval $[R_1,\infty)$ if they are
accomplished at $R=R_1$. Then these two conditions hold if and
only if
\begin{eqnarray}
\left[1+\left(\frac{R_1}{R_0}\right)^2\right]^{n+2}>1+(n+2)\left(\frac{R_1}{R_0}\right)^2+(n+1)(2n+1)\left(\frac{R_1}{R_0}\right)^4\,.
\end{eqnarray}
Note that condition {\bf 3} in that section is straightforwardly
satisfied if as considered, parameters $\lambda$ and $n$ are
positive. On the other hand, this model also satisfies the required
conditions to provide a matter dominated era at $R\gg R_0$ and
does not possess the already mentioned
Dolgov-Kawasaki instability. The
remaining condition {\bf 4} is easily accomplished if $R_0$ is
considered much smaller than $R$ at recent epochs.

A simple choice of parameters $\lambda$, $n$ and $R_0$ shows that
this model obeys the conditions \eqref{viable_matter} and
\eqref{acceleration_viable_2} and therefore according to the
analysis in the previous section it possess a matter dominated
epoch and a de Sitter late time acceleration.
%
%
%

{\it b)} {\bf $f(R)\,=\,-\alpha m_1\left(\frac{R}{\alpha}\right)^{n}\left[1+\beta\left(\frac{R}{\alpha}\right)^n\right]^{-1}$}
%

This model was first proposed in reference \cite{Hu&Sawicki_May_2007} in order to mimic $\Lambda\text{CDM}$ evolution in the high-redshift regime and to accelerate at low redshift with an expansion history close to $\Lambda\text{CDM}$ model. In this model, the parameter $n$ was considered to be positive and for convenience the mass scale $\alpha$ was given by
\begin{eqnarray}
\alpha\equiv \frac{8\pi\,G\overline{\rho}_0}{3}\,=\,(8315\,\text{Mpc})^{-2}\left(\frac{\Omega_{M} h^2}{0.13}\right)
\label{mass_scale_Hu}
\end{eqnarray}
where $\overline{\rho}_0$ is the average density today. This model does not have a bare cosmological constant since $f(0)=0$ and its parameters $m_1$, $\beta$ and $n$ may be rewritten as \cite{Amendola&al:2009}
\begin{eqnarray}
\frac{m_1}{\beta}\,\approx\,6\frac{1-\Omega_{M}}{\Omega_{M}}
\end{eqnarray}
\begin{eqnarray}
\frac{m_1}{\beta^2}\,=\,-\frac{f_{R}(R_0)}{n}\left(\frac{12}{\Omega_{M}}-9\right)^{n+1}
\end{eqnarray}
where $\Omega_{M}$ is the effective matter energy density at the present time.
Finally the constraint $|f_{R}(R_0)|<0.1$ was imposed and
$R_0$ is the scalar curvature today as would be obtained from
$\Lambda\text{CDM}$ model, i.e.
\begin{eqnarray}
R_0\approx \alpha\,m_{1}\left(\frac{12}{\Omega_{M}}-9\right).
\end{eqnarray}
Setting the values $n=1$ and $\Omega_M=0.3$, it was proven in
reference \cite{Amendola&al:2009} that this model belongs to the
Class {\bf II} presented in the previous section.
%

{\it c)} {\bf  $f(R)=
-\alpha\,R_{*}\,\text{log}\left(1+\frac{R}{R_{*}}\right) $}

This two-parameter $f(R)$ model presented in \cite{quartin}, where parameters $\alpha$ and $R_{*}$ are positive, has claimed to be
cosmologically viable and different from $\Lambda\text{CDM}$. In fact, it does satisfy both cosmological conditions \eqref{viable_matter} and \eqref{acceleration_viable_2} presented in previous Section \ref{sect:fR:Cosmological viability}, provided $\alpha>1$ and regardless of the value of $R_{*}$.

Conditions {\bf 1}, {\bf 2} and {\bf 3} given in Section
\ref{sec:Int:fR:Constraints} are also satisfied but
concerning the condition {\bf 4} also given in that section, it
will be shown in Section \ref{sec:Perturbations:Techniques to rule
out} that this $f(R)$ theory does not satisfy this
condition, showing that this model is indeed distinct from
$\Lambda\text{CDM}$. This fact and its consequences will be
studied in Section 3.4.
\section{$f(R)$ with no cosmological constant}
\label{sect:fR:Reconstruction}
Now that the modified Einstein equations have been
presented for FLRW metric, some interesting results
at cosmological scales will be obtained in this section. The presented
approach tries to mimic cosmological well-known GR results in different
cosmological eras employing adequate $f(R)$ functions.
Reconstruction procedures of this kind have been widely
studied in the literature \cite{Odintsov:2006, Saez&al1, Saez&al2}
where by rewriting the involved equations in new variables and assuming a given
cosmological solution, mainly in vacuum, the required $f(R)$ gravity is obtained.

In this section the addressed issue will be to find a $f(R)$
gravity able to reproduce the current cosmic speed-up appearing in
standard $\Lambda\text{CDM}$ cosmology. This function is required
to be analytical at $R=0$ and to have $R=0$ as a vacuum solution,
therefore it will not contain any cosmological constant
contribution. From a more formal point of view we are seeking for
some $f(R)$ gravity model having the same FLRW solution as the
standard EH action with cosmological constant for nonrelativistic
matter (dust matter, i.e. $P_{M}=0$).
For the searched cosmological constant absence it is clear that
the $f(R)$ expansion at $R=0$ must start at the $R^2$ term to
avoid, on the one hand, having cosmological constant and, on the
other hand, redefining the gravitational constant.
%
\subsection{Cosmological evolution in $\Lambda\text{CDM}$ model}
\label{subsect:fR:GRplusLambda}
Let us solve Einstein's equations \eqref{LCDM_fieldequations} for the standard EH action plus a cosmological constant $\Lambda$
with dust matter in the energy-momentum tensor side.

The most recent cosmological data quoted in reference \cite{WMAP}
are compatible at late times with a cosmological model based on a
spatially flat FLRW metric like \eqref{metric_t} together with
Einstein's equations with a cosmological constant $\Lambda_{}\neq0$
and dust matter (including dark matter). In this case, the equations
\eqref{LCDM_fieldequations} will be valid and matter content will
be written in terms of a pressureless perfect fluid
\begin{eqnarray}
T^{\mu}_{\,\,\,\nu}=\,\text{diag}\,(\rho_{M0}(t),\,0,\,0,\,0).
\end{eqnarray}

Equation $\mu=\nu=0$ (time-time component) in \eqref{LCDM_fieldequations} becomes
\begin{equation}
\left(\frac{\dot{a_{0}}(t)}{a_{0}(t)}\right)^2\,=\,\frac{8\pi G_{}}{3}\rho_{M0}(t)+\frac{\Lambda_{}}{3}
\label{00con}
\end{equation}
where $\rho_{M0}(t)$ in previous equation is given by expression
\eqref{matterdensity} if $\omega_{\alpha=M}$ is fixed to 0. Thus
\begin{equation}
\rho_{M0}(t)\,=\,\rho_{M0}(t_{0})\left(\frac{a_0(t_{0})}{a_0(t)}\right)^{3}
\label{matterdensity_M}
\end{equation}
where
the $0$ subindex means that the standard EH
equations \eqref{LCDM_fieldequations} with a cosmological constant
are being considered. This notation will be relevant later on when
standard $\Lambda\text{CDM}$ cosmology will be compared with the
results coming from the action \eqref{action} for the $f(R)$
function that we shall find in this section.

Substituting the expression \eqref{matterdensity_M} in the
equation \eqref{00con}, the solution $a_{0}(t)$, using the
notation introduced in Section 2.2 in this chapter, is found to
be:
\begin{eqnarray}
a_{0}(t)\,&=&\,\left(\frac{\Omega_{M}}{\Omega_{\Lambda_{}}}\right)^{1/3}\text{sinh}^{2/3}\left(\frac{3\sqrt{\Omega_{\Lambda_{}}}}{2}H_{0}\,t\right).
\label{00_EH_resuelta}
\end{eqnarray}

On the other hand, by taking the trace of \eqref{LCDM_fieldequations} in this case, i.e. cosmological constant and dust matter, it is found that
\begin{equation}
R_{0}(t)-4\Lambda_{}\,=\,8\pi G_{}\,\rho_{M0}(t).
\label{traza_con_f_nula}
\end{equation}
\subsection{$f(R)$ case with no cosmological constant}
Now let us consider the equations \eqref{fieldtensorialequation} but
in the case where no cosmological constant is considered and
 the energy-momentum tensor for dust matter will be
\begin{eqnarray}
T^{\mu}_{\,\,\,\nu}\,=\,\text{diag}\,(\rho_{M}(t),\,0,\,0,\,0).
\end{eqnarray}
%
Then the equation \eqref{00sin} becomes
\begin{equation}
3(1+f_R)\frac{\ddot{a}}{a}-\frac{1}{2}
(R+f(R))-3\frac{\dot{a}}{a}\dot{R}f_{RR}\,= \,-8\pi G_{}\rho_{M}
\label{00sin_dust}
\end{equation}
where we have eliminated the subindex $0$ in the different
quantities to avoid any confusion with the previous case presented in
Subsection \ref{subsect:fR:GRplusLambda}. As was already mentioned, it is clear
that the solutions for equation \eqref{00sin_dust} will strongly depend on
the function $f(R)$: different choices for this
function will lead to different evolutions of the universe for the
same initial conditions.  However, our approach to the problem will
be to find a function $f(R)$ so that the solution $a(t)$ of
the equation \eqref{00sin_dust} will be exactly the same as the solution provided by the
expression \eqref{00_EH_resuelta} that we obtained by using GR with nonvanishing
cosmological constant and which seems to fit the present cosmological data. In
other words, we want to find the $f(R)$ model such that the solution for the
equation \eqref{00sin_dust} is exactly the scale factor \eqref{00_EH_resuelta}, i.e.:
\begin{equation}
a(t)\equiv a_0(t)
\label{aa0}
\end{equation}
for the same initial (or present, i.e. $t=t_0$)
conditions. If it were possible to find such a function $f(R)$
then, it would be possible to avoid the necessity for introducing
any cosmological constant just by considering a gravitational
action such that given in the expression \eqref{action}.
In the following it will be shown that such a function happens to exist and its precise form will be provided. In
order to do that one first notices that accomplishing the condition \eqref{aa0}
after radiation-matter equality clearly implies
\begin{eqnarray}
R\,=\,R(t)\,\equiv\,R_{0}(t)
 \end{eqnarray}
and then $R(t)$ and $R_0(t)$ may be used indistinctly.
On the other hand we shall write the matter density
as the former matter density provided by expression \eqref{matterdensity} plus a new contribution, i.e.
\begin{eqnarray}
\rho_{M}(t)\,=\,\rho_{M0}(t) +\Delta\rho(t),
\label{density}
\end{eqnarray}
accounting for a slight variation with respect to the density provided in the Subsection \ref{subsect:fR:GRplusLambda}.
Assuming that matter for arbitrary $f(R)$ is still
nonrelativistic in this cosmological era we have
\begin{equation}
\Delta\rho(t)\,=\,\Delta\rho(t_{0})\left(\frac{a_{0}(t_{0})}{a_{0}(t)}\right)^{3}
\label{delta_densidad_escala}
\end{equation}
where according to the expression \eqref{matterdensity} particularized for the constraint \eqref{aa0}
and considering the relation given in \eqref{traza_con_f_nula} we can write

\begin{equation}
\left(\frac{a_{0}(t_{0})}{a_{0}(t)}\right)^3=\frac{R_{}(t)-4\Lambda_{}}{8\pi G_{}\,\rho_{M0}(t_{0})}
\label{factor_de_escala}
\end{equation}
and then \eqref{delta_densidad_escala} becomes
\begin{equation}
\Delta\rho(t)\,=\,-\eta\frac{R_{}(t)-4\Lambda}{\kappa}
\label{delta_rho_final}
\end{equation}
where we have introduced the parameter
\begin{eqnarray}
\eta\,\equiv\,-\frac{\Delta\rho(t_{0})}{\rho_{M0}(t_{0})}
\end{eqnarray}
so that matter density \eqref{density} is rewritten as
\begin{eqnarray}
\rho_{M}(t;\eta)=(1-\eta)\rho_{M0}(t)
\end{eqnarray}
 Finally the last term on the l.h.s. of equation
\eqref{00sin_dust} can be written in terms of the scalar curvature
by differentiating expression \eqref{traza_con_f_nula} with respect to cosmic time $t$
and using the conservation equation
\eqref{eq_dif_densidad}. Hence, we get
\begin{eqnarray}
&&3\left(R_{}-3\Lambda\right)\left(R_{}-4\Lambda\right)f_{RR}  +
 \left(-\frac{1}{2}R_{}+3\Lambda\right)f_{R} -\frac{1}{2}f(R)-\Lambda\nonumber\\
&&-\,\eta \left(R_{}-4\Lambda\right)\,=\,0
\label{e_omega=0}
\end{eqnarray}
where the time dependence of $R_{}$ is implicit. This last equation can be considered as
a second order linear differential equation for the function $f(R)$, so two initial conditions are needed to
solve it: the natural choice that has been judged more convenient and physically meaningful is the following:

\begin{enumerate}
 \item   Firstly, the absence of any cosmological constant in the gravitational action is required, so that
$f(0)=0$.

\item  Secondly, the standard EH action behaviour should be recovered for low
scalar curvatures without redefining the Newton constant, i.e. $f_{R}(0)=0$.
\end{enumerate}

Moreover, $f(R)$ function is wanted to be analytical at the origin
so that $R=0$ should be a solution for the field equations in
vacuum. This is an extremely important requirement since it allows
both Minkowski and Schwarzschild solutions to be vacuum solutions.

With these initial conditions, the equation \eqref{e_omega=0} can be solved by using standard methods. One particular solution is:
\begin{equation}
f_p(R)\,=\,-\eta R+2\left(\eta-1\right)\Lambda_{}.
\label{sol_particular_completa}
\end{equation}
The homogeneous equation associated with \eqref{e_omega=0} is a
Gauss-type equation solved in terms of hypergeometric functions.
The general solution of the homogeneous equation can be written
as:
\begin{equation}
f_h(R)\,= \Lambda_{}\left(K_{+} f_{+}(R)+ K_{-} f_{-}(R)\right)
\end{equation}
where
\begin{equation}
f_{\pm}(R)  \,=\,  \left(3-\frac{R}{\Lambda}\right)^{-a_{\pm}}
\,_{2}F_{1}\left[a_{\pm},\,1+a_{\pm}-c,\,1+a_{\pm}-a_{\mp};-\left(3-\frac{R}{\Lambda}\right)^{-1}\right]
\label{sol_homogenea}
\end{equation}
and the symbol $_{2}F_{1}$ holds for hypergeometric functions
and the constants
\begin{equation}
\begin{array}{l}
    a_{\pm}\,=\,-\frac{1}{12}\left(7 \pm \sqrt{73}\right)\,\,\,\,;\,\,\,\,c\,=\,-1/2
\end{array}
\label{ac}
\end{equation}
have been introduced. The $\eta$-dependent constants $K_{+}$ and $K_{-}$ must be
determined from the initial conditions given above. Numerically it is found that:
\begin{eqnarray}
K_{+}\,=\,0.6436 \,( -0.9058\,\eta+0.0596)\,\,\,\,;\,\,\,\,
K_{-}\,=\,0.6436\,(-0.2423\,\eta+3.4465).
\end{eqnarray}
The hypergeometric functions given in \eqref{sol_homogenea} are
generally defined in the whole complex plane. However, a
real gravitational action is wanted. In principle this requirement is very easy
to achieve since the coefficients in the equation
\eqref{e_omega=0} and the  constants $K_{\pm}$ are all
real. Then it is obvious that the real part of the functions appearing
in the expression \eqref{sol_homogenea} is a proper solution of the
homogeneous equation associated with \eqref{e_omega=0}. Thus the function we are
seeking can be written as:
\begin{eqnarray}
f(R)\,\equiv\, f_p(R)+\text{Re}\left[f_h(R)\right].
\label{sol_definition}
\end{eqnarray}
Nevertheless, the situation is more complicated. The homogeneous
equation has three regular singular points at $R_{1}\,=\,3\Lambda_{}$,
$R_{2}\,=\,4\Lambda_{}$ and $R_{3}\,=\,\infty$. This results in the
solution $f_h(R)$ having two branch points $R_1$ and $R_2$. More
concretely there are two cuts along the real axis: one from minus
infinity to $R_1$ and another from $R_2$ to infinity. Thus one must
be quite careful when interpreting \eqref{sol_definition}. From
minus infinity to $R_1$ there is only one Riemann sheet of $f_h(R)$
where $f(0)$ and $f_{R}(0)$ vanish and therefore this is the one that
we have to use to define  $f(R)$. From $R_1$ to $R_2$ the real part
of $f_h(R)$ is well defined. Finally from $R_2$ to infinity there is
only one Riemann sheet producing a smooth behaviour of $f(R)$. To
reach this sheet one must understand $R$ in the above equation as
$R+i\epsilon$.
At the present moment we do not know if this analytical structure
has any fundamental meaning or it is just an artefact of our
construction. Much more important is the fact that the function
$R+f_p(R)+f_h(R)$, which is the analytical extension of our
Lagrangian, is analytical at $R=0$, having at this point the local
behaviour $R+\Od(R^2)$. Therefore our generalized gravitational
Lagrangian $R+f(R)$ does guarantee that $R=0$ is a vacuum solution
as can be seen from expression \eqref{trace_eqn}. At
the same time, this Lagrangian reproduces the current evolution of the universe
without any cosmological constant.

Now that the $f(R)$ function in \eqref{sol_definition} has been
obtained, it is possible to check our result out by solving
\eqref{00sin} in terms of $a(t)$ for the $f(R)$ given in
\eqref{sol_definition}. This is done by rewriting the equation
\eqref{00sin_dust} in terms of $a(t)$ by using \eqref{R} and
\eqref{matterdensity} together with the dust matter density in
terms of $\eta$ and $a(t)$ as provided by expression
\eqref{density}. At this stage, the consistency of our results has
been checked out by introducing the scale factor
\eqref{00_EH_resuelta} and $f(R)$ given in \eqref{sol_definition}
in \eqref{00sin} and both sides of the equation turned to be equal
for all $\eta$ values.
%
%
Thus it has been guaranteed that our gravitational action proportional to $R+f(R)$ provides the same cosmic evolution --in the required cosmological eras -- as the EH action with cosmological constant $\Lambda_{}$ in a dust matter universe. Therefore, our model does verify, in the same range of precision, all the experimental tests that the standard cosmological model does in the present era. Notice also that in principle this can be achieved for any value of $\eta$, i.e. for any desired amount of matter. Nevertheless, some restrictions should be imposed on the parameter $\eta$. For instance it is obvious that in a
dust matter dominated universe  $\rho_{M}(t; \eta) \geq 0$ implies $\eta\leq1$.

Much more stringent bounds can be set on parameter $\eta$ by demanding that this model works properly back in time up to Big Bang Nucleosynthesis (BBN) era. Observations indicate that the cosmological standard model fits correctly primordial light elements abundances during BBN , that means that the expansion rate $H(t)$ cannot deviate from that of standard cosmology $H_{0}(t)$ in more than $10\%$  for the background evolution (see for instance \cite{Kolb&Turner} for further details).
Therefore by the time of BBN, departure of our model from the standard cosmology must not be too large and the equation
\eqref{00sin} should provide a similar behaviour to the one given by the standard Friedmann equation \eqref{00con} where now the density will include both dust and radiation contributions.

At BBN the DE contribution is negligible compared with dust and radiation densities. The scalar curvature
is of order $10^{-39}\,\text{eV}^2$ (with $\hbar\,=\,c\,=\,1$ for these calculations) and by that time dust
and radiation densities are of the order of  $10^{16}\,\text{eV}^{4}$ and $10^{21}\,\text{eV}^{4}$ respectively.
Since $R \simeq R_{0}$ we can rewrite the equation \eqref{00sin} as a modified Friedmann equation as follows
\begin{equation}
H^2(t)\,=\,H_{0}^2(t)\left\{ \frac{10^{5}R-\eta
R+\frac{1}{2}\left(Rf_{R}-f(R)\right)}{10^{5}R\left[1+f_{R}-3f_{RR}(1-\eta)R\right]}
\right\}.
\label{00sin_hubble_final}
\end{equation}
\\
As was commented above, to reproduce light elements abundances
it is required that $H^2(t)\,=\,H_{0}^2(t)(1\pm0.2)$ for
curvatures of order $R_{\text{BBN}}$. This implies that the second factor
on the r.h.s. of the previous expression \eqref{00sin_hubble_final}
should be between $0.8$ and $1.2$ by that period. Thus in order to match our $f(R)$
gravity model with the standard cosmology at the BBN times we need
to tune  $\eta$ to a value about $0.065$ with a stringent fine tuning. Therefore the matter
content of our model is not too different from the one in the
standard cosmology and the difference is in fact smaller than
experimental precision in \cite{WMAP}.

Concerning the problem of viability for this particular $f(R)$
model, if conditions {\bf 2}, {\bf 3} and {\bf 4} given in
Section \ref{sec:Int:fR:Constraints} are required to hold, the
parameter $\eta$ has to be fixed to a fine tuned value
$\eta\approx -1.4311 $ but for this value $f_{RR}$ reverse its
sign at high enough curvatures and therefore the condition {\bf 1}
in that section is not accomplished. Therefore it may be stated
that the $f(R)$ model given by expression \eqref{sol_definition} should be
considered as an effective model able to reproduce
$\Lambda\text{CDM}$ model cosmological expansion after
radiation-matter equality but not as a consistent gravitational
theory valid for all scales.
\section{Effective fluid description of $f(R)$ gravities}
\label{sect:fR:Arbitrary fluid}
In this section we shall be interested in finding those $f(R)$
functions such that the corresponding modified Einstein
equations in vacuum exactly reproduce the cosmological evolution
of EH gravity with a given perfect fluid, i.e., the introduced
modifications of the EH action through a function $f(R)$ will play
the role of the fluid source.

Let us thus consider such a perfect fluid 'f' obeying the following barotropic state equation
\begin{eqnarray}
P_{\text{f}}\,=\,\omega_{\text{f}}\,\rho_{\text{f}}
\end{eqnarray}
with constant $\omega_{\text{f}}$ and whose density
$\rho_{\text{f}}$ scales according to the conservation equation
\eqref{matterdensity} with the scale factor $a\equiv a(t)$ as
\begin{eqnarray}
\rho_{\text{f}}^{}(a)\,=\,
\rho_{\text{f}}(a(t_0))\left(\frac{1}{a(t)}\right)^{3(1+\omega_{\text{f}})}\,=\,
\rho_{\text{f}}(a(t_0))\,x^{1+\omega_{\text{f}}} \label{DE_density_0}
\end{eqnarray}
where $\rho_{\text{f}}(a(t_0))$ is the value of the fluid density for a
given value of the scale factor $a(t_0)\equiv1$. For the sake of simplicity
a new variable $x$ has been introduced in the previous expression
\eqref{DE_density_0} defined as follows
\begin{eqnarray}
x\equiv\frac{1}{a^3} \label{x}.
\end{eqnarray}
With this new variable $x$, if the only present (or at least the
dominant one) fluid is the defined above, the standard EH
Friedmann and acceleration equations in cosmic time $t$ are
respectively
\begin{eqnarray}
H^2\,&=&\,\frac{8\pi G_{} \rho_{\text{f}}}{3}\,=\,
\,H_{0}^2 x^{1+\omega_{\text{f}}} \nonumber\\
\frac{\ddot{a}}{a}\,&=&\,-\frac{8\pi
G_{}}{6}\rho_{\text{f}}(1+3\omega_{\text{f}})\,=\,-\frac{1}{2}\,H_{0}^2
\,(1+3\omega_{\text{f}})x^{1+\omega_{\text{f}}}
\label{EH_Eqns_fluid}
\end{eqnarray}
where as usual $8\pi G\rho_{\text{f}}(a(t_0))\,\equiv\,3H_{0}^2$.

Analogously to the procedure in the previous section of this
chapter, we shall consider the modified Einstein equations where
there will not be any fluid contribution. The solution for these
equations is wanted to be the same scale factor as the one which
is the solution of equations \eqref{EH_Eqns_fluid}. In other
words, the presence of the fluid in the EH equations wants to be replaced by
the contribution of some $f(R)$ function in the modified Einstein
equations.

%

To do so, note that the scalar curvature according its expression
\eqref{R} for spatially flat FLRW metric may be rewritten as
\begin{eqnarray}
R\,\equiv\,6\left[\left(\frac{\dot{a}}{a}\right)^2+\frac{\ddot{a}}{a}\right]\,=\,8\pi G_{}(1-3\omega_{\text{f}})\rho_{\text{f}}
\label{Scalar_Curvature}
\end{eqnarray}
and
\begin{eqnarray}
\dot{R}\,\equiv\,\frac{\text{d}R}{\text{d}t}\,=\,- 3H\,8\pi
G_{}(1+\omega_{\text{f}})(1-3\omega_{\text{f}})\rho_{\text{f}}(a(t_0)).
\end{eqnarray}
At this stage, let us introduce a dimensionless variable
$\tilde{R}\,\equiv\,R/H_{0}^2$. Hence with this notation in
variables $\tilde{R}$ and $x$  we get from expression
\eqref{Scalar_Curvature} that
\begin{eqnarray}
\tilde{R}\,=\,3
(1-3\omega_{\text{f}})\left(\frac{1}{a}\right)^{3(1+\omega_{\text{f}})}\,=\,3
(1-3\omega_{\text{f}})x^{1+\omega_{\text{f}}}.
\label{Rtilde_fluid}
\end{eqnarray}

Note that for the specific choices $\omega_{\text{f}}\,=\,-1,\,1/3$ no relation between variables $\tilde{R}$ and $x$ may be
straightforwardly established through \eqref{Rtilde_fluid}. For those cases $x$ has to be determined by solving the equations \eqref{EH_Eqns_fluid}.

%
%
Since $\dot{x}\,=\,-3\,H x$ the following equalities can be written down
\begin{eqnarray}
\frac{H^2}{H_{0}^2}\,=\,
x^{1+\omega_{\text{f}}}
\label{H2_x}
\end{eqnarray}
\begin{eqnarray}
\dot{\tilde{R}}\,=\,-9H\,
(1-3\omega_{\text{f}})(1+\omega_{\text{f}})x^{1+\omega_{\text{f}}}
\label{R_dot_tilde_x}
\end{eqnarray}

\begin{eqnarray}
\frac{\text{d}\tilde{R}}{\text{d}x}\, \equiv\,\tilde{R}_{x}
\,=\,3
(1-3\omega_{\text{f}})(1+\omega_{\text{f}})x^{\omega_{\text{f}}}
\label{D_x}
\end{eqnarray}

\begin{eqnarray}
\frac{\text{d}}{\text{d}x}\left(\frac{1}{\tilde{R}_{x}}\right)\,=\,-\frac{1}{\tilde{R}_{x}^2}3
(1-3\omega_{\text{f}})(1+\omega_{\text{f}})\omega_{\text{f}}\,x^{\omega_{\text{f}}-1}
\label{1_D_x}
\end{eqnarray}
\begin{eqnarray}
\frac{\text{d}\tilde{f}}{\text{d}\tilde{R}}\,=\,\frac{\text{d}\tilde{f}(x)}{\text{d}x}\frac{1}{\tilde{R}_{x}}
\,\,\,;\,\,\,
H_{0}^2\,\frac{\text{d}^2 f}{\text{d}R ^2}\,=\,
\frac{\text{d}^2 \tilde{f}(x)}{\text{d}x^2}\frac{1}{\tilde{R}_{x}}-
\frac{\text{d}\tilde{f}(x)}{\text{d}x}\frac{3}{\tilde{R}_{x}^3}
(1-3\omega_{\text{f}})(1+\omega_{\text{f}})\omega_{\text{f}}\,x^{\omega_{\text{f}}-1}
\label{derivatives_f_s}
\end{eqnarray}
%
%
\begin{eqnarray}
\frac{\dot{a}}{a}\dot{R}\,=\,-3\,H^2 x \tilde{R}_{x}\,H_{0}^2
\label{adot_Rdot_over_a}
\end{eqnarray}
where $\tilde{f}(\tilde{R})\equiv f(R)/H_{0}^2$.

Therefore, by considering a fluid density given by expression
\eqref{DE_density_0}, the modified Friedmann equation given by
\eqref{00sin} may be rewritten as a second order differential equation
for $f(R)$. Analogously, this equation may be expressed as a
differential equation in $x$ variable for $f(R(x))\equiv f(x)$ and it becomes
\begin{eqnarray}
&&\frac{\text{d}\tilde{f}(x)}{\text{d}x}\frac{1}{\tilde{R}_{x}}\left\{3\frac{\ddot{a}}{a}\tilde{R}_{x}-9H^2\left[3
(1-3\omega_{\text{f}})(1+\omega_{\text{f}})\omega_{\text{f}}x^{\omega_{\text{f}}}\right]\right\}-\frac{1}{2}H_{0}^2\,\tilde{f}(x)+9H^2\frac{x}{\tilde{R}_{x}}\frac{\text{d}^2
\tilde{f}(x)}{\text{d}x^2}\nonumber\\
&&\,=\,
3
\,H_{0}^2\,
x^{1+\omega_{\text{f}}}
\label{general_equation}
\end{eqnarray}
where expressions \eqref{EH_Eqns_fluid} and \eqref{D_x} have to be substituted in the previous equation.

The general solution of \eqref{general_equation} is then:
\begin{eqnarray}
\tilde{f}(x)\,=\,-3 
(1-3 \omega_{\text{f}})
x^{1+\omega_{\text{f}}}+
c_{+} x^{\omega_{\text{f}}^{+}}+c_{-} x^{\omega_{\text{f}}^{-}}
\label{f(x)_fluid}
\end{eqnarray}
where
\begin{eqnarray}
\omega_{\text{f}}^{\pm}\,=\,\frac{1}{12}\left[9\,\omega_{\text{f}}+7 \pm \left(9 \omega_{\text{f}} ^2+78 \omega_{\text{f}} +73\right)^{1/2}\right].
\end{eqnarray}
Requiring $\omega_\text{f}\neq -1,\,1/3$ to avoid possible indeterminacies, it is possible to rewrite \eqref{f(x)_fluid} in terms of $\tilde{R}$ as
\begin{eqnarray}
\tilde{f}(\tilde{R})\,=\,-
\tilde{R}\,+\, c_{+}\left[\frac{\tilde{R}}{3
(1-3\omega_{\text{f}})}\right]^{\frac{\omega_{\text{f}}^{+}}{1+\omega_{\text{f}}}}\,+\,c_{-}\left[\frac{\tilde{R}}{3
(1-3\omega_{\text{f}})}\right]^{\frac{\omega_{\text{f}}^{-}}{1+\omega_{\text{f}}}}.
\label{f(R)_fluid}
\end{eqnarray}
%
\subsection{Some examples}
Some interesting cases for the fluid content are the dust matter, radiation and cosmological constant fluids,
, i.e., $\omega_{M,\,\text{Rad},\,\Lambda}\,=\,0,1/3,-1$
and $\eta_{\,\text{f}}\,=\,\eta_{M,\,\text{Rad},\,\Lambda}$ respectively.
For these three cases, the corresponding functions become
\begin{eqnarray}
\tilde{f}_{M}(x)\,&=&\,c_{+} x^{\frac{1}{12} \left(7 + \sqrt{73}\right)} +
    c_{-} x^{\frac{1}{12} \left(7 - \sqrt{73}\right)} - 3 
x
%
\nonumber\\
\tilde{f}_{\Lambda_{}}(x)\,&=&\,- 12 
+
%
c_{-} x^{-1/3} + c_{+}\nonumber\\
\tilde{f}_{\text{Rad}}(x)\,&=&\,c_{+} x^{5/3} + c_{-}
\label{approximated_solutions}
\end{eqnarray}
\\
where constants $c_{\pm}$ are arbitrary integration constants that can be fixed if either boundary or initial conditions are imposed.
\section{Conclusions}
\label{sect:fR:Conclusions}
In this chapter we have found the $f(R)$ gravity which
exactly reproduces the same evolution of the universe, from BNN up to the present time, as standard
$\Lambda\text{CDM}$ model does, but without the introduction of any form of DE or cosmological
constant. The gravitational Lagrangian $R+f(R)$ is analytical at the origin and consequently $R=0$
is a vacuum solution for the field equations.
Therefore Minkowski, Schwarzschild and other important $R=0$ GR solutions, with $\Lambda_{}=0$, are
also solutions for this $f(R)$ gravity. This result was originally presented in \cite{Dobado&Dombriz}.\\

The price that we have to pay for all those good properties is
that our Lagrangian, considered as a function of $R$, has a very
complicated analytical structure with cuts along the real axis
from infinity to $R=3\Lambda_{}$ and from $R=4\Lambda_{}$ to
infinity. Obviously the only reasonable interpretation of our
action is as some kind of effective action. In classical physics
one typically starts from some action principle, obtains the
corresponding field equations and finally solves them for some
initial or boundary conditions. In this work we have proceeded in
the opposite way: we started from solutions obtained in the
standard cosmological model and then we have searched for an
action that, possessing certain properties, gives rise to field
equations having the same solutions.

Classical actions are of course real but effective quantum actions
usually have a complex structure coming from loops and related to
unitarity. The presence of an imaginary part in the action,
evaluated on some classical configuration, indicates quantum lost of
stability by particle emission of this configuration \cite{MD}.
Therefore it is tempting to think that our action could have some
interpretation in terms of an effective quantum action. However, our
action determination procedure does not allow to make such a kind of
statement. Complications in the action could be just an artifact of
our construction. In any case it was shown that such action exists
and it reproduces the present universe evolution without DE having
$R=0$ as a vacuum solution. As a drawback of this result, we have shown
that this $f(R)$ function cannot be considered as a fully consistent gravitational
theory since it does not obey the viability conditions revised in Chapter 1 and that therefore it
should be regarded as an effective model to mimic $\Lambda\text{CDM}$ cosmological background evolution.

To conclude the chapter a completely general procedure to
reproduce EH gravity with an arbitrary perfect fluid by using
$f(R)$ theories has been implemented. It has been explicitly shown
that any perfect fluid when is described by a constant equation of state
can be mimicked by an appropriate $f(R)$ model. Standard cases
for perfect fluids such as dust matter, radiation and cosmological constant
have been presented in this analysis.

\chapter[Cosmological perturbations in $f(R)$ theories]{Cosmological perturbations in $f(R)$ theories}
\label{chap:Perturbations}
\section{Introduction}
This chapter will be devoted to a study of the evolution of scalar
cosmological perturbations in $f(R)$ theories. To do so, a
completely general procedure will be implemented and several
consequences will be analyzed.
%
The importance for addressing the present problem lies in the necessity to discriminate among different DE models, including $f(R)$ modified gravities, by using observations. It is well-known
that by choosing adequate $f(R)$ functions, one can mimic any expansion history, and in particular that of the $\Lambda\text{CDM}$ model. Accordingly, the exclusive use of 
observations from SNIa \cite{SN}, baryon acoustic oscillations \cite{BAO} or CMB shift factor \cite{WMAP}, based upon different distance measurements which are sensitive only to the cosmological expansion history, cannot settle the question of the DE nature \cite{Linder2005}.

However, there exists a different type of observations which are sensitive, not only to the cosmological expansion
history, but also to the evolution of matter density perturbations. Pioneering work 
on density perturbations in FLRW cosmological models was presented in \cite{Lifshitz}. The fact
that the evolution of perturbations depends on the specific gravity model, i.e., it differs in general
from that of Einstein's gravity even though the background evolution is the same, means that this kind of observations
will help distinguishing between different models capable to explain cosmic acceleration.
This is therefore the aim of the present chapter: to provide a completely exact method to determine
how cosmological perturbations grow in $f(R)$ theories and to settle the extracted consequences from
this result. Such a problem has been exhaustively considered in the literature \cite{perturbations}.
In this chapter we shall show that for $f(R)$ theories the
differential equation for the matter density contrast is a fourth
order differential equation
in the longitudinal (also called conformal or Newtonian) gauge but it reduces to a second order differential equation for sub-Hubble modes.

This general equation will be compared with the standard
simplification procedure (so-called quasi-static approximation)
widely used in the previous literature. This approximation
considers simplifications in the density equation determination
from the very beginning. We shall show that for general $f(R)$
functions, the quasi-static approximation is not justified.
However, for those $f(R)$ adequately describing the present phase
of accelerated expansion and satisfying local gravity tests, it
does give a correct description for the evolution of
perturbations.

Once the general results are presented, some immediate applications may be implemented. For instance, our analysis
may also be used to settle the validity of some proposed $f(R)$ models, by comparing the predicted matter spectra with
recent observations of LSS \cite{Tegmark}.

The present chapter is organized as follows: in Section
\ref{sec:Perturbations:Theory perturbations} we briefly review the
theory of cosmological perturbations for the standard
$\Lambda\text{CDM}$ model, introducing the gauge-invariant
variables and revising the well-known results for the EH theory in
order to establish a comparison with $f(R)$ gravities.
%
%
Next, Section \ref{sec:Perturbations:Perturbations in f(R)} will
be devoted to thoroughly study cosmological perturbations in
$f(R)$ theories. The perturbed modified Einstein equations are
obtained through a completely general procedure for those theories
in the Subsection \ref{subsec:Perturbations:Einstein's eqns f(R)
perturbations}. In Subsection \ref{subsec:Perturbations:Density
equation in f(R)} the density perturbations equation is obtained
whereas the Subsection \ref{subsec:Perturbations:Quasi-static
approximation} compares the obtained exact results with the ones
given by using the quasi-static approximation. Finally in this
section, in Subsection \ref{subsec:Perturbations:Some proposed
models} we shall study the growth of perturbations for some
particular $f(R)$ models. Section
\ref{sec:Perturbations:Techniques to rule out} will then show how
our previous results may be used to constrain or rule some $f(R)$
models out
and finally some general conclusions for the presented results will be given in Section \ref{sec:Perturbations:Conclusions perturbations}.

This chapter is based upon the results presented in references \cite{Cruz, Perturbations_proceedings, PRLcomment09}.
\section{Theory of cosmological perturbations}
\label{sec:Perturbations:Theory perturbations}
\subsection{Generalities}
To study cosmological perturbations, the 4-dimensional full line element
%
may be decoupled into background and perturbed parts as follows
\begin{eqnarray}
\text{d}s^2\,=\,g_{(0)\,\mu\nu}^{}\text{d}x^{\mu}\text{d}x^{\nu}+\delta g_{\mu\nu}\text{d}x^{\mu}\text{d}x^{\nu}
\label{metric_plus_perturbations}
\end{eqnarray}
$\mu,\,\nu\,=\,0,1,2,3$ with $g_{(0)\,\mu\nu}^{}$ representing the 
homogeneous FLRW background metric and $\delta g_{\mu\nu}$ describing a 
small perturbation. This perturbation may be split \cite{Mukhanov} in
three different types: scalar (S), vector (V) and tensor (T)
contributions as follows
\begin{eqnarray}
\delta g_{\mu\nu}\,=\,\delta g_{\mu\nu}^{S}+\delta g_{\mu\nu}^{V}+\delta g_{\mu\nu}^{T}.
\end{eqnarray}

This classification obviously refers to the way that fields
included in $\delta g_{\mu\nu}$ transform under three-space
coordinate transformations on a constant time hypersurface. Tensor
perturbations produce gravitational waves which do not couple to
energy density and pressure inhomogeneities and propagate freely.
Vector perturbations are dumped with cosmological expansion and
are therefore negligible today. On the contrary, scalar
perturbations may lead to growing inhomogeneities which will give
rise to the large scale structures and the CMB anisotropies which
are seen today.
Thus let us explicitly implement each type of
perturbations:
\\
\\
- \text{Scalar perturbations}: 
the most general metric perturbations of this type are given by
four scalar functions: $\phi$, $\psi$, $E$ and $B$ of the
space-time coordinates, as follows
\begin{eqnarray}
\delta g_{\mu\nu}^{S}\,=\,a^{2}(\tau)\begin{pmatrix} 2\phi & -B_{,i} \\ -B_{,i} & 2(\psi\delta_{ij}-E_{,\,ij}) \end{pmatrix}
\label{S_perturbations}
\end{eqnarray}
where $i,\,j=1,2,3$ from now on hold for spatial indices and
subindex $,i$ means ordinary derivative with respect to
$i^{th}$-coordinate.
\\
\\
- \text{Vector perturbations}: these perturbations can be
represented by two divergenceless three-vectors $F_{i}$ and
$S_{i}$ as follows:
\begin{eqnarray}
\delta g_{\mu\nu}^{V}\,=\,-a^{2}(\tau)\begin{pmatrix} 0 & -S_{i} \\ -S_{i} & F_{i,\,j}+F_{j,\,i} \end{pmatrix}
\label{V_perturbations}
\end{eqnarray}
where divergenceless conditions - Einstein's convention applied - mean
\begin{eqnarray}
F^{i}_{\,\,\,,i}=S^{i}_{\,\,\,,i}\,=\,0
\label{divergenceless_condition}
\end{eqnarray}
and shift from upper to lower indices --and viceversa-- is
performed through the spatial part of spatially flat background
metric tensor, i.e., $\delta_{ij}$ and its inverse $\delta^{ij}$.
\\
\\
- \text{Tensor perturbations}: tensor perturbations are given by a
symmetric three-tensor $h_{ij}$ satisfying the following
conditions
\begin{eqnarray}
h^{i}_{\,i}=0 \,\,\,;\,\,\, h^{ij}_{\,\,\,,\,j}=0
\label{hij_conditions}
\end{eqnarray}
i.e., traceless and transversality conditions respectively,
meaning that $h_{ij}$ does not contain any piece transforming as
scalars nor as vectors. Thus, the metric contribution $\delta
g_{\mu\nu}^{T}$ is simply given by
\begin{eqnarray}
\delta g_{\mu\nu}^{T}\,=\,-a^{2}(\tau)\begin{pmatrix} 0 & 0 \\ 0 & h_{ij} \end{pmatrix}.
\label{T_perturbations}
\end{eqnarray}
The number of independent functions introduced to define $\delta
g_{\mu\nu}$ without loss of generality is ten: four scalar
functions for scalar perturbations, two three-vectors for vector
perturbations with one constraint each and one symmetric
three-tensor with four conditions for tensor perturbations. This
number coincides with the number of independent components of
$\delta g_{\mu\nu}$ as a $4\times4$ symmetric tensor.

\subsection{Gauge-invariant variables and gauge choice}
Metric perturbations, as the ones defined above, are gauge-dependent, i.e. an infinitesimal coordinates transformation could give rise to two apparently different perturbations whereas they indeed represent the same physical perturbation. This is the reason why Bardeen introduced \cite{Bardeen} gauge-invariant quantities that are explicitly invariant under infinitesimal coordinate transformations.
The starting point is to consider infinitesimal coordinate transformations
\begin{eqnarray}
x^{\mu}\rightarrow\tilde{x}^{\mu}\,=\,x^{\mu}+\xi^{\mu}(x).
\label{coordinate_transformations}
\end{eqnarray}
It can be proven very easily that
in the new coordinates $\{\tilde{x}^{\mu}\}$ the metric $g_{\mu\nu}(x)$  can be written as:
\begin{eqnarray}
\tilde{g}_{\mu\nu}(x)\,=\,g_{\mu\nu}(x)+\mathcal{L}_{\xi}g_{\mu\nu}(x)+\Od(\xi^2)
\label{metric_transformed}
\end{eqnarray}
what proves that two metrics $g_{\mu\nu}$ and $\tilde{g}_{\mu\nu}$ differing on a Lie derivative represent the same physical perturbation
\footnote{Lie derivative of a twice covariant tensor $g_{\mu\nu}$ with respect to 
$\xi$ is given by
$\mathcal{L}_{\xi}g_{\mu\nu}\,\equiv\,-g^{\lambda}_{\mu}\xi_{\lambda;\nu}-g^{\lambda}_{\nu}\xi_{\lambda;\mu}+g_{\mu\nu;\lambda}\xi^{\lambda}$
where $g_{\mu\nu;\lambda}=0$ if Levi-Civita connection is
considered.}.
Let us consider a coordinate transformation given by the parameters
$(\xi^{0},\xi^{i})$, i.e.,
\begin{eqnarray}
\tilde{\tau}\,&=&\,\tau+\xi^{0}\nonumber\\
\tilde{x}^{i}\,&=&\,x^{i}+\xi^{i}=x^{i}+\overline{\xi}^{i}+\delta^{ij}\xi,_{j}
\label{coord_transformation}
\end{eqnarray}
where prime holds for derivative with respect to  $\tau$,
and $\xi^{i}$ is decomposed as $\xi^{i}=\overline{\xi}^{i}+\xi,_{j}\delta^{ji}$, i.e., it is given by a solenoidal part, $\overline{\xi}^{i}$, and
an irrotational part $\xi_{,j}\delta^{ji}$ according to Helmholtz's theorem.
Therefore $\text{d}\tau$, $\text{d}x^{i}$ and $a(\tau)$ can be expressed in terms of the new coordinates $\{\tilde{x}^{\mu}\}$ as:
\begin{eqnarray}
\text{d}\tau\,&=&\,\text{d}\tilde{\tau}-\xi^{0'}\text{d}\tilde{\tau}-\xi^{0},_{i}\text{d}x^{i}\nonumber\\
\text{d}x^{i}\,&=&\,\text{d}\tilde{x}^{i}-\xi^{i'}\text{d}\tilde{\tau}-\xi^{i},_{j}\text{d}\tilde{x}^{j}\nonumber\\
&=&\text{d}\tilde{x}^{i}-\left(\overline{\xi}^{'}+\delta^{ij}\xi^{i'},_{\,\,j}\right)\text{d}\tilde{\tau}-(\overline{\xi}^{i},_{j}+\delta^{ik}\xi,_{\,kj})\text{d}\tilde{x}^{j}   \nonumber\\
a(\tau)\,&=&a(\tilde{\tau})-\xi^{0}a'(\tilde{\tau})\,.
\label{gauge_transformation_coordinates}
\end{eqnarray}
Therefore if identities \eqref{gauge_transformation_coordinates} are applied to the expression \eqref{metric_plus_perturbations}, the obtained metric should have the aspect of the original line element provided the involved quantities defined in expressions \eqref{S_perturbations}, \eqref{V_perturbations} and \eqref{T_perturbations} transform as follows:\\

- Scalar perturbations:
\begin{eqnarray}
\tilde{\Phi}\,&=&\,\phi-\mathcal{H}\xi^{0}-\xi^{0'}\nonumber\\
\tilde{\Psi}\,&=&\,\psi+\mathcal{H}\xi^{0}\nonumber\\
\tilde{B}\,&=&\,B+\xi^{0}-\xi^{'}\nonumber\\
\tilde{E}\,&=&\,E-\xi
\label{S_transfomations}
\end{eqnarray}
where only scalar contributions $\xi^{0}$ and $\xi$ are present.
\\

- Vector perturbations:
\begin{eqnarray}
\tilde{F}_{i}\,&=&\,F_{i}-\overline{\xi}^{i}\nonumber\\
\tilde{S}_{i}\,&=&\,S_{i}+\overline{\xi}^{i'}
\label{V_transfomations}
\end{eqnarray}
where only vector contribution $\overline{\xi}_{i}$ is present.
\\

- Tensor perturbations:
\begin{eqnarray}
\tilde{h}_{ij}\,&=&\,h_{ij}
\label{T_transfomations}
\end{eqnarray}
which turn out to be gauge-invariant.

From previous results, gauge-invariant quantities can be constructed. For scalar perturbations, since four scalar functions were introduced and two scalar gauge parameters ($\xi^0$ and $\xi$) are present, two gauge-invariant quantities could be, for instance:
\begin{eqnarray}
\Phi\,=\, \phi+\frac{1}{a}[(B-E')a]^{'}\,\,\,\,\,;\,\,\,\,\,\Psi\,=\,\psi+\mathcal{H}(B-E')
\label{S_invariant_quantities}
\end{eqnarray}
where by construction $\Phi=\tilde{\Phi}$ and $\Psi=\tilde{\Psi}$ are known as the Bardeen's potentials in \cite{Bardeen}.

For vector perturbations one gauge-invariant quantity could be
\begin{eqnarray}
\mathcal{S}_{i}=S_{i}+F_{i}^{'}.
\label{V_invariant_quanmtities}
\end{eqnarray}
With all previous results in mind, one can choose, i.e. one may
specify in which coordinate system the scalar perturbations are
going to be studied.
There exist several possibilities for the gauge choice. Among them we can mention synchronous and longitudinal (or conformal-Newtonian) gauges.
\\
\\
$\it{\,\,Synchronous\,\, gauge}$: This gauge is defined by the conditions $\phi\,=\,B\,=\,0$ \cite{Lifshitz}.
However, it can be shown that the required synchronous coordinates are not completely fixed since a residual coordinate freedom remains, what renders the interpretation of calculations in this gauge difficult.
\\
\\
$\it{\,\,Longitudinal\,\, gauge}$: This gauge is defined by the
conditions $B\,=\,E\,=\,0$ and, in this gauge, coordinates are
totally fixed since $E\,=\,0$ determines $\xi$ uniquely. Using
this result, $B\,=\,0$ allows to fix determines $\xi^{0}$ without
any uncertainty. We draw the important conclusion that in this
gauge $\phi$ and $\psi$ coincide with the gauge invariant
variables \eqref{S_invariant_quantities} $\Phi$ and $\Psi$
respectively which have a simple physical interpretation as the
amplitudes of the metric perturbations in the usually so-called
conformal-Newtonian coordinate system.

\subsection{Equations for cosmological perturbations in EH gravity}
In $\Lambda\text{CDM}$ model within the metric formalism
it is possible to obtain a second order differential equation for the growth of matter density perturbation. Let us previously define the density contrast $\delta$ as follows:
\begin{eqnarray}
\delta\,\equiv\,\frac{\delta\rho}{\rho_{0}}\,\equiv\,\frac{\rho-\rho_0}{\rho_0}
\label{delta_definition}
\end{eqnarray}
where $\rho_0$ holds for the unperturbed mean cosmological energy density for a fluid and $\rho$ for the perturbed energy density of the same cosmological fluid.

In the following, as was mentioned in the beginning of the
chapter, the longitudinal gauge will be considered to perform our
calculations. Thus, the flat FLRW $D=4$ metric tensor with scalar
perturbations expressed in this gauge and by using conformal time
$\tau$ is written as:
\begin{equation}
\text{d}s^2\,=\,a^2(\tau)\left[(1+2\Phi)\text{d}\tau^2-(1-2\Psi)(\text{d}r^2+r^2\text{d}\Omega_{2}^2)\right]
\label{perturbed_metric}
\end{equation}
where $\Phi\equiv\Phi(\tau,\,\vec{x})$ and
$\Psi\equiv\Psi(\tau,\,\vec{x})$ are the well-known Bardeen's potentials \cite{Bardeen}.
From this metric, the first order perturbed standard Einstein equations are obtained:
\begin{eqnarray}
\delta G^{\mu}_{\;\;\nu}=-8\pi G_{}\,\delta T^{\mu}_{\;\;\nu}
\end{eqnarray}
Now that the metric with scalar perturbations is known, $\delta G^{\mu}_{\nu}$ is straightforwardly determined through Einstein's tensor $G^{\mu}_{\nu}$ definition. Therefore the first order perturbed standard Einstein equations, i.e., for EH gravity read:\\
\begin{eqnarray}
\delta G^{0}_{0}\,&=&\,-6\mathcal{H}^2\Phi-6\mathcal{H}\Psi'+2\nabla^2\Psi\,=\,8\pi G_{} a^2\delta T^{0}_{\,0}\nonumber\\
\delta G^{0}_{i}\,&=&\,(2\mathcal{H}\Phi+2\Psi'),_{i}\,=\,8\pi G_{} a^2\delta T^{0}_{\,i}\nonumber\\
\delta G^{i}_{j}\,&=&\,(-4\mathcal{H}'\Phi-2\mathcal{H}^2\Phi-2\mathcal{H}\Phi'-2\Psi''-4\mathcal{H}\Psi'-\nabla^2 D)\delta^{i}_{j}+D,^{\,i}_{\,\,j}\,=\,8\pi G_{} a^2\delta T^{i}_{\,j}\nonumber\\
&&
\label{Eqns_perturbations_$EH$}
\end{eqnarray}
where $D\equiv\Phi-\Psi$, the prime denotes derivative with
respect to conformal time $\tau$ and the subindex $,i$ is the
usual derivative with respect to the $i^{th}$-spatial coordinate.

To study the growth rate of cosmological perturbations we shall consider models with conventional hydrodynamical matter described by a perfect fluid energy-momentum tensor as given in Section
\ref{sec:Int:fR:Action and field equations}.
It should be reminded that $u^{\mu}\equiv \text{d}x^{\mu}/\text{d}s$ is the mean 4-velocity of the fluid. Unperturbed 4-velocity in FLRW conformal coordinates becomes:
\begin{eqnarray}
u_{(0)}^{\mu}\,=\,\frac{1}{a(\tau)}\left(1,\vec{0}\right)
\label{u0_velocity}
\end{eqnarray}
and to first order in scalar perturbations, it can be shown that
\begin{eqnarray}
u^{\mu}\,=\,a(\tau)^{-1}(1-\Phi,\delta u^{i})
\end{eqnarray}
where $\delta u^{i}$ can be decomposed as follows:
\begin{eqnarray}
\delta u^{i}=\overline{u}^{i}+v,^{\,i}
\label{delta_u}
\end{eqnarray}
with $\overline{u}^{i}$ and $v,^{\,i}$ being the solenoidal and irrotational components respectively. Note at this stage that $\overline{u}^{i}$ only contributes to the vector perturbations but not to the scalar ones and
$v$ is usually referred to as the potential for velocity perturbations.

Taking into account the previous digression, the perturbed energy-momentum tensor
components are proven to be:
%
%
\begin{eqnarray}
\delta T^{0}_{\;\;0}\,&=&\,\delta\rho \,=\, \rho_{0} \delta \nonumber\\
\delta T^{i}_{\;\;j}\,&=&\, -(\delta P) \delta^{i}_{\;\;j}
 \nonumber\\
\delta T^{0}_{\;\;i}\,&=&\,-\delta T^{i}_{\;\;0}\,=
-\,(\rho_0 + P_0)\partial_{i}v
\label{perturbed_stress_energy_tensor}
\end{eqnarray}
with $\delta P$ pressure fluctuation. Substituting expressions \eqref{perturbed_stress_energy_tensor}
in \eqref{Eqns_perturbations_$EH$}, they become

\begin{eqnarray}
-3\mathcal{H}^2\Phi-3\mathcal{H}\Psi'+\nabla^2\Psi\,&=&\,4\pi G_{} a^2\delta\rho\nonumber\\
(\mathcal{H}\Phi+\Psi'),_{i}\,&=&\,4\pi G_{} a^2(\rho_{0}+P_{0})v,_{i}\nonumber\\
(-2\mathcal{H}'\Phi-\mathcal{H}^2\Phi-\mathcal{H}\Phi'-\Psi''-2\mathcal{H}\Psi'-\frac{1}{2}\nabla^2 D)\delta^{i}_{j}+\frac{1}{2} D,^{\,i}_{\,\,j}\,&=&\,-4\pi G_{} a^2\delta P\delta^{i}_{\,j}\nonumber\\
&&
\label{Eqns_perturbations_$EH$_substituted}
\end{eqnarray}

For $i\neq j$ considered, $D,_{\,\,\,j}^{\,\,i}=0$ which in
Fourier space means $k_i k_j D_{k}=0$ for any $i,j$ values and
then $D$ is identically null and thus
\begin{eqnarray}
\Phi(\tau,\,\vec{x})\equiv\Psi(\tau,\,\vec{x}).
\end{eqnarray}
Such a result permits to simplify the previous equations
\eqref{Eqns_perturbations_$EH$_substituted} to become:
\begin{eqnarray}
\nabla^{2}\Phi-3\mathcal{H}^2\Phi-3\mathcal{H}\Phi'\,&=&\,4\pi G_{} a^2\delta\rho\nonumber\\
(a\Phi),^{'}_{i}\,&=&\,4\pi G_{} a^3(\rho_{0}+P_{0})v_{,i}\nonumber\\
\Phi''+3\mathcal{H}\Phi'+(\mathcal{H}^2+2\mathcal{H}')\Phi\,&=&\,4\pi G_{} a^2\delta P.
\label{Eqns_perturbations_$EH$_simplified}
\end{eqnarray}

At this stage, a short digression about the pressure $P$
dependence may be valuable: the pressure is, in principle, a
quantity depending on energy density and entropy per baryon ratio.
Thus, a pressure fluctuation $\delta P$ can be expressed in terms
of density and entropy perturbations as follows
\begin{eqnarray}
\delta P\,=\,\left(\frac{\partial P}{\partial \rho}\right)_{S}\delta\rho+\left(\frac{\partial P}{\partial S}\right)_{\rho}\delta S\,\equiv\,c_{S}^2\delta\rho+\left(\frac{\partial P}{\partial S}\right)_{\rho}\delta S
\label{Pressure_dependence}
\end{eqnarray}
where $c_{S}^2$ can be understood as the squared sound velocity of
the fluid perturbations. In a single component perfect fluid with
constant equation of state there are no entropy perturbations.
However, if the perturbation description needs to include more than
one component, 
entropy perturbations may be present.

In the following we shall restrict ourselves to adiabatic
perturbations, i.e.  $\delta S=0$ and therefore
\begin{eqnarray}
\delta P = c_{S}^2\,\delta\rho
\label{Pressure_density}
\end{eqnarray}
and equation of state for the fluid will be considered constant,
i.e. $P=\omega\rho$ with $\omega$ constant. Thus, perturbed and
unperturbed content matter are assumed to have the same equation
of state, i.e.  $\delta P/\delta\rho\equiv c_{S}^2 \equiv
P_{0}/\rho_{0}$, where $c_S=0$ for dust matter adiabatic
perturbations.

With the previous assumptions, the equations \eqref{Eqns_perturbations_$EH$_simplified} can be combined to obtain the growth rate $\delta$ evolution in Fourier space \footnote{In the rest of the present chapter it must be understood that symbols $\delta$, $\Phi$, $\Psi$ and $v$ will hold for the Fourier corresponding quantities but the subindex $k$ will be omitted in order to simplify the notation.}. For instance, for dust matter the resulting differential equation is
\\
\begin{eqnarray}
\delta''+\mathcal{H}\frac{k^4 -6\tilde{\rho}k^2 -18\tilde{\rho}^2}
{k^4 - \tilde{\rho}(3 k^2 +9\mathcal{H}^2)}\,\delta'
-\tilde{\rho}\frac{k^4 + 9\tilde{\rho}(2\tilde{\rho}-3\mathcal{H}^2)
-k^2 (9\tilde{\rho}-3\mathcal{H}^2)}{k^4 - \tilde{\rho}(3 k^2
+9\mathcal{H}^2)}\,\delta\,=\,0
\label{delta_0_general_LambdaCDM}
\end{eqnarray}
\\
where $\tilde{\rho} \equiv 4\pi G_{} \rho_{0} a^2 \,=\,
-\mathcal{H}'+\mathcal{H}^2$ according to the background standard
Einstein equations, as seen for instance from
\eqref{bg_density_plus_pression} setting $P_0\equiv 0$ and $f(R)$
constant. We point out that in order to obtain the equation
\eqref{delta_0_general_LambdaCDM} it is not necessary to calculate
the potentials $\Phi$ and $\Psi$ explicitly, but algebraic
manipulations in the equations
\eqref{Eqns_perturbations_$EH$_simplified} are enough to get this
result.

Some limits can be taken in the previous equation: for instance,
it is of particular interest to consider those $k$ modes whose
wavelength is much smaller than the Hubble radius. These modes are
known as sub-Hubble modes and are identified by the condition
$k\gg\,\mathcal{H}$ or equivalently $k\tau\gg \, 1$. In this
approximation the equation \eqref{delta_0_general_LambdaCDM}
reduces to the well-known expression:
\begin{equation}
\delta''+\mathcal{H}\delta' - 4\pi G_{} \rho_0 a^2\delta\,=\,0.
\label{delta_0_SubHubble}
\end{equation}

In this regime and at early times, the matter energy density dominates over the cosmological constant term and it is easy to show
that $\delta$ solutions for \eqref{delta_0_SubHubble} grow as
$a(\tau)$. At late times (near today) the cosmological constant contribution is not negligible and
thus the equation \eqref{delta_0_SubHubble} does not admit any more power-law solutions of the type $\delta\,\propto\,a(\tau)^{\gamma}$ for some $\gamma$.
It is necessary in this case to
assume an ansatz for $\delta$: one which works very well is
that proposed in references \cite{Linder2005} and \cite{Linder2006} namely
\begin{equation}
\frac{\delta(a)}{a}\,= \, \exp\left[\int_{a_{i}}^{a}\left(\Omega_{M}(a)^{\gamma}-1\right)\text{d}\text{ln} a\,\right]
\label{modelo_delta_Linder}
\end{equation}
where
$\Omega_{M}(a)\,\equiv\,\frac{\Omega_{M}\mathcal{H}_{0}^2}{a\mathcal{H}^2}$
and $\Omega_{M}$ was defined in expression \eqref{omegas}.
Expression \eqref{modelo_delta_Linder} turns out to fit with high precision 
the numerical solution for  $\delta$  with a constant parameter $\gamma\,=\,6/11$.
\section{Cosmological perturbations in $f(R)$ theories}
\label{sec:Perturbations:Perturbations in f(R)}
\subsection{Perturbed Einstein equations in $f(R)$ theories}
\label{subsec:Perturbations:Einstein's eqns f(R) perturbations}
Using the perturbed metric \eqref{metric_plus_perturbations} and
the perturbed energy-momentum tensor
\eqref{perturbed_stress_energy_tensor}, the first order perturbed
equations for $f(R)$ theories in the metric formalism, assuming that
the background equations given in expression
\eqref{fieldtensorialequation} in Chapter 1 hold, may be written
as:
\begin{eqnarray}
&&(1+f_{R})\delta \text{G}^{\mu}_{\nu}+((R_{(0)})^{\,\mu}_{\nu} +\nabla^{\mu}\nabla_{\nu}
-\delta^{\mu}_{\nu} \square)f_{RR}\delta R +
\,[(\delta g^{\mu\alpha}) \nabla_{\nu}\nabla_{\alpha} - \delta^{\mu}_{\nu}
(\delta g^{\alpha\beta})\nabla_{\alpha}\nabla_{\beta}]f_{R}\nonumber\\
&&-\,\left[g_{(0)}^{\alpha\mu}(\delta\Gamma^{\gamma}_{\alpha\nu})
-\delta^{\mu}_{\nu} g_{(0)}^{\alpha\beta}
(\delta\Gamma^{\gamma}_{\beta\alpha})\right] \partial_{\gamma}f_{R}\,=
\,-8\pi G_{} \delta T^{\mu}_{\nu}
\label{tensorial_bg_eqns}
\end{eqnarray}
where $(R_{(0)})^{\mu}_{\nu}$ will denote here the Ricci tensor components corresponding
to the unperturbed FLRW metric \eqref{metric_tau} in comoving coordinates whose trace provides
the scalar curvature already given in equation \eqref{R}.
Note that $f(R)$ derivatives with respect to $R_{(0)}$ have been expressed as usually,
i.e., $f_{R}\,\equiv\,\text{d}f(R_{})/\text{d}R_{}|_{R_{(0)}}$, $f_{RR}\,=\,\text{d}^{2}f(R_{})/\text{d}R^2|_{R_{(0)}}$
and again $\square\,\equiv\,\nabla_{\alpha}\nabla^{\alpha}$ and $\nabla$
is the usual covariant derivative with respect to the unperturbed FLRW metric.
Notice also that unlike the ordinary EH case where Einstein's equations are second order, the equations \eqref{tensorial_bg_eqns} constitute a set of fourth order differential equations.

For the linearized modified Einstein equations, the components
$(00)$, $(ii)$, $(0i)\equiv (i0)$ and $(ij)$, where $i,j\,=\,1,2,3$,
$i\neq j$, in  Fourier space, read respectively:
\begin{eqnarray}
&&(1+f_{R})[-k^2(\Phi+\Psi)-3\mathcal{H}(\Phi'+\Psi')+(3\mathcal{H}'
-6\mathcal{H}^2)\Phi-3\mathcal{H}'\Psi]\nonumber\\
&&\,+\,3f'_{R}[\mathcal{H}(-3\Phi+\Psi)-\Psi']\,=\,2\tilde{\rho}\delta
\label{00_pert}
\end{eqnarray}
\begin{eqnarray}
&&(1+f_{R})[\Phi''+\Psi''+3\mathcal{H}(\Phi'+\Psi')
+3\mathcal{H}'\Phi+(\mathcal{H}'+
2\mathcal{H}^2)\Psi]
+f'_{R}(3\mathcal{H}\Phi-\mathcal{H}\Psi+3\Phi') \nonumber\\
&& + f''_{R}(3\Phi-\Psi)\,=\,2c_{s}^{2} \tilde{\rho}\delta
\label{ii_pert}
\end{eqnarray}
\begin{eqnarray}
(1+f_{R})[\Phi'+\Psi'+\mathcal{H}(\Phi+\Psi)]+f'_{R}(2\Phi-\Psi)\,=\,- 2\tilde{\rho}(1+c_{S}^2) v
\label{0i_pert}
\end{eqnarray}
\begin{eqnarray}
\Phi-\Psi\,=\,-\frac{f_{RR}}{1+f_{R}}\delta R
\label{ij_pert}
\end{eqnarray}
where $\delta R\,\equiv\,R-R_{(0)}$ is given by:
\begin{eqnarray}
\delta R \,=\, -\frac{2}{a^2}\Big[3\Psi''+6(\mathcal{H}'+\mathcal{H}^2)\Phi+3\mathcal{H}(\Phi'+3\Psi')-k^2(\Phi-2\Psi)\Big]
\label{deltaRdefinition}
\end{eqnarray}
and $\delta P=c_{S}^{2}\delta\rho$ has been again assumed.

By computing the covariant derivative with respect to the
perturbed metric $\tilde{\nabla}$ of the perturbed energy-momentum
tensor $\tilde{T}^{\mu}_{\nu}$, we find the conservation
equations:
\begin{equation}
\tilde{\nabla}_{\mu}\tilde{T}^{\mu}_{\nu}\,=\,0
\label{cons}
\end{equation}
which do not depend on $f(R)$ explicitly. To first order, the equations
\eqref{cons} read
\begin{eqnarray}
3\Psi'(1+c_{S}^2)-\delta'+k^2(1+c_{S}^2)v\,=\,0
\label{Nabla0}
\end{eqnarray}
and
\begin{eqnarray}
\Phi+\frac{c_{S}^2}{1+c_{S}^2}\delta+v'+\mathcal{H}v(1-3c_{S}^2)\,=\,0
\label{Nablaj}
\end{eqnarray}
for the temporal and spatial components respectively.

In a dust matter dominated universe,
\eqref{Nabla0} and \eqref{Nablaj} can be combined to give
\begin{eqnarray}
\delta''+\mathcal{H}\delta'+k^2\Phi-3\Psi''-3\mathcal{H}\Psi'\,=\,0.
\label{Nabla}
\end{eqnarray}
\subsection{Equation for density perturbations in $f(R)$ theories}
\label{subsec:Perturbations:Density equation in f(R)}
In this subsection, we are going to obtain the differential
equation obeyed by $\delta$ in a dust matter dominated universe
when the longitudinal gauge and the metric formalism are
considered to study first order scalar perturbations of $f(R)$
theories.

To do so, let us consider equations \eqref{00_pert} and
\eqref{0i_pert} for a dust matter dominated universe, and combine
them to express the potentials $\Phi$ and $\Psi$ in terms of  $\{
\Phi', \Psi'; \delta , \delta' \}$ by means of algebraic
manipulations. The resulting expressions are the following
\begin{eqnarray}
\Phi\,&=&\,\frac{1}{\mathcal{D}(\mathcal{H}, k)}\Big\{[3(1 +f_{R})\mathcal{H}(\Psi'+\Phi')
+ f_{R}'\Psi' +2\tilde{\rho} \delta](1+f_{R})(\mathcal{H}-f_{R}')\nonumber\\
&+&[(1+f_{R})(\Phi'+\Psi')+\frac{2\tilde{\rho}}{k^2}(\delta'-3\Psi')][(1+f_{R})(-k^2-3\mathcal{H}') + 3f_{R}'\mathcal{H}]\Big\}
\label{Phi_desp}
\end{eqnarray}
and
\begin{eqnarray}
\Psi\,&=&\,\frac{1}{\mathcal{D}(\mathcal{H},k)}\Big\{[-3(1+f_{R})\mathcal{H}(\Psi'+\Phi')-3f_{R}'\Psi'
-2\tilde{\rho}\delta][(1+f_{R})\mathcal{H}+2f_{R}']-[(1+f_{R})\nonumber\\
&&\times(\Phi'+\Psi')+\frac{2\tilde{\rho}}{k^2}(\delta'-3\Psi')][(1+f_{R})(-k^2+3\mathcal{H}'-6\mathcal{H}^2) -9\mathcal{H}f_{R}']\Big\}
\label{Psi_desp}
\end{eqnarray}
where
\begin{eqnarray}
&&\mathcal{D}(\mathcal{H},k)\,\equiv\,-6(1+f_{R})^2\mathcal{H}^{3}+3\mathcal{H}[f_{R}'^{2}+2(1+f_{R})^2\mathcal{H}']+
3(1+f_{R})f_{R}'(-2\mathcal{H}^{2}+k^2+\mathcal{H}').\nonumber\\
&&\label{Phi&Psi_denominator}
\end{eqnarray}
The second step will be to derive equations \eqref{Phi_desp} and
\eqref{Psi_desp} with respect to  $\tau$ and thus $\Phi'$ and
$\Psi'$ may be rewritten algebraically in terms of
$\{\Phi'',\Psi'';\delta,\delta',\delta'' \}$. These last results
can be substituted in equations \eqref{00_pert} and
\eqref{0i_pert} to obtain the potentials $\Phi$ and $\Psi$ just in
terms of  $\{\Phi'',\Psi'';\,\delta,\delta',\delta''\}$. So at
this stage, let us summarize that we have been able to express the
following quantities
\begin{eqnarray}
\Phi \,&=&\,\Phi(\Phi'',\Psi''; \delta,\delta',\delta'') \nonumber\\
\Psi\,&=&\,\Psi(\Phi'',\Psi''; \delta,\delta',\delta'') \nonumber\\
\Phi'\,&=&\,\Phi'(\Phi'',\Psi''; \delta,\delta',\delta'') \nonumber\\ \Psi'\,&=&\,\Psi'(\Phi'',\Psi'';\delta,\delta',\delta'')
\label{Todos}
\end{eqnarray}
but we do not do here explicitly. By
the previous expressions we mean that the functions on the l.h.s.
depend on the functions inside the parenthesis on the r.h.s. in an algebraic way.

The natural reasoning at this point would be to try to obtain the
potentials second derivatives $\{\Phi'',\Psi''\}$ in terms
of $\{\delta,\delta',\delta''\}$ by an algebraic process. The
chosen equations to do so will be \eqref{Nabla}
and the first derivative of \eqref{ij_pert} with respect to $\tau$.
In \eqref{Nabla}  it is necessary to substitute $\Phi$
and $\Psi'$ by the expressions obtained in \eqref{Todos}
whereas \eqref{ij_pert} first derivative may be sketched as follows
\begin{equation}
\Phi'-\Psi'= - \frac{f_{RR}}{1+f_{R}}\delta R'
+ \left[\frac{f_{RR}f_{R}'-f_{RR}'(1+f_{R})}{(1+f_{R})^2}\right]\delta R.
\label{ij_pert_derivative}
\end{equation}
Before deriving, we are going to substitute
$\Psi''$ that appears on \eqref{ij_pert} by
lower derivatives potentials $\{\Phi,\Psi,\Phi', \Psi'\}$, $\delta$
and its derivatives. To do so we consider
\eqref{00_pert} and \eqref{0i_pert} first derivatives with
respect to $\tau$ where the quantity $v$ has been previously
substituted by its expression in \eqref{Nabla0}. Following
this process we may express $\Psi''$ as follows
\begin{eqnarray}
\Psi''\,=\,\Psi''(\Phi, \Psi, \Phi', \Psi'; \delta, \delta', \delta'')
\label{Psi2_function}
\end{eqnarray}
and now substituting the previous result \eqref{Psi2_function} in
$\delta R$ definition given by expression \eqref{ij_pert}, we can
derive equation \eqref{ij_pert} with respect to $\tau$. Solving a
two algebraic equations system with equations \eqref{Nabla} and
\eqref{ij_pert_derivative} and introducing \eqref{Todos} we are
able to express $\{\Phi'',\Psi''\}$ in terms of
$\{\delta,\delta',\delta'',\delta'''\}$.
\begin{eqnarray}
\Phi''\,=\,\Phi''(\delta,\delta',\delta'',\delta''') \,\,;\,\,  \Psi''\,=\,\Psi''(\delta,\delta',\delta'',\delta''').
\label{Phi2&Psi2_fuentes}
\end{eqnarray}

Thus we substitute the results obtained in
\eqref{Phi2&Psi2_fuentes} straightforwardly in \eqref{Todos} in
order to express $\{\Phi, \Psi, \Phi ', \Psi'\}$ in terms of
$\{\delta , \delta', \delta'', \delta'''\}$. With the two
potentials and their first derivatives as algebraic functions of
$\{\delta, \delta', \delta'', \delta'''\}$, we perform the last
step: We consider $\Phi(\delta, \delta, \delta'', \delta''')$ and
derive it with respect to $\tau$. The result should be equal to
$\Phi'(\delta, \delta', \delta'', \delta''')$ so we only need to
express together these two results obtaining a fourth order
differential equation for $\delta$.
Once this fourth order differential equation has been solved we
may go backwards and by using the results for $\delta$  we obtain
$\{\Phi'', \Psi''\}$ from \eqref{Phi2&Psi2_fuentes} as conformal
time $\tau$ functions. Analogously from \eqref{Todos} the behaviour
of the
 potentials $\{\Phi, \Psi\}$ and their first derivatives could be
determined.

The resulting equation for $\delta$  can be written as follows:
\begin{eqnarray}
\beta_{4,f}\delta^{iv}+\beta_{3,f}\delta'''
+(\alpha_{2,\text{EH}}+\beta_{2,f})\delta''
+(\alpha_{1,\text{EH}}+\beta_{1,f})\delta'+
(\alpha_{0,\text{EH}}+\beta_{0,f})\delta \,=\,0
\label{delta_equation_separated}
\end{eqnarray}
where the coefficients $\beta_{i,f}$  $(i\,=\,0,...,4)$
involve terms with $f_{R}'$ and $f_{R}''$, i.e.
terms disappearing if the choice $f_{R}$ constant is made.
Equivalently,  $\alpha_{i,\text{EH}}$ $(i\,=\,0,1,2)$ contain
terms coming from EH term and the linear part in $R_{(0)}$ of $f(R)$.

It is very useful to define the parameter $\epsilon \equiv
\mathcal{H}/k$ since it will allow us to perform a perturbative
expansion of the previous coefficients $\alpha$'s and $\beta$'s in
the sub-Hubble limit. Other dimensionless parameters which will be
used are the following:
\begin{eqnarray}
\kappa_{i} \equiv \frac{\mathcal{H}^{(i)}}{\mathcal{H}^{i+1}}\,\,\,\,\, i=1,2,3\,\,\,\,;\,\,\,\,
f_{i}\equiv \frac{f_{R}^{(i)}}{\mathcal{H}^{i} f_{R}}\,\,\,\,\, i=1,2.
\label{dimensionless_parameters}
\end{eqnarray}
where superindex $(i)$ means $i^{th}$ derivative with respect to time $\tau$.
Expressing now the $\alpha$'s and $\beta$'s coefficients as parameter $\epsilon$ expansions,
we may write
\begin{eqnarray}
\alpha_{i,\text{EH}}\,&=&\,\sum_{k=1}^{3}\alpha^{(k)}_{i,\text{EH}}\,\,\,\,\, i=0,1,2 \nonumber\\
\beta_{i,f}\,&=&\,\sum_{k=1}^{7}\beta^{(k)}_{i,f} \,\,\,\,\, i=3,4
\nonumber\\
\beta_{i,f}\,&=&\,\sum_{k=1}^{8}\beta^{(k)}_{i,f} \,\,\,\,\, i=0,1,2
\end{eqnarray}
where two consecutive terms in each series differ in a $\epsilon^2$
factor. The expressions for the coefficients are too long to be
written explicitly. Instead, in the following sections we shall
show different approximated formulae which are proven to be useful
in certain limits.
As a consistency check, we find that, both in a matter dominated
universe and in $\Lambda \text{CDM}$, all $\beta$ coefficients are
absent since $f_1$ and  $f_2$ defined by expression
\eqref{dimensionless_parameters} vanish identically. For these
cases, equation \eqref{delta_equation_separated} becomes equation
\eqref{delta_0_SubHubble} as expected. For instance, in the pure
matter dominated case, coefficients $\kappa$'s are constant and
they take the following values $\kappa_1\,=\,-1/2$,
$\kappa_2\,=\,1/2$, $\kappa_3\,=\,-3/4$ and $\kappa_4\,=\,3/2$.

Another important feature from our results is that, in general,
without imposing $\vert f_{R}\vert\ll 1$,
the quotient
\begin{eqnarray}
\frac{\alpha_{1,\,\text{EH}}+\beta_{1,\,f}}{\alpha_{2,\,\text{EH}}+\beta_{2,\,f}}
\end{eqnarray}
is not always equal to $\mathcal{H}$. In fact only the quotients
\begin{eqnarray}
\frac{\beta^{(1)}_{1,\,f}}{\beta^{(1)}_{2,\,f}} \,\,\,\text{
and}\,\,\, \frac{\alpha^{(1)}_{1, \text{EH}}}{\alpha^{(1)}_{2,
\text{EH}}}
\end{eqnarray}
are identically equal to $\mathcal{H}$. This last result, namely
$\alpha^{(1)}_{1, \text{EH}}/\alpha^{(1)}_{2, \text{EH}}$, is in
perfect agreement with $\delta'$ coefficient in expression
\eqref{delta_0_SubHubble} when one is studying sub-Hubble modes in
$\Lambda\text{CDM}$ theory, i.e., when $\beta_{i,\,f}$ $i =0,...,4$ are not
present.
\subsection{Evolution of sub-Hubble modes and the quasi-static approximation}
\label{subsec:Perturbations:Quasi-static approximation} We are
interested in the possible effects on the growth of density
perturbations once they enter the Hubble radius in the matter
dominated era. In this case $\mathcal{H}\ll k$ and therefore the
sub-Hubble limit $\epsilon\ll 1$ can be considered. It can be seen
that the $\beta_{4,f}$ and $\beta_{3,f}$ coefficients are
suppressed by $\epsilon^2$ with respect to $\beta_{2,f}$,
$\beta_{1,f}$ and $\beta_{0,f}$, i.e., in this limit the equation
for perturbations reduces to the following second order
expression:
\begin{eqnarray}
\delta''+\mathcal{H}\delta'+\frac{(1+f_{R})^{5} \mathcal{H}^{2}
(-1+\kappa_1)(2\kappa_1-\kappa_2)-\frac{16}{a^8}
 f_{RR}^{4}(\kappa_2-2)k^{8}8\pi G_{} \rho_{0}a^2}
{(1+f_{R})^{5}(-1+\kappa_1)+\frac{24}{a^8}f_{RR}^{4}(1+f_{R})
(\kappa_2-2)k^{8}}\delta \,=\,0
\label{eqn_ours}
\end{eqnarray}
where we have taken only the leading terms in the $\epsilon$
expansion for both $\alpha$ and $\beta$ coefficients.

This expression can be compared with the one usually
considered in the literature by performing the so-called quasi-static approximation, obtained after performing strong
simplifications in the perturbed equations - \eqref{00_pert},
\eqref{ii_pert}, \eqref{0i_pert}, \eqref{ij_pert},
\eqref{Nabla0} and  \eqref{Nablaj} - by
neglecting time derivatives of $\Phi$ and $\Psi$ potentials,
(see  \cite{Starobinsky}).
Thus, for instance in references \cite{Qstatic_Zhang} and \cite{Qstatic_Tsujikawa}, the quasi-static approximation is given by:

\begin{eqnarray}
\delta^{''}+\mathcal{H}\delta^{'}-\frac{1+4\frac{k^2}{a^2}
\frac{f_{RR}}{1+f_{R}}}{1+3\frac{k^2}{a^2}\frac{f_{RR}}{1+f_{R}}}
\frac{4\pi G\rho_{0}a^2}{1+f_{R}}\delta \,=\,0.
\label{Qstatic_strong_equation}
\end{eqnarray}

This approximation has been nonetheless considered as too
aggressive in \cite{Qstatic_Bean} since neglecting time
derivatives can remove important information about the evolution
of perturbations.

Note also that in equation \eqref{eqn_ours} there exists a difference in a
power $k^8$ between those terms coming from the $f$-part and those coming
from the EH-part.
This result differs from that in the quasi-static
approximation where difference is in a power $k^2$ according to expression \eqref{Qstatic_strong_equation}.

\begin{figure}[h]
\begin{center}
\resizebox{8.8cm}{6.4cm}
{\includegraphics{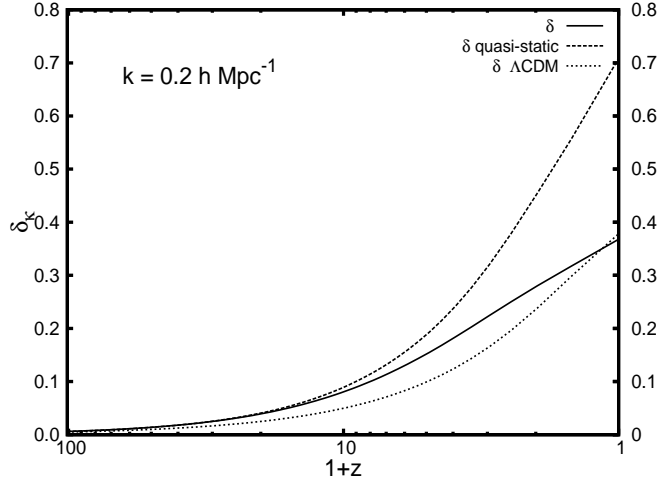}}
\caption {\footnotesize $\delta_{k}$ with $k = 0.2\,h\,\text{Mpc}^{-1}$
for $f_{test}(R)$ model and $\Lambda\text{CDM}$. Both standard
quasi-static evolution and equation \eqref{eqn_ours} have
been plotted in the redshift range from 100 to 0.}
\label{Figure_Models_delta_Falso}
\end{center}
\end{figure}

In order to compare the evolution for both equations,
we have considered a specific function
\begin{eqnarray}
f_{test}(R)=-4\,R^{0.63}
\label{f_test}
\end{eqnarray}
where  $H_0^2$ units have been used, which gives rise to a matter
era followed by  a late time accelerated phase with the correct
deceleration parameter today. In fact this model belongs to Class
{\bf II} $f(R)$ models presented in Section 2.4 and
it is therefore cosmologically viable.

Initial conditions in the matter era were given at redshift $z
=485$ where the EH-part was  dominant. Results, for
$k=0.2\,h\,\text{Mpc}^{-1}$ are presented in Figure
\ref{Figure_Models_delta_Falso}. It can be seen that, as expected,
both expressions give rise to the same evolutions at early times
(large redshifts) where they also agree with the standard
$\Lambda\text{CDM}$ evolution. However, at late times the
quasi-static approximation fails to describe the evolution of
perturbations correctly.

Notice that this $f(R)$ model given by expression \eqref{f_test} satisfies all the viability conditions
described in Section \ref{sec:Int:fR:Constraints}
except for the local gravity tests implemented by the condition {\bf 4} in that section.
As is proven below, it is precisely this last condition namely
\begin{eqnarray}
\vert f_{R}\vert\ll 1
\end{eqnarray}
which in fact ensures the validity of the quasi-static approximation. Therefore we shall now
restrict ourselves to $f(R)$ models satisfying all the viability conditions, including
$\vert f_{R}\vert\ll 1$.

In Appendix \ref{Appendix I} we have reproduced  all the $\alpha$'s  and
the first four $\beta$'s coefficients for each
$\delta$ term in \eqref{delta_equation_separated}.
When the sub-Hubble modes are studied
and the condition $\vert f_{R}\vert\ll 1$ is imposed, it
can shown that the dominant contributions are the first four $\beta$
coefficients of the $f$-part plus the first $\alpha$ coefficient of the EH-part
for each term in equation \eqref{delta_equation_separated}. Thus in this case, the full differential
equation \eqref{delta_equation_separated} can be simplified as
\begin{eqnarray}
c_{4}\delta^{iv}+c_{3}\delta^{'''}+c_{2}\delta^{''}+c_{1}\delta^{'}+c_{0}\delta\,=\,0
\label{eq_delta_cs}
\end{eqnarray}
where coefficients $c_{0,1,...,4}$ are:
\begin{eqnarray}
c_{i}\,&\equiv&\,\lim_{\vert f_{R}\vert\ll 1}\left(\alpha^{(1)}_{i,\,\text{EH}}+\sum_{j=1}^{4}\beta^{(j)}_{i,\,f}\right)\,\,\, i=0,1,2\nonumber\\
c_{k}\,&\equiv&\,\lim_{\vert f_{R}\vert\ll 1}\sum_{j=1}^{4}\beta^{(j)}_{k,\,f}\,\,\,k=3,4
\end{eqnarray}
once the condition $\vert f_{R}\vert\ll 1$ has been imposed on the corresponding $\alpha$'s and $\beta$'s contributions. These
coefficients $c_{0,...,4}$ have been explicitly written in Appendix \ref{Appendix II}.
We see that indeed in the sub-Hubble limit the $c_4$ and $c_3$
coefficients are negligible and the equation can be reduced to a
second order expression. Moreover, for our approximated expression
\eqref{eq_delta_cs}  it is true that $c_1/c_2\,\equiv\,\mathcal{H}$ as can
be seen in Appendix \ref{Appendix II} straightforwardly.

From those expressions in Appendix \ref{Appendix II}, the
second order equation for $\delta$ becomes
\begin{eqnarray}
&&\delta''+\mathcal{H}\delta'-\frac{4}{3}\frac{\left[\frac{6 f_{RR}k^2}{a^2}+\frac{9}{4}\left(1-\sqrt{1-\frac{8}{9}\frac{2\kappa_1-\kappa_2}{-2+\kappa_2}}\right)\right]\left[\frac{6 f_{RR}k^2}{a^2}+\frac{9}{4}\left(1+\sqrt{1-\frac{8}{9}\frac{2\kappa_1-\kappa_2}{-2+\kappa_2}}\right)\right]}{
\left[\frac{6f_{RR}k^2}{a^2}+\frac{5}{2}\left(1-\sqrt{1-\frac{24}{25}\frac{-1+\kappa_1}{-2+\kappa_2}}\right)\right]\left
[\frac{6 f_{RR}k^2}{a^2}+\frac{5}{2}\left(1+\sqrt{1-\frac{24}{25}\frac{-1+\kappa_1}{-2+\kappa_2}}\right)\right]}\nonumber\\
&&\times(1-\kappa_1)\mathcal{H}^{2}\delta\,=\,0
\label{Qstatic_ours}
\end{eqnarray}
which can also be written as:
\begin{eqnarray}
\delta''+\mathcal{H}\delta'-\frac{4}{3}\frac{\big(\frac{6f_{RR}k^2}{a^2}+\frac{9}{4}\big)^{2}-\frac{81}{16}+\frac{9}{2}\frac{2\kappa_1-\kappa_2}{-2+\kappa_2}}{\big(\frac{6f_{RR}k^2}{a^2}+\frac{5}{2}\big)^{2}-\frac{25}{4}+6\frac{-1+\kappa_1}{-2+\kappa_2}}(1-\kappa_1)\mathcal{H}^{2}\delta\,=\,0.
\label{Qstatic_ours_2}
\end{eqnarray}
Note that the quasi-static
expression \eqref{Qstatic_strong_equation}
is only recovered in the dust matter era (i.e. for $\mathcal{H}=2/\tau$) or for
a pure  $\Lambda\text{CDM}$ evolution  for the background dynamics.
Nonetheless, in the considered limit $|f_{R}|\ll1$
it can be proven, using the background equations of motion, that
\begin{eqnarray}
1+\kappa_1-\kappa_2 \approx 0
\end{eqnarray}
and therefore
$2\kappa_1-\kappa_2 \approx -2+\kappa_2 \approx -1
+ \kappa_1$ what allows to simplify the
expression \eqref{Qstatic_ours_2} to
become \eqref{Qstatic_strong_equation}.
This is nothing but the fact that for viable models
the background evolution resembles that of $\Lambda\text{CDM}$
\cite{Silvestri_D77_2008}.

In other words, although for general $f(R)$ functions the quasi-static approximation is not justified, for
those viable $f(R)$ functions describing the present phase of accelerated expansion and satisfying local gravity tests,
it does give a correct description for the evolution of perturbations. This result has been here stated for the first time shedding some light about the controversy which remained about the validity of the quasi-static approximation.
\begin{figure}[h]
\begin{center}
\resizebox{8.8cm}{6.4cm}
{\includegraphics{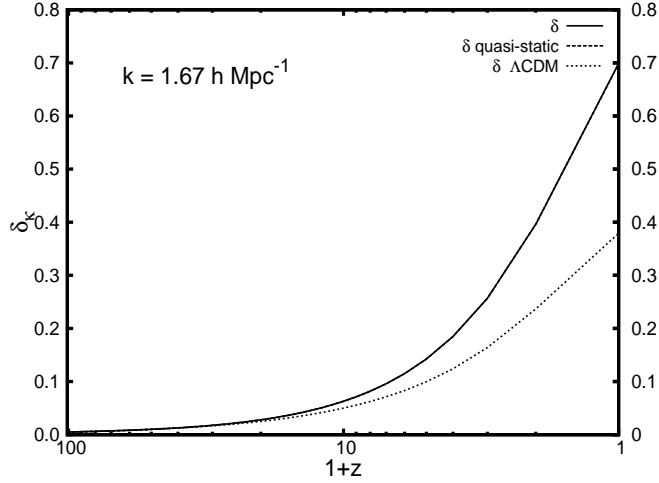}}
\caption {\footnotesize $\delta_{k}$ with $k = 1.67\,
h\,\text{Mpc}^{-1}$ for $f(R)$ model $\bf{A}$ evolving
according to equation
\eqref{Qstatic_ours},
$\Lambda\text{CDM}$  and quasi-static approximation
given by equation \eqref{Qstatic_strong_equation}
in the redshift range from 100 to 0. The quasi-static
evolution is indistinguishable from that coming from the equation
(\ref{Qstatic_ours}), but diverges
from $\Lambda\text{CDM}$ behaviour as $z$ decreases.}
\label{Figure_Model_2_k5000}
\end{center}
\end{figure}
\begin{figure}[h]
\begin{center}
\resizebox{8.8cm}{6.4cm}
{\includegraphics{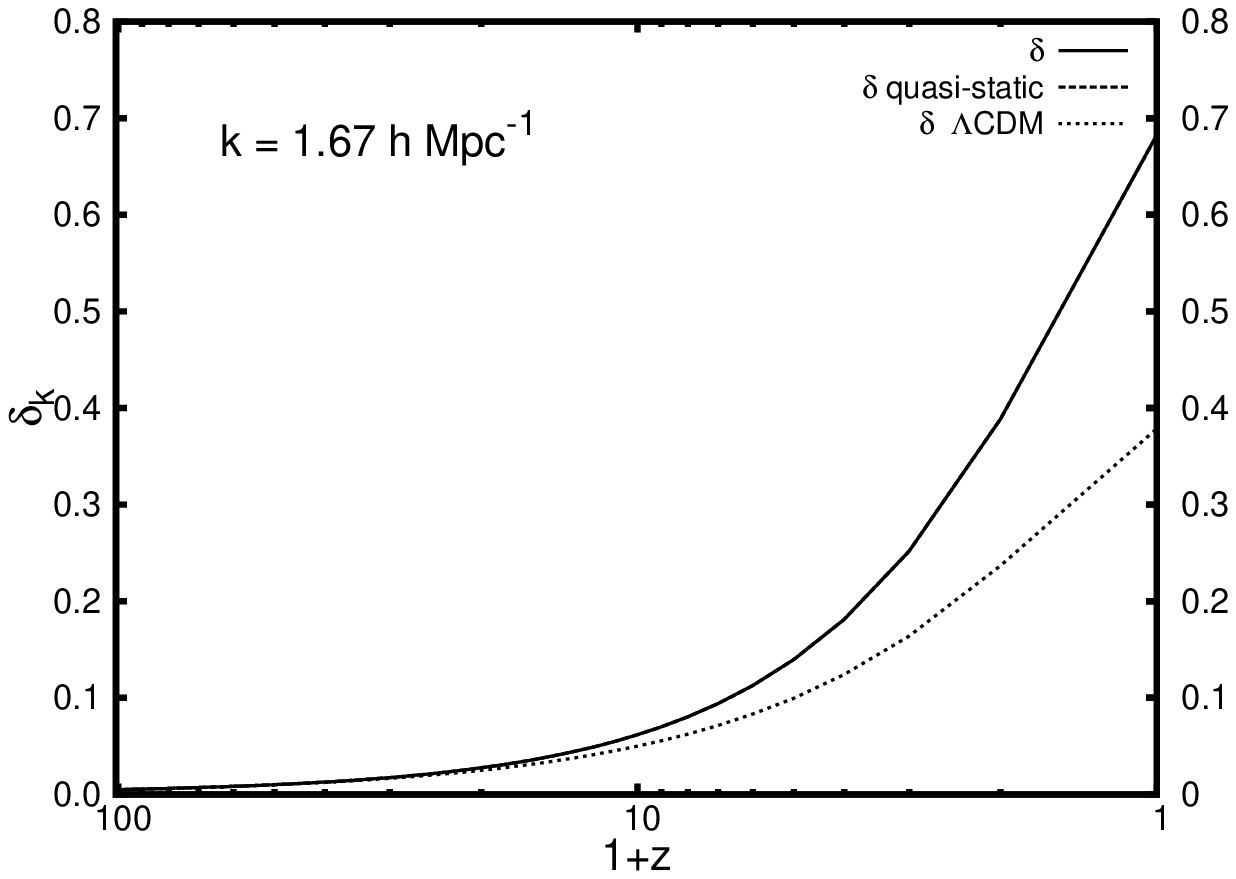}}
\caption {\footnotesize $\delta_{k}$ with $k = 1.67
\,h\,\text{Mpc}^{-1}$ for $f(R)$ model $\bf{B}$ evolving according to
equation \eqref{Qstatic_ours},
$\Lambda\text{CDM}$  and quasi-static evolution
given by equation \eqref{Qstatic_strong_equation}
 in the redshift range from 100 to 0. The quasi-static
evolution is indistinguishable from that coming from the equation
(\ref{Qstatic_ours}), but diverges
from $\Lambda\text{CDM}$ behaviour as $z$ decreases.}
\label{Figure_Model_3_k5000}
\end{center}
\end{figure}
\begin{figure}[h]
\begin{center}
\resizebox{8.8cm}{6.4cm}
{\includegraphics{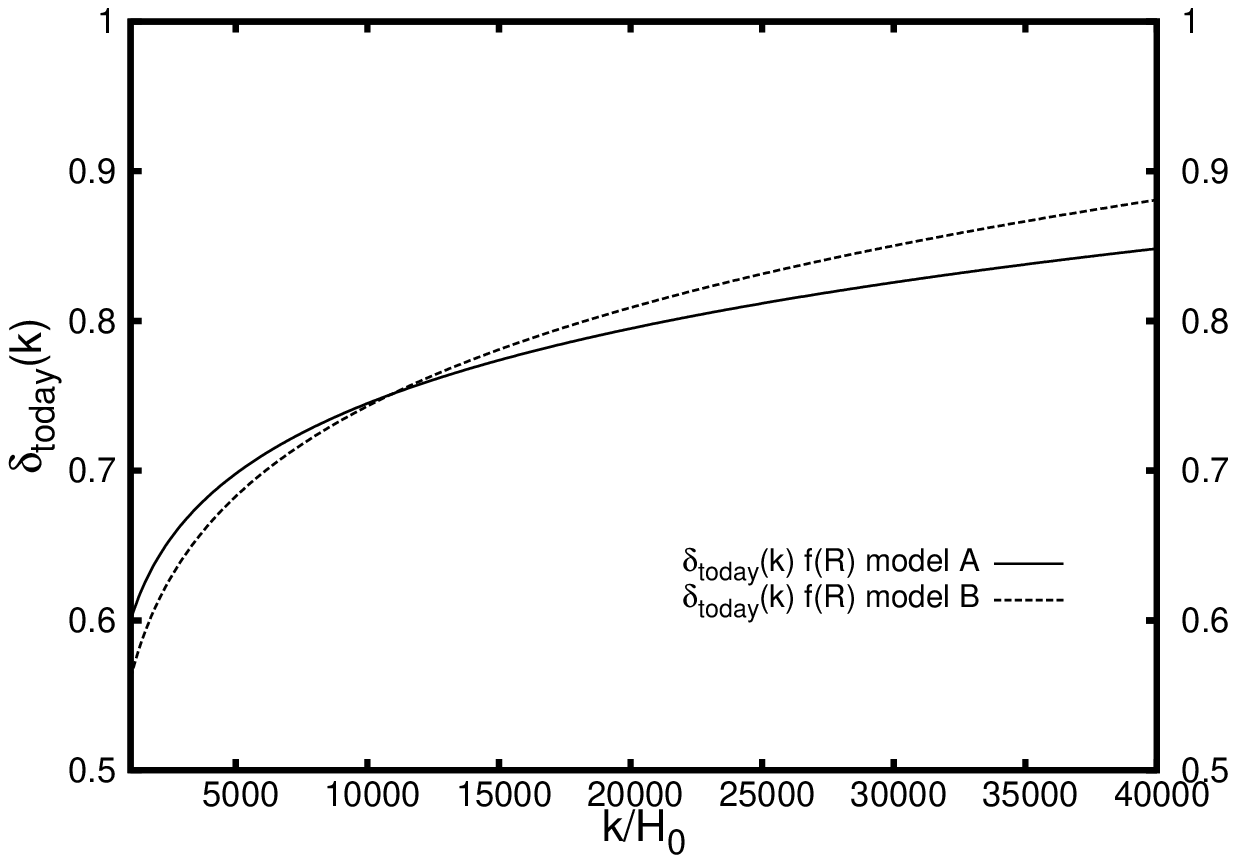}}
\caption {\footnotesize Scale dependence of $\delta_{k}$ evaluated
today $(z=0)$ for $k/H_{0}$ in the  range from $10^3$ to $4\cdot 10^4$.}
\label{Figure_Models_delta_vs_k}
\end{center}
\end{figure}

\subsection{Some proposed models}
\label{subsec:Perturbations:Some proposed models}
In order to illustrate the results obtained in the previous section, we propose two particular $f(R)$ theories which
allow us to determine - at least numerically - all  the quantities involved in the calculations and therefore to obtain  solutions for the equation \eqref{Qstatic_ours}.

As was proven before, the background evolution for viable $f(R)$ models resembles that of $\Lambda \text{CDM}$
at low redshifts  and that of a matter dominated universe at high redshifts.
Nevertheless, the $f(R)$ contribution gives the dominant contribution to the gravitational action \eqref{action} for small curvatures and therefore it may explain the cosmological acceleration. For the sake of concreteness, the models parameters have been fixed by imposing a deceleration parameter today $q_{0} \approx -0.6$.

Thus, our first model $\bf{A}$ will be:
\begin{eqnarray}
f(R) \,=\, c_{1} R^p.
\end{eqnarray}
According to the results presented in references \cite{Amendola&al:2007} and \cite{Sawicki} viable models
of this type belong to Class {\bf II} introduced in Chapter 2. As was mentioned there, they both include matter
dominated and late time accelerated eras provided the parameters satisfy
$c_{1} < 0$ and $0 < p < 1$. We have chosen $c_1 \,=\,-4.3$ and $p\,=\,0.01$
in $H_0^2$ units.
This choice does  verify all the viability conditions, including
$\vert f_{R} \vert \ll 1$ today.\\
\\
As a second model $\bf{B}$ we have chosen:
\begin{eqnarray}
f(R)\,=\,\frac{1}{c_1 R^{e_1}+c_2}
\end{eqnarray}
with values $c_1\,=\,2.5\cdot 10^{-4}$\,,\,$e_1\,=\,0.3$  and $c_2\,=\,-0.22$ also in the same units.

For each model, we compare our result \eqref{Qstatic_ours} with
the standard $\Lambda\text{CDM}$ evolution and the quasi-static approximation
\eqref{Qstatic_strong_equation} by plotting $\delta$ evolution in Figures
\ref{Figure_Model_2_k5000} and \ref{Figure_Model_3_k5000}
for models {\bf A} and {\bf B} respectively.
In both cases, the initial conditions are given at redshift
$z = 1000$ where $\delta$ is assumed to behave as in a
matter dominated universe, i.e. $\delta_{k}(\tau)\propto a(\tau)$
with no $k$-dependence.  We see that for both models,
the quasi-static approximation gives a correct
description for the evolution which clearly deviates from the $\Lambda\text{CDM}$
case.

In Figure \ref{Figure_Models_delta_vs_k} the density contrast
evaluated today was plotted  as a function of $k$ for both models.
The growing dependence of $\delta$ with respect to $k$ is
verified. This modified $k$-dependence with respect to the
standard $\Lambda\text{CDM}$ model could give rise to observable
consequences in the matter power spectrum, as shown in
\cite{Starobinsky2, PRLcomment09}, and could be used to constrain
or even to discard $f(R)$ theories for cosmic acceleration as will
be done in the next section.
\section{A viable $f(R)$ model different from $\Lambda\text{CDM}$?}
\label{sec:Perturbations:Techniques to rule out}
Some modified $f(R)$ gravity models have recently been proposed (see for instance \cite{quartin}) claiming to be
cosmologically viable in spite of
having a cosmological behaviour clearly distinguishable from $\Lambda\text{CDM}$.
Contrary to already mentioned opinions which consider that self-consistent $f(R)$ gravity models distinct from $\Lambda\text{CDM}$ are almost ruled out, authors in \cite{quartin} seem to claim that their proposed model would be cosmologically viable. We have shown \cite{PRLcomment09} that although that model does satisfy some consistency conditions, precisely because of its departure from $\Lambda\text{CDM}$ behaviour, it does not satisfy local gravity constraints and, in addition, the predicted matter power spectrum conflicts with SDSS data provided in reference \cite{Tegmark}.
The proposed $f(R)$ model reads
\begin{eqnarray}
f(R)\,=\,-\alpha\,R_{*}\text{log}\left(1+\frac{R}{R_{*}}\right).
\label{Quartin_model}
\end{eqnarray}
This model does satisfy three of the usual viability conditions
for $f(R)$ theories provided in \cite{Silvestri_D77_2008} and
specified in Section \ref{sec:Int:fR:Constraints}.
However, the model \eqref{Quartin_model} does not satisfy  the
fourth of those conditions, namely, $|f_{R}|\ll 1$  at recent
epochs, imposed by local gravity tests \cite{Hu&Sawicki_May_2007}
from solar system. Although it is still not clear what is the
actual limit on this parameter, certain estimations give $\vert
f_{R}\vert <10^{-6}$ today. This condition also ensures that the
cosmological evolution at late times resembles that of
$\Lambda\text{CDM}$. However, for the model \eqref{Quartin_model},
$\vert f_R\vert\sim 0.2$ today for $\alpha=2$ and $\Omega_{M}\sim
0.25$.

If we are only interested in considering large scales, local gravity inconsistencies could be ignored, but still deviations from $\Lambda\text{CDM}$ can have drastic cosmological consequences on the evolution of density perturbations, as discussed by several authors \cite{Starobinsky2,Cruz,Qstatic_Bean}.

Thus, the linear evolution of  matter density perturbations for sub-Hubble ($k\gg \mathcal{H}$) modes in $\Lambda\text{CDM}$ is given by equation \eqref{delta_0_SubHubble}.
Notice that in this equation the evolution of the Fourier modes does not depend on $k$. This means that once the density contrast starts growing after matter-radiation equality, the mode evolution only changes the overall normalization of the matter power-spectrum, 
but not its shape.

However, in $f(R)$ theories for sub-Hubble modes, as was thoroughly studied in the previous section, the corresponding equation reads as \eqref{eqn_ours}. Notice from this equation the $k^8$ dependence in the $\delta$ term which appears due to the fact that $f_{RR}\neq 0$.
Moreover, a careful calculation of involved contributions in
equation \eqref{eqn_ours} shows that in the $\delta$
term\footnote{Let us consider the numerator of this term for
simplicity. This does not mean any loss of generality since the denominator
behaviour is completely analogous.} there exist two contributions:
one term is proportional to $k^8$, which is coming from the $\beta$'s contribution, i.e. $f$-part, and another term coming from $\alpha$'s contribution, i.e. EH-part. At high enough redshift the EH-part term dominates, whereas at low redshift the situation is
reversed and the $f$-part term becomes dominant. This is the crucial point that explains why $k$-independent terms both in numerator and denominator cannot be straightforwardly removed from $\delta$ coefficient in equation \eqref{eqn_ours} but they have to be preserved for a correct sub-Hubble modes study.
\begin{table}
\centering \small{
\begin{tabular}{||c|c|c|c||}
\hline Redshift & EH term (in $\mathcal{H}_0$ units) & $f$ term (in $\mathcal{H}_0$ units) & $R f_{RR}$\\
\hline
\hline
$100$ & $61.35$ & $1.90\cdot10^{-4}$  & $1.45\cdot10^{-6}$\\
\hline
$50$ & $30.97$ & $5.34$ & $1.12\cdot10^{-5}$\\
\hline
$20$ & $12.69$ & $2.86\cdot10^6$ & $1.58\cdot10^{-4}$\\
\hline
$0$ & $-0.45$ & $1.01\cdot10^{17}$  & $0.079$\\
\hline
\end{tabular}
} \caption{\footnotesize{Values (in $\mathcal{H}_0$ units) for both the first EH term and the first $f$ term (the one proportional to $k^8$) in the numerator of the $\delta$ coefficient in equation \eqref{eqn_ours} for this model \eqref{Quartin_model}.
Different redshifts have been considered and the studied scale was $k=0.33\,h\,\text{Mpc}^{-1}$.
The EH term, which is $k$ independent, cannot be ignored at high redshift. In fact, the redshift at which
EH and $f$ terms contributions (for this value of $k$) equal is around $z=45$.
The strong suppression with the redshift of the $f$-part term -- which is proportional to
$k^8$ in the $\delta$ coefficient in equation \eqref{eqn_ours} --
comes from the rapid suppression of $f_{RR}^4$ factor as the redshift increases. The dimensionless quantity $R\,f_{RR}$ has also been included in the last column of the table and it is seen to grow as the redshift decreases ($z\rightarrow0$).
}}
\label{table_Quartin_model}
\end{table}
\begin{figure}[h]
\begin{center}
\begin{center}
{\epsfxsize=10.7cm\epsfbox{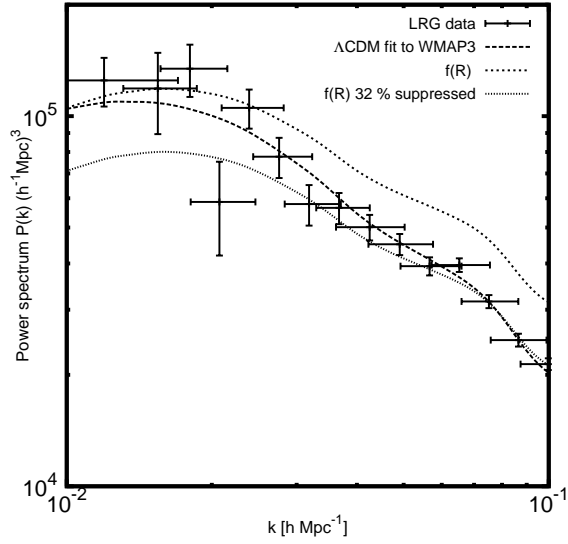}} \end{center}
\caption {\footnotesize Linear matter power-spectra for $\Lambda\text{CDM}$ and
$f(R)$ in \cite{quartin} with $\alpha=2$. Data were taken from SDSS \cite{Tegmark}.}
\label{Figure_PRL}
\end{center}
\end{figure}
\begin{figure}[h]
\begin{center}
\begin{center}
{\epsfxsize=10.7cm\epsfbox{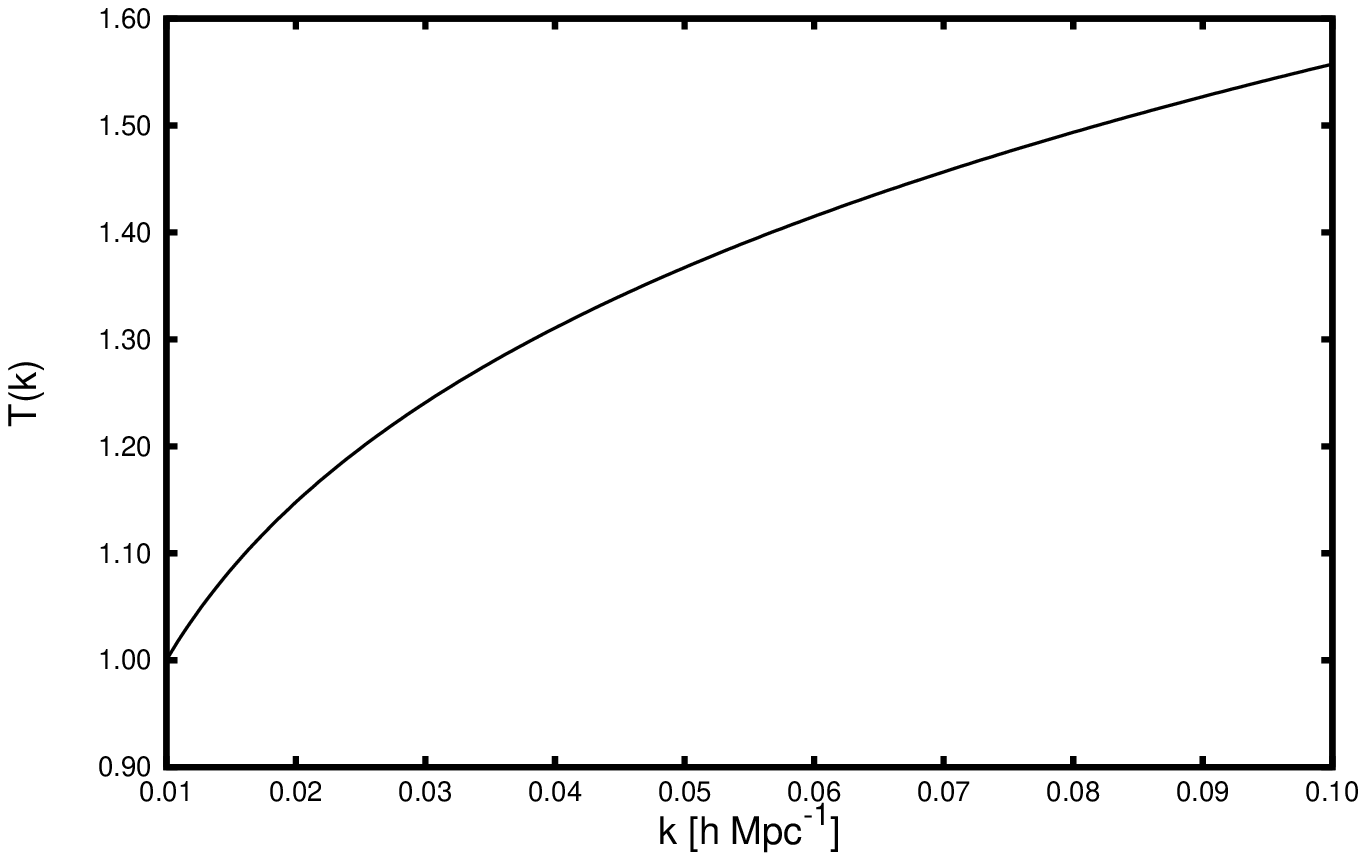}} \end{center}
\caption {\footnotesize Transfer function $T(k)$ for $f(R)$ model given in equation \eqref{Quartin_model} with $\alpha=2$ and $\Omega_M\sim0.25$. The $k$ dependence for $T(k)$ with this parameters choice was seen to be $T(k)\propto k^{0.19}$ in the plotted range of scales. Both the numerical $T(k)$ and the power law proportional to $k^{0.19}$ were plotted in scales
$k\,=\, 0.01-0.10\,h\,\text{Mpc}^{-1}$ and no difference was observed.}
\label{Figure_PRL_2}
\end{center}
\end{figure}
Thus for instance, for $k=0.33 \,h\,\text{Mpc}^{-1}$, we give explicit values in Table \ref{table_Quartin_model} for the terms in the numerator of the $\delta$ coefficient in equation \eqref{eqn_ours}.

As a consequence the matter power-spectrum $P_{k}^{f(R)}$ is further processed after equality
and would differ today from the standard $\Lambda\text{CDM}$ power spectrum $P_{k}^{\Lambda\text{CDM}}$. These two quantities would be related by a linear transfer function $T(k)$ given by:
\begin{eqnarray}
P_{k}^{f(R)}=T(k)\,P_{k}^{\Lambda\text{CDM}}\,.
\label{transfer}
\end{eqnarray}
This fact changes the shape of the matter power spectrum dramatically, as shown in Figure \ref{Figure_PRL}, where normalization to WMAP3 \cite{WMAP3} was imposed.
In this figure, SDSS data from luminous red galaxies \cite{Tegmark}
and the $\Lambda\text{CDM}$ power spectrum from the linear perturbation theory with
WMAP3 cosmological data \cite{WMAP3} are also shown.
Notice that $\Lambda\text{CDM}$ gives an excellent fit to data with $\chi^2=11.2$, whereas for the $f(R)$ theory
$\chi^2=178.9$, i.e. $13\sigma$ out. Even if the overall normalization is drastically reduced by
a 20$\%$, which is the present uncertainty over this parameter, the discrepancy would still remain at the 7$\sigma$ level.
Actually, leaving the power spectrum normalization as a free parameter, the best fit would require a $32\%$ normalization reduction and still would be $4.8\sigma$ away as seen in Figure \ref{Figure_PRL}.

The linear transfer function \eqref{transfer} for this model has been plotted in Figure \ref{Figure_PRL_2} and it has been seen that $T(k)$ follows with a nice fit a power law in the plotted interval $T(k)=(k/k_{eq})^{0.19}$ where $k_{eq}\simeq10^{-2}\,h\,\text{Mpc}^{-1}$ corresponds to the physical scale entering the Hubble radius when matter-radiation equality happened.
\newpage
\section{Conclusions}
\label{sec:Perturbations:Conclusions perturbations}
In this chapter we have studied the evolution of matter density perturbations in $f(R)$ theories of gravity. Thus we have presented a completely general procedure to obtain the exact fourth order differential equation for the evolution of scalar perturbations in the longitudinal gauge. This expression is valid for any general $f(R)$ theory and applicable at any scale $k$. If the EH gravitational action -- both with and without cosmological constant -- is considered in this general expression, well-known standard results are recovered.

We have also shown that for sub-Hubble modes, the obtained expression
reduces to a second order differential equation. Hence, we have
been able to compare this result with that obtained within the
quasi-static approximation, widely used in the literature. Our research has explicitly
shown that
for arbitrary $f(R)$ functions such an
approximation is not justified.

However, if we limit ourselves to $f(R)$ theories for which $\vert f_R\vert \ll 1$ today, then the perturbative calculation for sub-Hubble
modes requires to take into account, not only the first terms,  but also higher-order terms in the $\epsilon\equiv{\cal H}/k$ parameter. In that case,
the resummation of such terms modifies the equation. Thus, this equation can be seen to be equivalent to the quasi-static case but only if
the universe expands approximately as in a matter dominated phase or in a $\Lambda\text{CDM}$ model. Finally, the fact
that for $f(R)$ models with $\vert f_R\vert \ll 1$ the background behaves today precisely as that of $\Lambda\text{CDM}$ makes the quasi-static approximation correct in those cases.

We have finally applied our results to prove that no significant
departure of $f(R)$ theories from $\Lambda\text{CDM}$ is allowed for
those models that intend to be cosmologically viable according to recent data.
In fact, the strong $k$-dependence appearing in the evolution of
perturbations has allowed to rule out $f(R)$ gravities which have recently
been claimed to be cosmologically viable.
\chapter[Black holes in $f(R)$ theories]{Black holes in $f(R)$ theories}
\label{chap:Black holes}
\section{Introduction}\par
\label{sec:BH:Introduction} We finish the exposition of our
research on $f(R)$ modified gravities by considering some aspects
derived from the study of BHs in these theories.

Following the motivation already explained in Section \ref{sec:Int:Motivation},
$f(R)$ models
may present BH solutions as GR and other alternative gravity theories do.
Therefore it is quite natural to ask about BHs features in those
gravitational theories since, on the one hand, some BHs signatures
may be peculiar to Einstein's gravity and others may be robust
features of all generally covariant theories of gravity. On the
other hand, the results obtained may lead to rule out some models
which will be in disagreement with expected physical results. For
those purposes, research on thermodynamical quantities of BHs is of
particular interest.

These attempts to detect particular
signatures from these objects could be experimentally
implemented, as was explained in Section
\ref{sec:Int:BW:BH}, at the LHC in the coming years. Therefore the
generation of mini BHs could provide important
information about the correct underlying gravity theory.

%
Previous literature on $f(R)$ theories \cite{Whitt}  proved, by
previously performing a conformal transformation in the gravitational action,
that  Schwarzschild solution is the only  static spherically
symmetric solution for an action  of the form $R+aR^2$ in $D=4$.
Also by using this conformal transformation,
uniqueness theorems of spherically symmetric solutions for
general polynomial actions in arbitrary dimensions 
were proposed in \cite{Mignemi}
(see also \cite{Multamaki} for additional results and \cite{olmo} for
spherical solutions with sources).

Using the Euclidean action method \cite{Witten, Hawking&Page} in
order to determine different thermodynamical quantities, anti-de
Sitter ($AdS$) BHs in $f(R)$ models have been studied
\cite{Cognola}. In  \cite{Briscese} the entropy of
Schwarzschild-de Sitter BHs was calculated for some particular
cosmologically viable models in vacuum and their cosmological
stability was discussed.

BH properties have also been widely studied in other modified gravity theories: for
instance, \cite{cvetic,Cai_GaussBonet_AdS} studied BHs
in Einstein's theory with a Gauss-Bonnet
term plus a cosmological constant. Different results were found depending on
the dimension $D$ and the sign of the constant horizon curvature $k$.
For $k=0,-1$, the Gauss-Bonnet term does not
modify $AdS$ BHs thermodynamics at all (only the horizon position is modified with
respect to the EH theory) and BHs are not only
locally thermodynamically stable, but
also globally preferred. Nevertheless, for $k=+1$ and $D=5$ (for $D\geq6$ thermodynamics
is again essentially that for $AdS$ BH) there exist some features not present in
the absence of Gauss-Bonnet term. Gauss-Bonnet and/or Riemann squared interaction terms
were studied in \cite{Cho}, where the authors concluded that in this case phase transitions may occur
with $k=-1$ .

Another approach is given by Lovelock gravities, which are free of ghosts and where
the field equations contain no more than second derivatives of the metric. These theories
were for instance studied in \cite{Matyjasek} and the corresponding entropy was evaluated.

The layout of this chapter is as follows: In Section \ref{sec:BH:Constant curvature black hole solutions}
some generalities about BHs in $f(R)$ theories such as several aspects of constant curvature solutions
for static spherically symmetric cases with and without electric charge are presented.
Then in Section \ref{sec:BH:Perturbative results} a perturbative
approach around the EH action, with no previous imposition of constant curvature, is performed. There we shall find
that up to second order in perturbations only BH solutions of the Schwarzschild-$AdS$ type
are present. Explicit expressions for the effective cosmological constant are obtained
in terms of the $f(R)$ function. To deal with such differential equations we have used
the algebraic manipulation package $\tt Mathematica$~\cite{mathematica}.

Finally, we shall consider in Section \ref{sec:BH:Black hole thermodynamics} the BHs thermodynamics in $AdS$ space-time. There
it will be found that this kind of solutions can only exist provided the
theory satisfies $R_0+f(R_0)<0$ where $R_0$ holds for the constant curvature solution.
Interestingly, this expression is proved to be related to
the condition which guarantees the positivity of the effective Newton's constant
in this type of theories. In addition, it also ensures that the thermodynamical properties
in $f(R)$ gravities are qualitatively similar to those of
standard GR. Then, some consequences for local and global stability
for some particular $f(R)$ models will be provided in Section 4.5 and figures of thermodynamical regions will be shown
in Section 4.6. The present chapter is finished
by Section \ref{sec:BH:Conclusions} remarking some conclusions of the presented results.

This chapter is mainly based upon the results that have been presented
in references \cite{BH_Dombriz} and \cite{BH_Dombriz_Proceeding}.
\section{Constant curvature black-hole solutions}
\label{sec:BH:Constant curvature black hole solutions}
The most general static and spherically symmetric $D\geq 4$
dimensional metric can be written as (see \cite{ortin}):
\begin{eqnarray}
\text{d}s^2\,=\,e^{-2\Phi(r)} A(r)\text{d}t^2-A^{-1}(r)\text{d}
r^2-r^2\text{d}\Omega_{D-2}^2
\label{metric_D_v1}
\end{eqnarray}
or alternatively
\begin{eqnarray}
\text{d}s^2\,=\,\lambda(r)\text{d}t^2-\mu^{-1}(r)\text{d} r^{2}-r^2\text{d} \Omega_{D-2}^2
\label{metric_D_v2}
\end{eqnarray}
where $\text{d}\Omega_{D-2}^2$ is the metric on the $S^{D-2}$ sphere. The
identification $\lambda(r)=e^{-2\Phi(r)}A(r)$ and $\mu(r)=A(r)$ can be straightforwardly
established. For obvious reasons, the $\Phi(r)$ function is called the
{\it anomalous redshift}: notice that a photon emitted at $r$ with proper
frequency $\omega_0$ is measured at infinity with frequency
$\omega_{\infty}= \omega_{0}e^{-\Phi(r)}\sqrt{A(r)}$.


Since the metric is static, the scalar curvature $R$ in $D$ dimensions depends only on  $r$ and
it is given, for the metric parametrization \eqref{metric_D_v1}, by:
\begin{eqnarray}
R(r)\,  &=& \,\frac{1}{r^2}\left\{(D-2)\left[(D-3)(1-A(r))+2r(A(r)\Phi'(r)-A'(r))\right]\right.\nonumber\\
&&\left. +\,r^{2}\left[3A'(r)\Phi'(r)-A''(r)
-2A(r)(\Phi'^{2}(r)-\Phi''(r))\right]\right\}
%
%
%
\label{Dcurv}
\end{eqnarray}
where the prime denotes derivative with respect to $r$.
At this stage it is interesting to ask about which are the most
general static and spherically symmetric metric tensors with constant
scalar curvature $R_{0}$. This metric tensor can be found by solving the
equation $R(r)=R_0$. Thus, it is immediate to see that for a $\Phi(r)=\Phi_{0}$
constant, this equation becomes
\begin{eqnarray}
R(r)\equiv R_{0}\,=\,\frac{1}{r^2}\Big[(D^2-5D+6)(1-A(r))+rA'(r)(-2D+4)-r^{2}A''(r)\Big]
\label{eq_A_Phi_constant}
\end{eqnarray}
whose general solution is
\begin{eqnarray}
A(r)\,=\,1+a_{1}r^{3-D}+a_{2}r^{2-D}-\frac{R_0}{D(D-1)}r^2
\label{A_solution_R_constant_Dobado_procedure}
\end{eqnarray}
with $a_{1}$ and $a_{2}$ arbitrary integration constants. In fact,
for the particular case $D=4$, $R_{0}=0$ and $\Phi_{0}=0$, the
metric can be written exclusively in terms of the function:
\begin{eqnarray}
A(r)\,=\,1+\frac{a_{1}}{r}+\frac{a_{2}}{r^{2}}.
\label{RN_solution_R_constant_Dobado_procedure}
\end{eqnarray}
By establishing the identifications $a_{1}=-2G_{}M$ and
$a_{2}=Q^2$, this solution corresponds to a Reissner-Nordstr\"{o}m
solution, i.e. a charged massive BH solution with mass $M$
and charge $Q$. Further comments about this result will be made
at the end of the section.

Now that the general static and spherically symmetric solution for constant curvature has been obtained as given by expression \eqref{A_solution_R_constant_Dobado_procedure}, let us insert the metric \eqref{metric_D_v1} into the general
$f(R)$ gravitational action \eqref{action}, and let us also perform variations with
respect to the functions $A(r)$ and $\Phi(r)$, in order to find the corresponding modified Einstein equations for this parametrization. Thus, we obtain
\begin{eqnarray}
(2-D) (1+f_{R}) \Phi'(r)-r\left[f_{RRR}
R'(r)^2+f_{RR}\left(\Phi'(r)R'(r)+ R''(r)\right)\right]\,=\,0
\label{eqn_A}
\end{eqnarray}
and
\begin{eqnarray}
&&2 r A(r) f_{RRR}\,R'(r)^2+ f_{RR}\left[2(D-2)A(r)R'(r)+ 2 r A(r) R''(r)+A'(r) r R'(r)\right]
\nonumber\\&&+\,(1+f_R)\left[2r(A(r)\Phi''(r)- A(r) \Phi'(r)^{2}) + 2(D-2)A(r)\Phi'(r)-r A''(r)\right.\nonumber\\ &&\left.
+\,A'(r)(2 - D  + 3 r \Phi'(r))\right]- r(R+f(R))\,=\,0
\label{eqn_phi}
\end{eqnarray}
where as in previous chapters $f_{R}\equiv \text{d}f(R)/\text{d}R$, $f_{RR}\equiv \text{d}^2f(R)/\text{d}R^2$ and
$f_{RRR}\equiv \text{d}^3 f(R)/\text{d}R^3$.
The above equations look in principle quite difficult to solve. For this reason
we shall firstly consider the case with constant scalar curvature
$R=R_0$ solutions. In this simple case the two previous equations reduce to:
\begin{equation}
(2-D)\,(1+f_{R})\Phi'(r)=0
\label{eq_motion_phi}
\end{equation}
and
\begin{eqnarray}
&&R+f(R)+(1+f_R)\left[A''(r)+\frac{(D-2)}{r}\left(A'(r)-2A(r)\Phi'(r)\right)
-3A'(r)\Phi'(r)\right.\nonumber\\&&\left.+2A(r)\left(\Phi'(r)^2-\Phi''(r)\right)\right]\,=\,0
\label{eq_motion_A}
\end{eqnarray}
As was commented in the Section \ref{sec:Int:fR:Geometrical results},
the constant curvature solutions in vacuum of $f(R)$ gravities are given by the equation
\eqref{dif} which can be rewritten as:
\begin{equation}
R_0=\frac{D\,f(R_0)}{2(1+f_{R}(R_0))-D}
\label{const}
\end{equation}
whenever $2(1+f_{R}(R_0))\neq D$. Thus from \eqref{eq_motion_phi}  one obtains $\Phi'(r)=0$ and then, for
a constant scalar curvature $R_0$, the equation \eqref{eq_motion_A} becomes
\begin{equation}
R_{0}+f(R_0)+(1+f_{R}(R_0))\left[A''(r)+(D-2)\frac{A'(r)}{r}\right]\,=\,0\,.
\label{eqn_A_determination}
\end{equation}

By inserting expression \eqref{const} in the previous equation \eqref{eqn_A_determination}, we get
\begin{equation}
A''(r)+(D-2)\frac{A'(r)}{r}=-\frac{2}{D}R_{0}\,.
\label{A_eq_R_constant}
\end{equation}
This is a $f(R)$-independent linear second order inhomogeneous differential
equation which can be easily integrated to give the general solution:
\begin{equation}
A(r)\,=\,c_1\,+\,c_2r^{3-D}-\frac{R_0}{D(D-1)}r^2
\label{A_solution_R_constant}
\end{equation}
which depends on two arbitrary constants $c_1$ and $c_2$. However,
this solution has no constant curvature in
the general case since, as was found above, the constant curvature requirement
demands $c_{1}=1$. Then for negative $R_0$ this solution
is basically the $D$ dimensional generalization studied by Witten \cite{Witten}
of the BH in $AdS$ space-time
solution considered by Hawking and Page \cite{Hawking&Page}. With the natural choice $\Phi_0=0$, that
solution can be written as:
\begin{equation}
A(r)=1-\left(\frac{R_{S}}{r}\right)^{D-3}+\frac{r^2}{l^2}
\label{A(r)}
\end{equation}
where
\begin{equation}
R_S^{D-3}=\frac{16\pi G_D M}{(D-2)\mu_{D-2}} \label{BHmass}
\end{equation}
with
\begin{equation}
\mu_{D-2}=\frac{2\pi^{\frac{D-1}{2}}}{\Gamma\left(\frac{D-1}{2}\right)}
\end{equation}
being the area of the $D-2$ sphere,
\begin{equation}
l^2\equiv-\frac{D(D-1)}{R_0}
\label{lR0}
\end{equation}
is the asymptotic $AdS$ space scale squared and $M$ is the mass
parameter usually found in the literature.

Thus we have concluded  that the only static and spherically
symmetric vacuum solutions with constant curvature of any $f(R)$
gravity ($R_0<0$ provided) is just the Hawking-Page BH in $AdS$ space. However, this
kind of solution is not the more general static and spherically
symmetric metric with constant curvature as can be seen by
comparison with the solutions found in expression
(\ref{A_solution_R_constant_Dobado_procedure}). Therefore we have
to conclude that there are constant curvature BH solutions that
cannot be obtained as vacuum solutions of any $f(R)$ theory.

Let us now consider the case of charged BHs in $f(R)$ theories. For the sake
of simplicity we shall limit ourselves to  the $D=4$ case. The action of the theory
will be now:
\begin{equation}
S_{f(R)\text{-Maxwell}}=\frac{1}{16 \pi G}\int \text{d}^{4}x\sqrt{\mid g\mid}\,\left(R+f(R)-F_{\mu\nu}F^{\mu\nu}\right)
\end{equation}
which is a generalization of the Einstein-Maxwell theory. The tensor $F_{\mu\nu}$ is defined as:
\begin{eqnarray}
F_{\mu\nu}\,\equiv\,\partial_\mu A_\nu-\partial_\nu A_\mu\,.
\end{eqnarray}

Considering an electromagnetic potential of the form: $A_\mu=\left(V(r),\,\,\vec{0}\right)$ and the  static
spherically symmetric metric (\ref{metric_D_v2}), we find that the solution with
constant curvature $R_0$ reads:
\begin{eqnarray}
V(r)&=&-\frac{Q}{r}\,,\nonumber \\
\lambda(r)&=&\mu(r)=1-\frac{2GM}{r}+\frac{(1+f_{R}(R_0))Q^2}{r^2}-\frac{R_0}{12}\,r^2\,.
\end{eqnarray}
Notice that unlike the EH case, the contribution of the BH charge
to the metric tensor is corrected by a $\left(1+f_{R}(R_0)\right)$ factor.
\section{Perturbative results}
\label{sec:BH:Perturbative results}
In the previous section we have considered static spherically
symmetric solutions with  constant curvature. In EH theory it is
trivial to show that the only static and spherically symmetric
solution possess constant scalar curvature. However, it is not
guaranteed this to be the case in $f(R)$ theories too. The
problem of finding the general  static spherically symmetric
solution in arbitrary $f(R)$ theories without imposing the
constant curvature condition is in principle quite complicated. For
that reason, we shall present in this section a perturbative
analysis of the problem, assuming that the modified action given
by the expression \eqref{action} is a small perturbation around EH
theory.

The computation we are about to sketch is quite an involved
one since,  in order to calculate the solutions to a given order, it
requires to introduce previous order results.
To deal with such large equations 
we have used the algebraic manipulation package
$\tt Mathematica$~\cite{mathematica}.

Therefore let us consider a $f(R)$ function  of the form
\begin{eqnarray}
f(R)\,=-(D-2)\Lambda_{D}+\alpha g(R)
\label{expansion_en_alpha_fR}
\end{eqnarray}
where $\alpha\ll 1$ is a dimensionless parameter and $g(R)$ is assumed to be analytic
when expanded for perturbative solutions.
Note that nonanalytic functions in $\alpha$ are therefore excluded
of this analysis.
\\
\\
By using the metric parametrization given by
\eqref{metric_D_v2} the equations of motion become:
\begin{eqnarray}
&&\lambda (r) (1+f_{R}) \left\{2 \mu (r) \left[(D -2) \lambda '(r)
+r \lambda ''(r)\right]+r \lambda '(r) \mu '(r)\right\}\nonumber \\
&-&2 \lambda (r)^2 \left\{2 \mu(r)[(D -2) R'(r) f_{RR}
+r f_{RRR}\,R'(r)^2+r R''(r) f_{RR}]+r R'(r) \mu '(r) f_{RR}\right\}\nonumber\\
&-&r \mu (r) \lambda '(r)^2 (1+f_{R})+2 r \lambda (r)^2(R+f(R))=\,0
\label{eqn_lambda}
\end{eqnarray}
and
\begin{eqnarray}
&&\mu (r) \Big\{2 \lambda (r) R'(r)
\left[2 (D -2) \lambda(r)+r\lambda '(r)\right] f_{RR}
+r(1+f_{R})(\lambda '(r)^2-2 \lambda (r) \lambda ''(r))\Big\}\nonumber\\
&&-\lambda (r) \mu '(r) \left[2 (D -2) \lambda (r)
+r \lambda '(r)\right] (1+f_{R})-  2 r \lambda (r)^2 (R+f(R))\,=\,0
\label{eqn_mu}
\end{eqnarray}
where
$R\equiv R(r)$ is given by expression \eqref{Dcurv}.
Now the $\lambda(r)$ and $\mu(r)$
functions appearing in the metric \eqref{metric_D_v2} can
be written as follows
\begin{eqnarray}
\lambda(r)\,&=&\,\lambda_{0}(r)+\sum_{i=1}^{\infty}\alpha^{i}\lambda_{i}(r)    \nonumber\\
\mu(r)\,&=&\,\mu_{0}(r)+\sum_{i=1}^{\infty}\alpha^{i}\mu_{i}(r).
\label{expansion_en_alpha_lambda&mu}
\end{eqnarray}
Notice that $\lambda_{0}(r)$ and $\mu_{0}(r)$ are the unperturbed solutions for the
EH action with cosmological constant given by
\begin{eqnarray}
\mu_{0}(r)\,&=&\,1+\frac{C_1}{r^{D-3}}-\frac{\Lambda_{D}}{(D-1)}r^2\nonumber\\
\lambda_{0}(r)\,&=&\,C_{2}\,\mu_{0}(r)
\label{mu0_lambda0}
\end{eqnarray}
which are the standard BH solutions in a $D$ dimensional $AdS$
spacetime ($\Lambda_D<0$ provided). Note that the factor 
$C_2$ can be chosen by performing
a coordinate $t$ reparametrization so that both functions, $\mu_{0}(r)$ and $\lambda_{0}(r)$, could be
identified. For the moment, we shall keep the background solutions
as given in \eqref{mu0_lambda0} and we shall discuss the
possibility of getting $\lambda(r)=\mu(r)$  in the
 perturbative expansion later on.

By inserting the expressions \eqref{expansion_en_alpha_fR} and
\eqref{expansion_en_alpha_lambda&mu} in the equations \eqref{eqn_lambda} and
\eqref{eqn_mu} we obtain the following
\\
\\
\textbf{First order equations}:
\\
\begin{eqnarray}
(D-3)\mu_{1}(r)+r\mu_{1}'(r)+\frac{2\Lambda_{D}g_{R}(R_0)-g(R_0)}{D-2}r^{2}\,=\,0
\label{lambda_eqn1}
\end{eqnarray}
\\
\begin{eqnarray}
&C_{2}& \left[C_{1} (D-1) r^{3-D}-\Lambda_{D} r^2+D-1\right]g(R_{0}) r^2+\left[C_{1} (D-3)
 r^{3-D}+\frac{2\Lambda_{D}}{D-1}r^{2}\right]\lambda_{1}(r)\nonumber\\
&+&C_{2} (D-2)(D-1)\left(\Lambda_{D} r^2-D+3\right)\mu_{1}(r)\nonumber\\
&+&\left(1+C_{1} r^{3-D}-\frac{\Lambda_{D} r^2}{D-1}\right)
 \left[2 C_{2} (1-D) r^2\Lambda_{D}g_{R}(R_0)+r\lambda_{1}'(r)\right]\,=\,0\nonumber\\
&&
\label{mu_eqn1}
\end{eqnarray}
where $g_{R}(R_0)\equiv\text{d}g(R)/\text{d}R|_{R=R_0}$ and whose solutions are:
\begin{eqnarray}
\lambda_{1}(r)\,&=&\,C_{4}(D-1)(D-2)+\frac{(C_{1}C_{4}-C_{2}C_{3})(D-2)(D-1)}{r^{D-3}}
\nonumber \\
&-&\left[C_{4}(D -2)
\Lambda_{D}+C_{2}\left(g(R_0)-2\Lambda_{D} g_{R}(R_{0})\right)\right] r^{2}
\nonumber\\
&&
\label{mu1}
\end{eqnarray}
\begin{eqnarray}
\mu_{1}(r)\,=\,\frac{C_{3}}{r^{D-3}}+\frac{\left(g(R_0)-2\Lambda_{D}g_{R}(R_0)\right)}
{(D-2)(D-1)}r^{2}\,.
\label{lambda1}
\end{eqnarray}
Up to second order in $\alpha$, we obtain the following
\\
\\
\textbf{Second order equations}:
\\
\begin{eqnarray}
&&(D-3)\mu_{2}(r)+r\mu_{2}'(r) +\frac{(g(R_0)-2 \Lambda_{D}g_{R}(R_0))}{D-2}\left(g_{R}(R_0)
-\frac{2D}{D-2}\Lambda_{D}g_{RR}(R_0)\right)r^2\,=\,0\nonumber\\
&&
\label{lambda_eqn2}
\end{eqnarray}
\\
\begin{eqnarray}
&&\left[-C_{1} (D-3) r^{3-D}-\frac{2\Lambda_{D} r^2}{D-1}\right]\lambda_{2}(r)
+C_{2}(D-2)(D-1)\left(-\Lambda_{D}r^2+D-3\right)\mu_{2}(r)\nonumber\\
&-&\left(C_{1} r^{4-D}+r-\frac{r^3\Lambda_{D}}{D-1}\right)
\lambda_{2}'(r)-C_{3}C_{4} (D-2) (D-1) \left(-\Lambda_{D} r^2+D-3\right) r^{3-D}\nonumber\\
&-&C_{2}\left[(D-1)(C_{1}r^{3-D}+1)-\Lambda_{D} r^2\right]
\Big[2\Lambda_{D}g_{R}(R_0)^2 +g(R_0)\left(\frac{2D\Lambda_{D}g_{RR}(R_0)}{D-2}-g_{R}(R_0)\right)\nonumber\\
&-&\frac{4D\Lambda_{D}^{2}g_{R}(R_0)g_{RR}(R_0)}{D-2}\Big]r^{2}
-C_{4}\left[C_{1}(D-1)r^{3-D}+2\right]\left[2\Lambda_{D}g_{R}(R_0)-g(R_0)\right]r^{2}\,=\,0\nonumber\\
&&
\label{mu_eqn2}
\end{eqnarray}
where $g_{RR}(R_0)\equiv\text{d}^{2} g(R)/\text{d}R^{2}|_{R=R_0}$ and whose solutions are:
\begin{eqnarray}
\lambda_{2}(r)\,&=&\,C_{6}+\frac{C_{6} C_{1}+(C_{3}C_{4}-C_{2}C_{5})(D-2)(D-1)}{r^{D-3}}
\nonumber\\
&+&\left[-\frac{C_{6}\Lambda_{D}}{D-1}+\left(g(R_0)-2 \Lambda_{D}g_{R}(R_0)\right)
\left(C_{4}+C_{2} g_{R}(R_0)-\frac{2 C_{2} D\Lambda_{D}g_{RR}(R_0)}{D-2}\right)\right] r^2
\nonumber\\
&&
\label{lambda2}
\end{eqnarray}
\begin{eqnarray}
\mu_{2}(r)\,=\,\frac{C_{5}}{r^{D-3}}+\frac{\left(g(R_0)-2\Lambda_{D}g_{R}(R_0)\right)
\left(2 D\Lambda_{D} g_{RR}(R_{0})-(D -2) g_{R}(R_0)\right)}{(D -2)^2 (D -1)}r^{2}.
\label{mu2}
\end{eqnarray}
\\
\\
Further orders in $\alpha^{3,4,...}$ can be obtained by inserting
the previous results in the subsequent differential equations
to get $\{\lambda_{3,4,...}(r),\mu_{3,4,...}(r)\}$.
Nevertheless, these equations become increasingly complicated
to be explicitly solved.

Notice that from the obtained results up to second order in $\alpha$,
the corresponding metric has constant scalar curvature for any
value of the parameters $C_{1, 2, \dots, 6}$. As a matter of fact,
this metric is nothing but the corresponding one to the standard Schwarzschild-$AdS$ geometry,
and can be easily rewritten in the usual form  by making a
trivial time reparametrization as follows:
\begin{eqnarray}
\overline{\lambda}(r)\,&\equiv&\,\lambda(r)\left[ -C_{2} (D ^2+3D-2)
+C_{4}\left(D^2-3D +2\right)\alpha +C_6\alpha^{2}+\Od(\alpha^{3})\right]\nonumber\\
\overline{\mu}(r)\,&\equiv&\,\mu(r).
\label{lambda_reparametrization}
\end{eqnarray}

Therefore, at least up to second order, the only static,
spherically symmetric solutions which are analytical
in $\alpha$ are the standard Schwarzschild-$AdS$ space-times.
\subsection{General expression to arbitrary order for constant curvature}
\label{subsec:BH:General expression constant curvature}

Let us now assume from the very beginning that the solutions for
the equations \eqref{eqn_lambda} and \eqref{eqn_mu} belong to Schwarzschild-$AdS$ BH type at
any order in the $\alpha$ expansion. Thus we can write ($J>0$ provided):
\begin{eqnarray}
\lambda(r)\,\equiv\,\mu(r)=\,1\,-\,\left(\frac{\overline{R}_{S}}{r}\right)^{D-3} \,+\, J r^2
\label{mu_lambda}
\end{eqnarray}
as solutions for the modified Einstein equations \eqref{eqn_lambda} and \eqref{eqn_mu}
derived when the modification in the gravitational Lagrangian is given by expression \eqref{expansion_en_alpha_fR}. If we expand the quantities $\overline{R}_S$ and $J$
in terms of parameter $\alpha$ we get:
\begin{eqnarray}
\overline{R}_{S}\,&\equiv&\,R_{S}+\sum_{i=1}^{\infty} C_{i}\alpha^{i}\nonumber\\
J\,&\equiv&\,-\frac{\Lambda_{D}}{(D-1)}+\sum_{i=1}^{\infty}J_{i}\alpha^{i}
\end{eqnarray}
where $R_S$ and $C_i$ are arbitrary constants. The $J_{i}$
coefficients can be determined from expression \eqref{dif}, which here becomes:
\begin{eqnarray}
R-(D-2)\Lambda_{D} +\alpha g(R)+2(D-1)J(1+\alpha g'(R))\,=\,0
\label{algebraic_eqn}
\end{eqnarray}
with $R\,=\,-D(D-1)J$ is the obtained result when calculating $R$ with
functions given by \eqref{mu_lambda}. Expanding equation
\eqref{algebraic_eqn} in powers of $\alpha$ it is possible to find
a recurrence equation for the $J_i$ coefficients, namely for the
$J_l$ (with $l>0$) coefficient, we find:
\begin{eqnarray}
&&(2-D)(D-1)J_{l}+\sum_{i=0}^{l-1}\sum_{cond.1}\frac{1}{i_{1}!i_{2}!
\ldots i_{l-1}!}(J_{1})^{i_{1}}(J_{2})^{i_{2}} \ldots (J_{l-1})^{i_{l-1}}g^{(i)}(R_{0})
\nonumber\\
&+&2(D-1)\sum_{k=0}^{l-1}J_{k}\sum_{i=0}^{l-k-1}\sum_{cond.2}
\frac{1}{i_{1}!i_{2}!\ldots i_{l-k-1}!}(J_{1})^{i_{1}}(J_{2})^{i_{2}}\ldots
(J_{l-k-1})^{i_{l-k-1}}g^{(i+1)}(R_{0}) \,=\,0\nonumber\\
&&
\end{eqnarray}
with $R_{0}=-D(D-1)J_{0}\,\equiv\,D\Lambda_{D}$ and
$g^{(i)}(R_0)\equiv \text{d}^{(i)}g(R)/\text{d}R^{(i)}|_{R=R_0}$ .
In the previous recurrence relation, the first sum  is done under
the condition 1 given by:
\begin{eqnarray}
\sum_{m=1}^{l-1}i_{m}=i, \,\, i_{m}\,\in \, \Bbb{N}\cup\{0\}\,\,\; \mbox{and}\,\,\;
\sum_{m=1}^{l-1} m \,i_{m}=l-1
\end{eqnarray}
and the second one under the condition 2:
\begin{eqnarray}
\sum_{m=1}^{l-k-1}i_{m}=i, \,\, i_{m}\,\in \, \Bbb{N}\cup\{0\}\,\,\; \mbox{and}\,\,\;
\sum_{m=1}^{l-k-1} m \,i_{m}=l-k-1.
\end{eqnarray}
For instance we have:
\begin{eqnarray}
J_{1}\,&=&\,
\frac{\mathcal{A}(g\,;\,D,\,\Lambda_{D})}{(D-2)(D-1)}\nonumber\\
J_{2}\,&=&\,
- \frac{\mathcal{A}(g\,;\,D,\,\Lambda_{D})[(D - 2) g_{R}(R_{0})
- 2 D\Lambda_{D}g_{RR}(R_{0})]}{(D - 2)^2 (D-1)}
\label{lambda&mu_expansions}
\end{eqnarray}
\\
where $\mathcal{A}(g\,;\,D,\,\Lambda_{D})\equiv g(R_0)-2\Lambda_{D} g_{R}(R_0)$

Now we can consider the possibility of removing $\Lambda_{D}$ from
the gravitational Lagrangian \eqref{expansion_en_alpha_fR}
from the very beginning and still getting an $AdS$ BH
solution with an effective cosmological constant depending on
$g(R)$ and its derivatives evaluated at $R_{0}\equiv0 $. In this case the results,
order by order in $\alpha$ up to order $\alpha^2$, are:
\begin{eqnarray}
J_{0}(\Lambda_{D}=0)\,&=&\,0,\nonumber\\
J_{1}(\Lambda_{D}=0)\,&=&\,\frac{g(0)}{(D -2) (D -1)},\nonumber\\
J_{2}(\Lambda_{D}=0)\,&=&\,-\frac{g(0) g_{R}(0)}{(D -2) (D -1)}.
\end{eqnarray}
\\
As we see, in the context of $f(R)$ gravities, it is therefore possible
to have a BH in an $AdS$ asymptotic space-time even if the initial
cosmological constant $\Lambda_D$ vanishes but $\alpha\,g(0)>0$ and $\alpha$ 
small enough as considered in the explained reasoning.

To end this section we can summarize by saying that in the context
of $f(R)$ gravities the only spherically symmetric and
static solutions 
in the general case (without imposing constant curvature) in perturbation theory up to second order
are the standard BHs in $AdS$ space.
However, the possibility of having static and spherically symmetric solutions with nonconstant curvature cannot be excluded
in the case of $f(R)$ functions which are not analytical in $\alpha$.
\section{Black-hole thermodynamics}
\label{sec:BH:Black hole thermodynamics}

In order to consider the different thermodynamic quantities for
the $f(R)$ BHs in $AdS$ space-time, we start from the temperature. In principle,
there are two different ways of introducing this quantity for the
kind of solutions that we are considering here. Firstly we can use the
definition coming from Euclidean quantum gravity \cite{HGG}. In this case one
introduces the Euclidean time $\tau=it$ and the Euclidean metric
$\text{d}s_E^2$ is defined as:
\begin{equation}
-\text{d}s_E^2=-\text{d}\sigma^2-r^2\text{d}\Omega^2_{D-2}
\end{equation}
where:
\begin{equation}
\text{d}\sigma^2=e^{-2\Phi(r)}A(r)\text{d}\tau^2+A^{-1}(r)\text{d}r^2.
\end{equation}
The metric corresponds only to the region $r>r_H$ where
$r_H$ is the outer horizon position with $A(r_H)=0$. Expanding
$\text{d}\sigma^2$ near $r_H$ we have:
\begin{equation}
\text{d}\sigma^2=e^{-2\Phi(r_H)}A'(r_H)\rho\,
\text{d}\tau^2+\frac{\text{d}\rho^2}{A'(r_H)\rho}
\label{tau_eqn}
\end{equation}
where $\rho=r-r_H$. Now we introduce the new coordinates $\tilde R$ and
$\theta$ defined as:
\begin{eqnarray}
\theta=\frac{1}{2}e^{-\Phi(r_H)}A'(r_H)\tau\,\,\,\,;\,\,\,\,
\tilde R=2\sqrt{\frac{\rho}{A'(r_H)}}
\end{eqnarray}
so that:
\begin{equation}
\text{d}\sigma^2=\tilde R^2\text{d}\theta^2+\text{d}R^2.
\end{equation}
According to the Euclidean quantum gravity prescription, $\tau$ coordinate in expression \eqref{tau_eqn}
is included in the interval defined by $0$ and $\beta_E=1/T_E$. On the
other hand, in  order to avoid conical singularities, $\theta$
must run between $0$ and $2\pi$. Thus it is found that
\begin{equation}
T_E=\frac{1}{4\pi} e^{-\Phi(r_H)}\,A'(r_H)\,.
\end{equation}

Another possible definition of temperature was firstly proposed in
\cite{Hawking1974} stating that temperature can be given in terms
of the horizon gravity $\mathcal{K}$ as:
\begin{eqnarray}
T_{\mathcal{K}}\equiv\frac{\mathcal{K}}{4\pi}
\end{eqnarray}
where $\mathcal{K}$ is given by:
\begin{eqnarray}
\mathcal{K}\,=\,\lim_{r\rightarrow
r_H}\frac{\partial_{r}g_{tt}}{\sqrt{|g_{tt}g_{rr}|}}.
\end{eqnarray}
Then it is straightforward to find:
\begin{eqnarray}
T_{\mathcal{K}}= T_E.
\end{eqnarray}
Therefore, both definitions give the same result for this kind of
metric tensors. Notice also that in any case the temperature depends
only on the behaviour of the metric near the horizon but it is
independent from the gravitational action. By this we mean that
different actions having the same solutions have also the same
temperature. This is not the case for other thermodynamic
quantities as we shall see later. Taking into account the results
in previous sections for Schwarzschild-$AdS$ BHs, we shall concentrate
for simplicity only on those solutions, i.e. for a metric as \eqref{metric_D_v2}
where $A(r)$ is given by expression
\eqref{A(r)} and $\Phi=0$ is adopted as a natural
choice.

Then, both definitions of temperature lead to:
\begin{equation}
\beta=\frac{1}{T}=\frac{4 \pi l^2 r_H}{(D-1)r_H^2+(D-3)l^2}\,.
\end{equation}
Notice that the temperature is a function of $r_H$ only, i.e. it
depends only on the BH size. In the limit $r_H$ going to zero the
temperature diverges as $T \sim 1/r_H$ and for $r_H$ going to
infinity $T$ grows linearly with $r_H$. Consequently $T$ has a
minimum at:
\begin{equation}
 r_{H0}=l\sqrt{\frac{D-3}{D-1}}
\end{equation}
  corresponding to a temperature:
\begin{equation}
T_0=\frac{\sqrt{(D-1)(D-3)}}{2 \pi l}.
\end{equation}
The existence of this minimum was established in \cite{Hawking&Page}
for $D=4$ by Hawking and Page long time ago and it is well-known.
More recently Witten extended this result to higher dimensions
\cite{Witten}. This minimum in the temperature is important in order to set the regions
with different thermodynamic behaviours and stability properties.
 For $D=4$, an
exact solution can be found for $r_H$:
\begin{eqnarray}
r_{H}\,=\,l\frac{\left(9 \frac{R_{S}}{l}+\sqrt{12+81 \frac{R_{S}^2}{l^2}}
\right)^{2/3}-12^{1/3}}{18^{1/3} \left(9\frac{R_S}{l}
+\sqrt{12+81\frac{R_S^2}{l^2}}\right)^{1/3}}.
\end{eqnarray}
Thus, in the $R_S\ll l$ limit, we find $r_H\backsimeq R_S$, whereas in the opposite
case $l\ll R_S$, we get $r_H \backsimeq (l^2 R_S)^{1/3}$. For the particular
case $D=5$, $r_H$ can also be exactly found to be:
\begin{equation}
r_H^2=\frac{l^2}{2}\left(\sqrt{1+\frac{4R_S^2}{l^2}}-1\right)
\end{equation}
which goes to $R_S^2$ for $R_S\ll l$ and to $lR_S$ for $l\ll R_S$.
Notice that for any $T > T_0$, we have two
possible BH sizes: one corresponding to the small BH phase with
$r_H < r_{H0}$ and the
other corresponding to the large BH phase with $r_H > r_{H0}$.

In order to compute the remaining thermodynamic quantities, the Euclidean action
\begin{equation}
S_E=-\frac{1}{16 \pi G_D}\int \text{d}^{D}x\sqrt{g_E}\,(R+f(R))
\end{equation}
is considered.
Extending to the $f(R)$ theories the computation by Hawking and
Page \cite{Hawking&Page} and Witten \cite{Witten},
we evaluate the gravitational Lagrangian in the Schwarzschild-$AdS$
scalar curvature solution times the difference between
the $AdS$ space-time volume minus the BH space-time volume. Thus,
we may write:
\begin{equation}
\Delta S_E=-\frac{R_0+f(R_0)}{16 \pi G_D}\Delta V
\end{equation}
where $R_0=-D(D-1)/l^2$ and $\Delta V$ is the volume difference
between $AdS$ and BH solutions, which is given by:
\begin{equation}
\Delta V=\frac{\beta \mu_{D-2}}{2(D-1)}\left(l^2r^{D-3}_H-r^{D-1}_H\right).
\end{equation}
Notice that from these expressions, it is straightforward to obtain the
free energy $F$ since $\Delta S_E =\beta F$ and therefore
\begin{equation}
\Delta S_E=-\frac{(R_0+f(R_0))\beta \mu_{D-2}}{36 \pi (D-1)
G_D}\left(l^2r^{D-3}_H-r^{D-1}_H\right)=\beta F.
\end{equation}
We see that provided $-(R_0+f(R_0))>0$, which is the usual
case in EH gravity, one has
$F>0$ for $r_H<l$ and $F<0$ for $r_H>l$. The temperature corresponding to the
horizon radius $r_H=l$ will be denoted $T_1$ and it is given by:
\begin{equation}
T_1=\frac{D-2}{2\pi l}.
\end{equation}
Notice that for $D>2$ we have $T_0<T_1$.

On the other hand, the total thermodynamical energy may now be obtained as:
\begin{equation}
E=\frac{\partial \Delta S_E}{\partial \beta}=-\frac{(R_0+f(R_0))M l^2}{2(D-1)}
\label{energy}
\end{equation}
where $M$ is the mass defined in \eqref{BHmass}. This is one of the possible
definitions for the BH energy for $f(R)$ theories, see for instance \cite{Multamaki2007}
for a more general discussion. For the EH
action with nonvanishing cosmological constant one has $f(R)=-(D-2)\Lambda_D$ and then it is immediate to
find $E=M$. However, this is not the case for general $f(R)$
actions. Notice that positive energy in $AdS$ space-time requires
$R_0+f(R_0)<0$. Now the entropy $S$ can be obtained from the well-known
relation:
\begin{equation}
S=\beta E- \beta F.
\end{equation}
Then one gets:
\begin{equation}
S=-\frac{(R_0+f(R_0))l^2 A_{D-2}(r_H)}{8 (D-1)G_D}
\label{entropy}
\end{equation}
where $A_{D-2}(r_H)$ is the horizon area given by
$A_{D-2}(r_H)\equiv r_H^{D-2}\mu_{D-2}$. Notice that once again
positive entropy requires $R_0+f(R_0)<0$.
For the EH action with nonvanishing cosmological constant one has $f(R)=-(D-2)\Lambda_D$ one
has $R_0+f(R_0)=-2(D-1)/l^2$ and then the 
Hawking-Bekenstein result \cite{Bekenstein}
\begin{equation}
S=\frac{ A_{D-2}(r_H)}{4G_D}\,.
\end{equation}
is recovered. Finally, one can compute the heat capacity $C$ which can be written
as:

\begin{equation}
C=\frac{\partial E}{\partial T}=\frac{\partial E}{\partial
r_H}\frac{\partial r_H}{\partial T}.
\end{equation}
Then it is easy to find
\begin{equation}
C=\frac{-(R_0+f(R_0))(D-2)\mu_{D-2}r^{D-2}_Hl^2}{8G_D(D-1)}
\frac{(D-1)r^2_H+(D-3)l^2}{(D-1)r^2_H-(D-3)l^2}.
\end{equation}
For the already mentioned case of the EH action with nonvanishing cosmological constant one finds:
\begin{equation}
C=\frac{(D-2)\mu_{D-2}r^{D-2}_H}{4G_D}\frac{(D-1)r^2_H+(D-3)l^2}{(D-1)r^2_H-(D-3)l^2}.
\end{equation}
In the Schwarzschild limit $l\rightarrow\infty$, this formula
gives:
\begin{equation}
C\,\backsimeq\,-\frac{(D-2)\mu_{D-2}r^{D-2}_H}{4G_D}< 0
\end{equation}
which is the well-known negative result for standard BHs of this type. In the
general case, assuming like in the EH case $(R_0+f(R_0))<0$, one
finds $C>0$ for $r_H > r_{H0}$ (the large BH region) and $C<0$ for
$r_H < r_{H0}$ (the small BH region). For $r_H \sim r_{H0}$ ($T$
close to $T_0$)   $C$ is divergent. Notice that in EH gravity, $C<0$ necessarily
implies $F>0$ since $T_0<T_1$.

In any case, for $f(R)$ theories with $R_0+f(R_0)<0$, we have found
an scenario similar to the one described in
full detail by Hawking and Page in \cite{Hawking&Page} long time ago
for the EH case.

For $T<T_0$, the only possible state of thermal equilibrium in  an
$AdS$ space is pure radiation with negative free energy and there
is no stable BH solutions. For $T>T_0$ we have two possible BH
solutions: the small (and light) BH and the large (heavy) BH. The
small one has negative heat capacity and positive free energy as
the standard Schwarzschild BH. Therefore this last configuration is unstable under
Hawking radiation decay. For the large BH we have two
possibilities: if $T_0<T<T_1$ then both the heat capacity and the
free energy are positive and the BH will decay by tunnelling into
radiation, but if $T>T_1$ then the heat capacity is still positive
but the free energy becomes negative. In this case the free energy
of the heavy BH will be less than that of pure radiation. Then
pure radiation will tend to tunnel or to collapse to the BH
configuration in equilibrium with thermal radiation.

In arbitrary $f(R)$ theories one could in principle consider the
possibility of having $R_0+f(R_0)>0$. However, in this case
both the energy and the entropy, given by expressions \eqref{energy} and \eqref{entropy} respectively,
would be negative and therefore in such theories the $AdS$ BH solutions would be unphysical.
Therefore, $R_0+f(R_0)<0$ can be regarded  as a necessary condition
for $f(R)$ theories in order to support the existence of $AdS$ BH solutions. Using
(\ref{dif}), this condition  implies $1+f_{R}(R_0)>0$. Let us remind that this
condition has a clear physical interpretation in $f(R)$ gravities
already presented as the condition {\bf 2} in Section 1.5.

\section{Particular examples}
\label{sec:BH:Particular examples}
In this section we are going to consider several $f(R)$ models in order to calculate the heat capacity $C$ and
the free energy $F$ since, as was explained in the previous section, these are
the relevant thermodynamical quantities for local and global
stability of BHs. For these particular models, $R_0$
can be calculated exactly by using the relation \eqref{dif} with $R=R_0$. In the following
we will fix the $D$ dimensional Schwarzschild radius in expression \eqref{BHmass}
as $R_S^{D-3}=2$ for simplicity.

The models that we consider in this section have been
previously studied in the literature, but attention there was drawn
in studying their cosmological viability according to conditions
provided in Section 1.5. Here we draw our attention on
thermodynamics for Schwarzschild-$AdS$ BH solutions for these $f(R)$
gravities.

Both free energy and heat capacity signs are studied for the different
values that parameters which appear in these $f(R)$ functions take.
Once these signs are known, both local or global thermodynamical stability
can be determined for these $f(R)$ theories
following the reasoning explained at the end of the previous section.
The considered models are:
%
%
%
%
%
%
%

{\bf Model I: $f(R)\,=\,\alpha (-R)^{\beta}$}

This model belongs to Class {\bf II} of $f(R)$ models presented in Section
\ref{sect:fR:Cosmological viability} if parameters satisfy $\alpha<0$ and $0<\beta<1$
as seen from Table 2.1. and it could be therefore cosmologically viable.

Substituting in expression \eqref{dif} for arbitrary dimensions we get
\begin{eqnarray}
R \left[\left(1-\frac{2}{D}\right)-\alpha(-R)^{\beta-1}
\left(1-\frac{2}{D}\beta\right)\right] \,=\,0\,.
\label{eqn_Model_I}
\end{eqnarray}

Since we are only considering nonvanishing curvature solutions, then we find:
\begin{eqnarray}
R_{0}\,=\,-\left[\frac{2-D}{(2\beta-D)\alpha}\right]^{1/(\beta-1)}\,.
\label{R0_Model_I}
\end{eqnarray}
Since $D$ is assumed to be larger than 2,
the condition $(2\beta-D)\alpha<0$ provides well defined scalar curvatures $R_{0}$.
Thus, two separated regions have to be studied: Region $1$ $\{\alpha<0,\,\beta>D/2\}$
and Region $2$ $\{\alpha>0,\, \beta<D/2\}$.
For this model we also get
\begin{eqnarray}
1+f_{R}(R_{0})\,=\,\frac{D(\beta-1)}{2\beta-D}\,.
\end{eqnarray}
Notice that in Region $1$, $1+f_{R}(R_0)>0$ for $D>2$, since in this case $\beta>1$
is straightforwardly accomplished.
In Region $2$, we find that for $D>2$, the requirement $R_0+f(R_0)<0$, i.e. $1+f_{R}(R_0)>0$,
fixes $\beta<1$, since this is the most stringent constraint over the
parameter $\beta$ in this region. Therefore the physical space of
parameters in Region $2$ is restricted to be $\{\alpha>0,\,\beta<1\}$.

\begin{figure}[!hbp]
\subfigure[ \hspace{1ex} Model I, $D=4$, Region 1, $\alpha<0$, $\beta>2$.]{
\begin{overpic}[width=7.0cm]{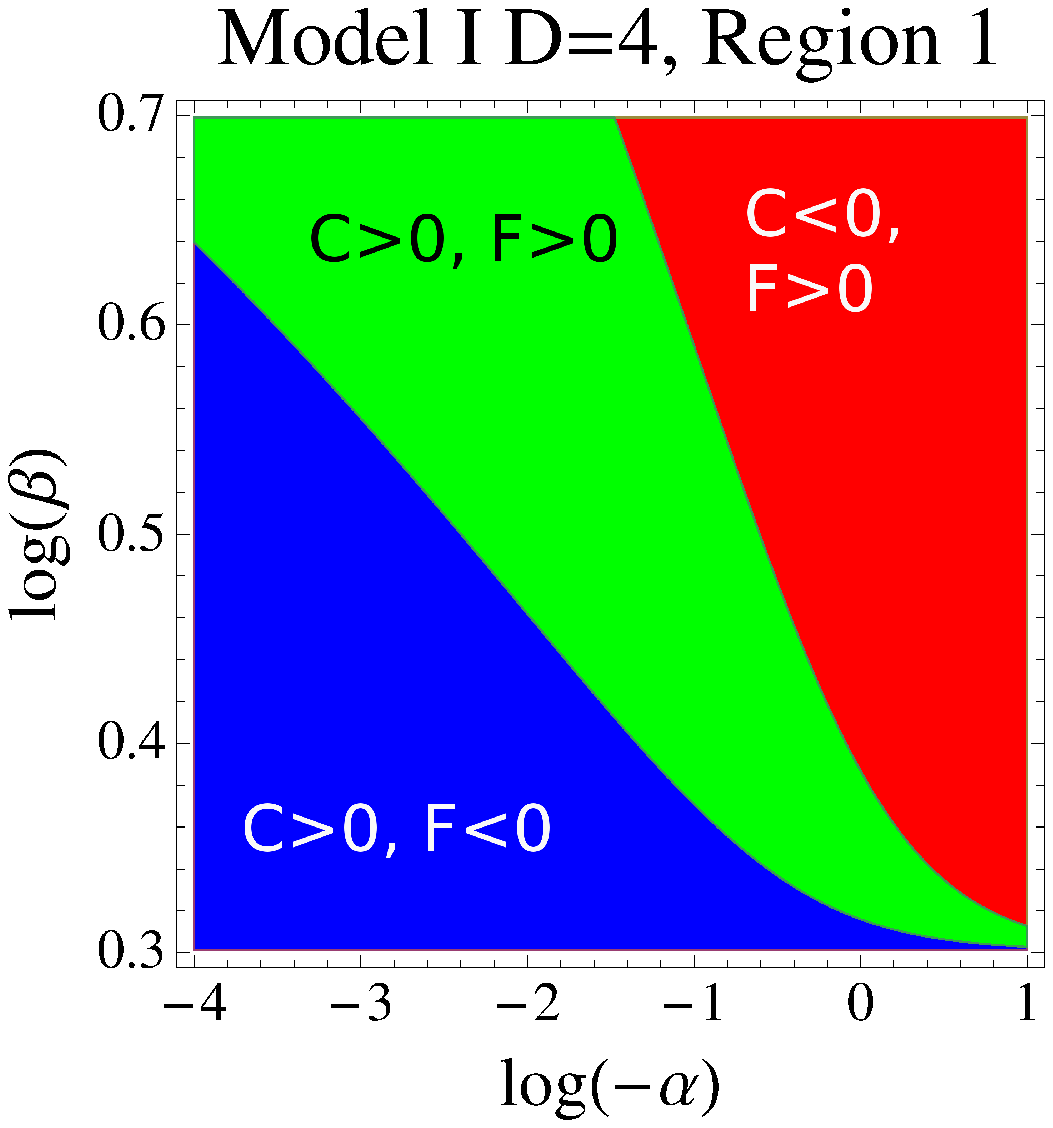}
\end{overpic}
}
\subfigure[ \hspace{1ex} Model I, $D=4$, Region 2, $\alpha>0$, $\beta<1$.]{
\begin{overpic}[width=7.70cm]{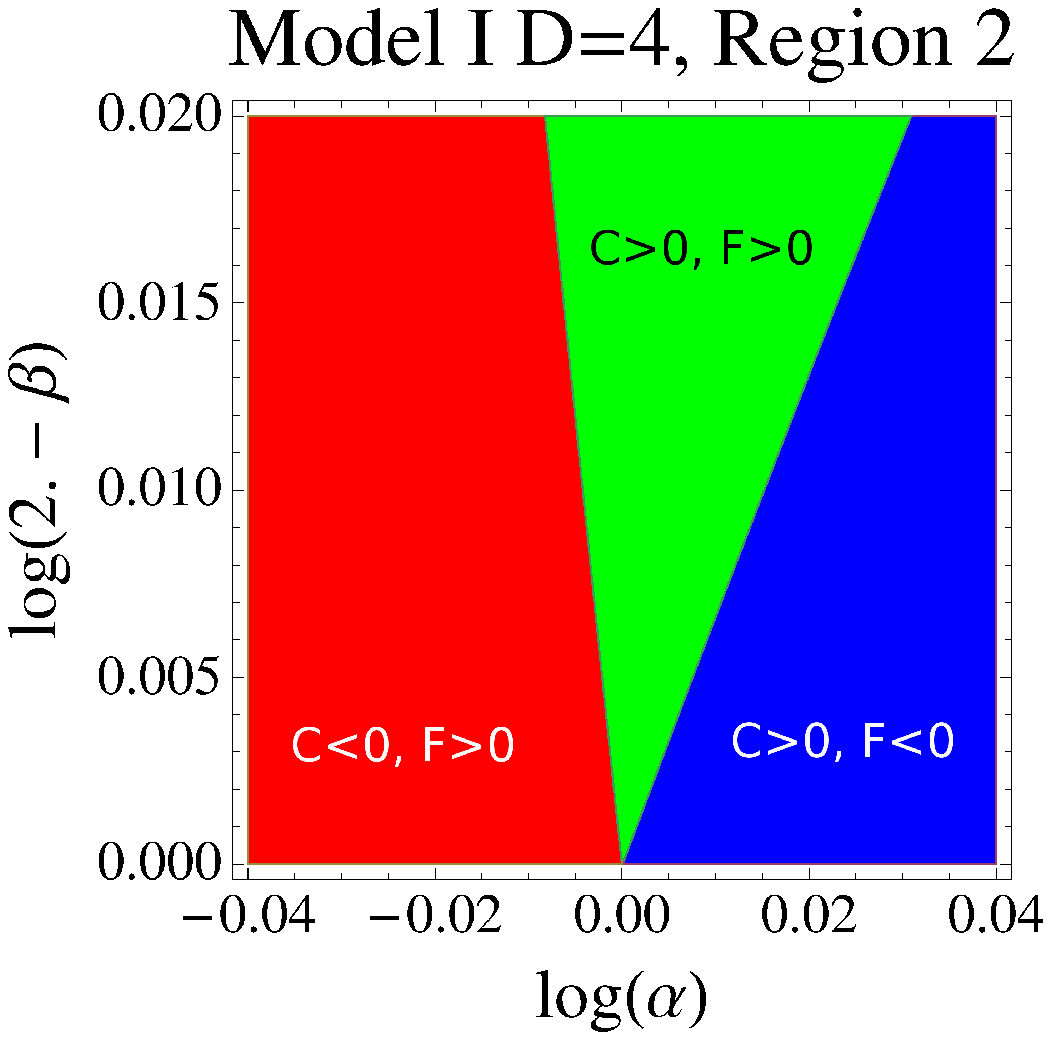}
\end{overpic}
}
\caption{Thermodynamical regions in the $(\alpha,\beta)$ plane for Model I in $D=4$.
Region 1(left),
Region 2 (right).}
\label{M1_figure1}
\end{figure}

\begin{figure}[!hbp]
\subfigure[ \hspace{1ex} Model I, $D=5$, Region 1, $\alpha<0$, $\beta>2.5$.]{
\begin{overpic}[width=7.20cm]{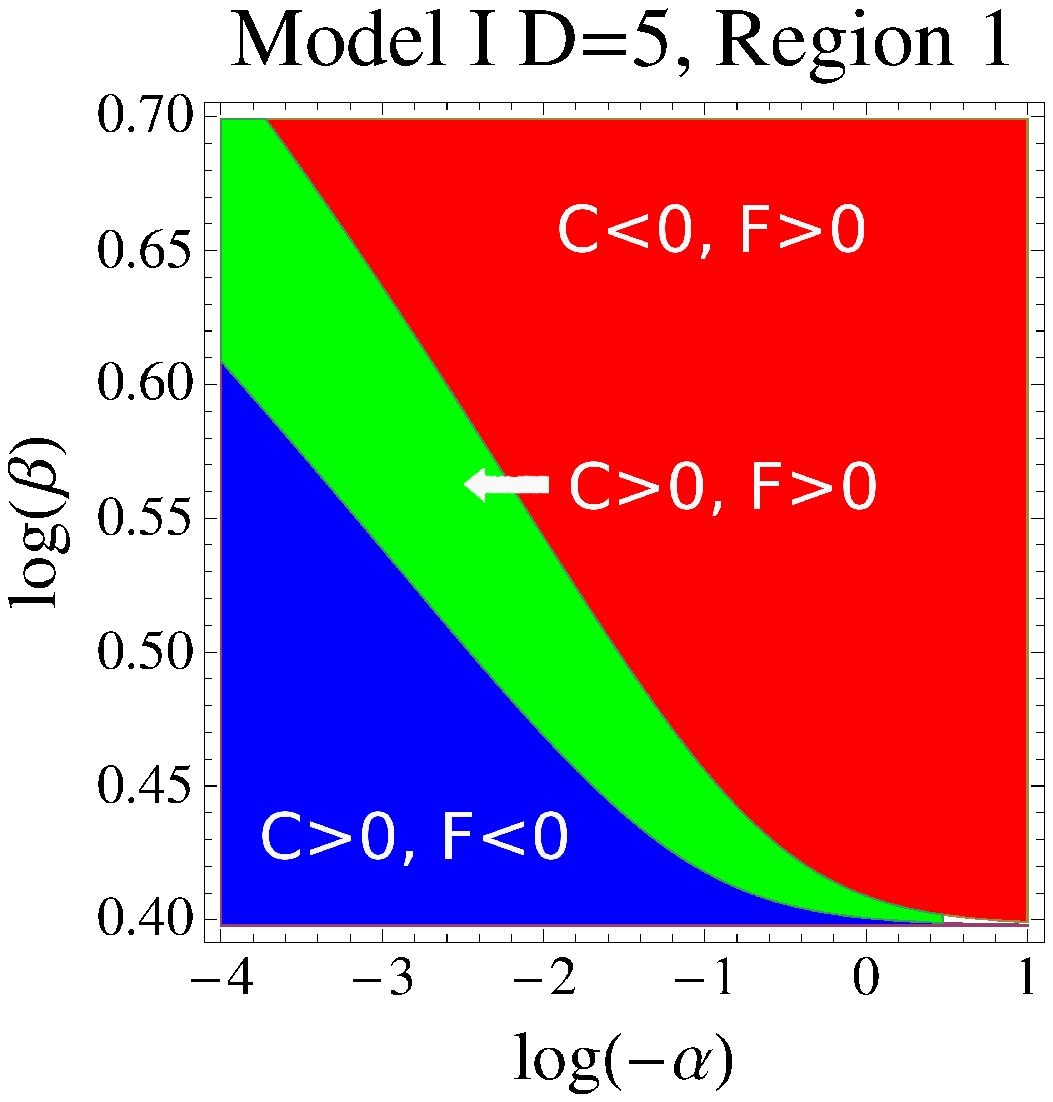}
\end{overpic}
}
\subfigure[ \hspace{1ex} Model I, $D=5$, Region 2, $\alpha>0$, $\beta<1$.]{
\begin{overpic}[width=7.70cm]{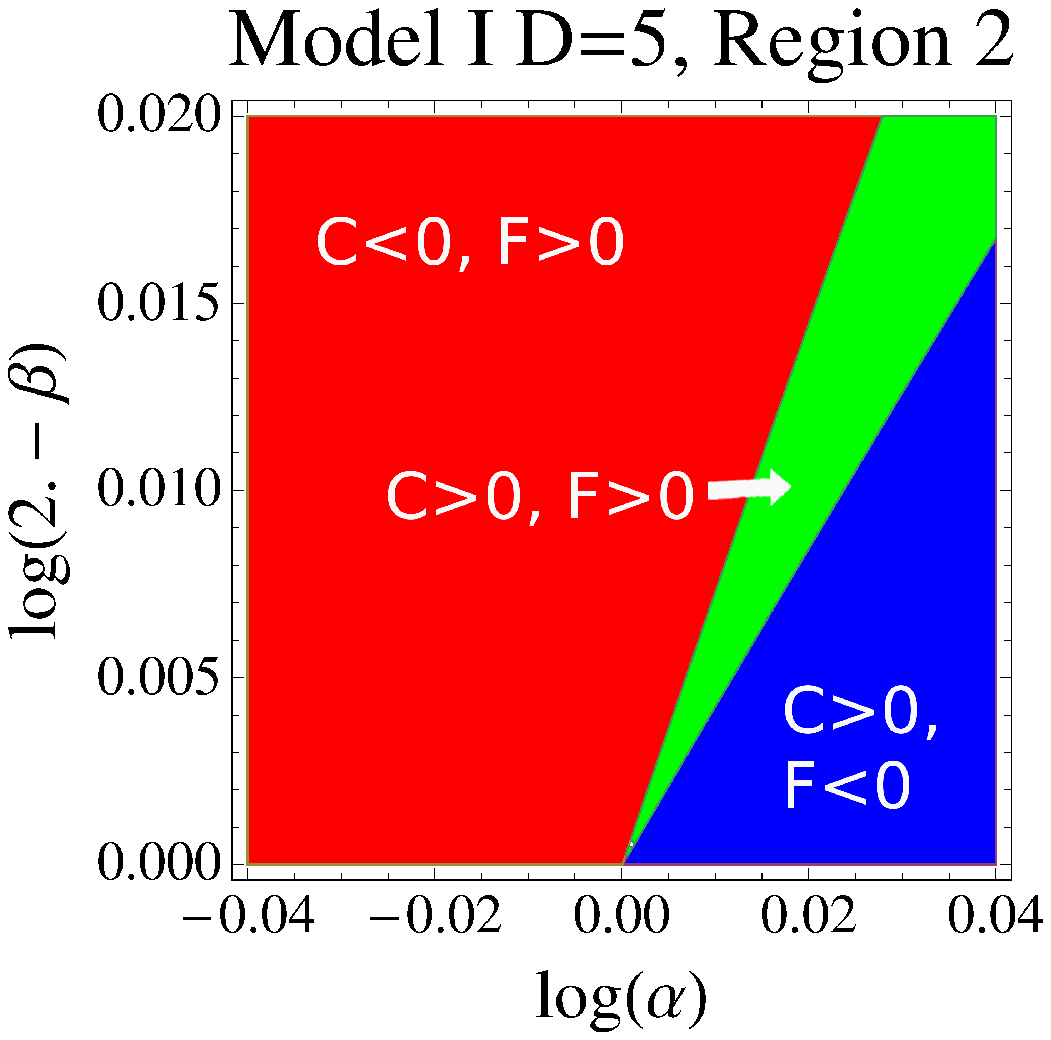}
\end{overpic}
}
\caption{Thermodynamical regions in the $(\alpha,\beta)$ plane for Model I in $D=5$.
Region 1(left),
Region 2 (right).}
\label{M1_figure2}
\end{figure}

\begin{figure}[!hbp]
\subfigure[ \hspace{1ex} Model I, $D=10$, Region 1, $\alpha<0$, $\beta>5$.]{
\begin{overpic}[width=7.70cm]{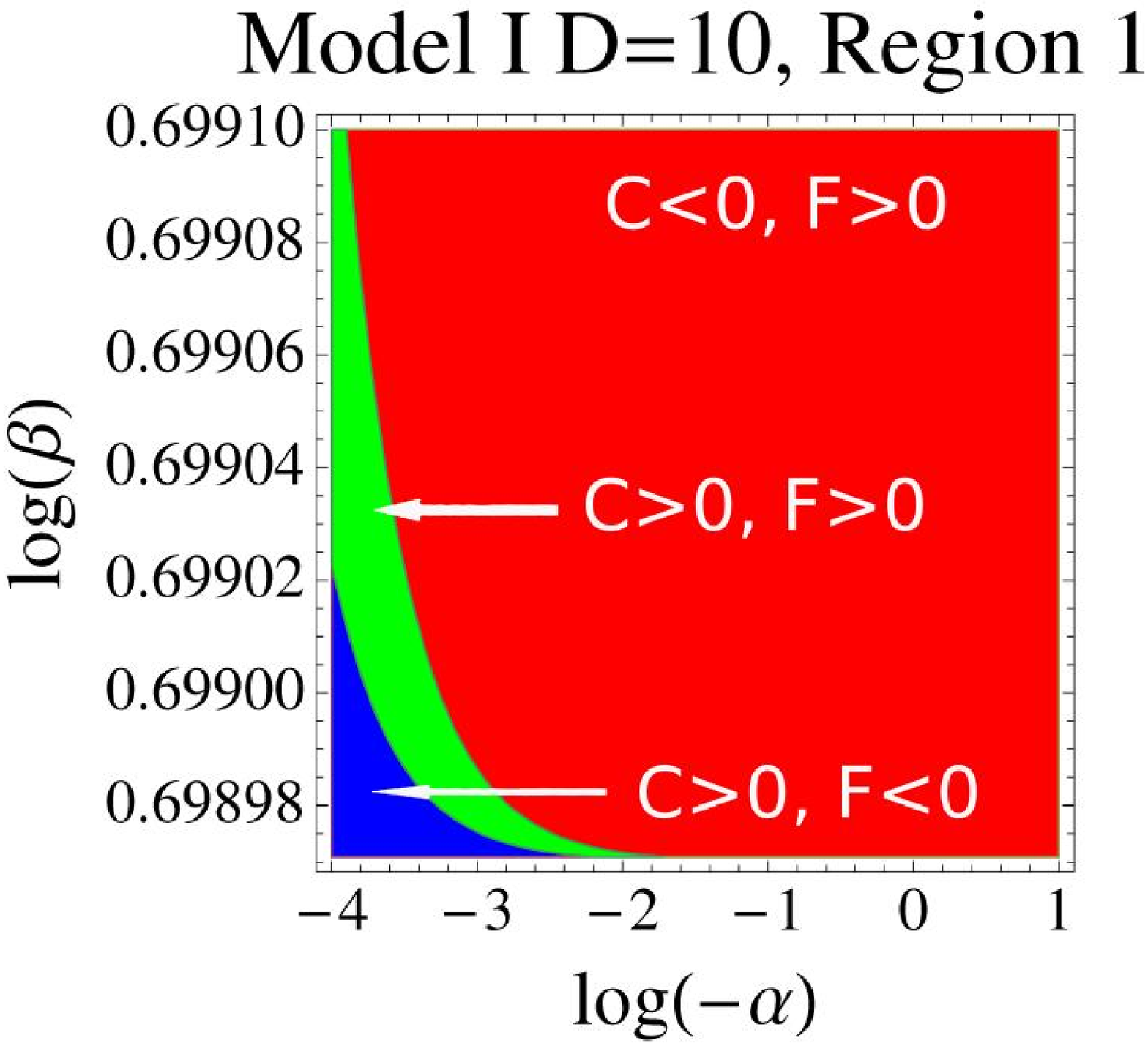}
\end{overpic}
}
\subfigure[ \hspace{1ex} Model I, $D=10$, Region 2, $\alpha>0$, $\beta<1$.]{
\begin{overpic}[width=7.0cm]{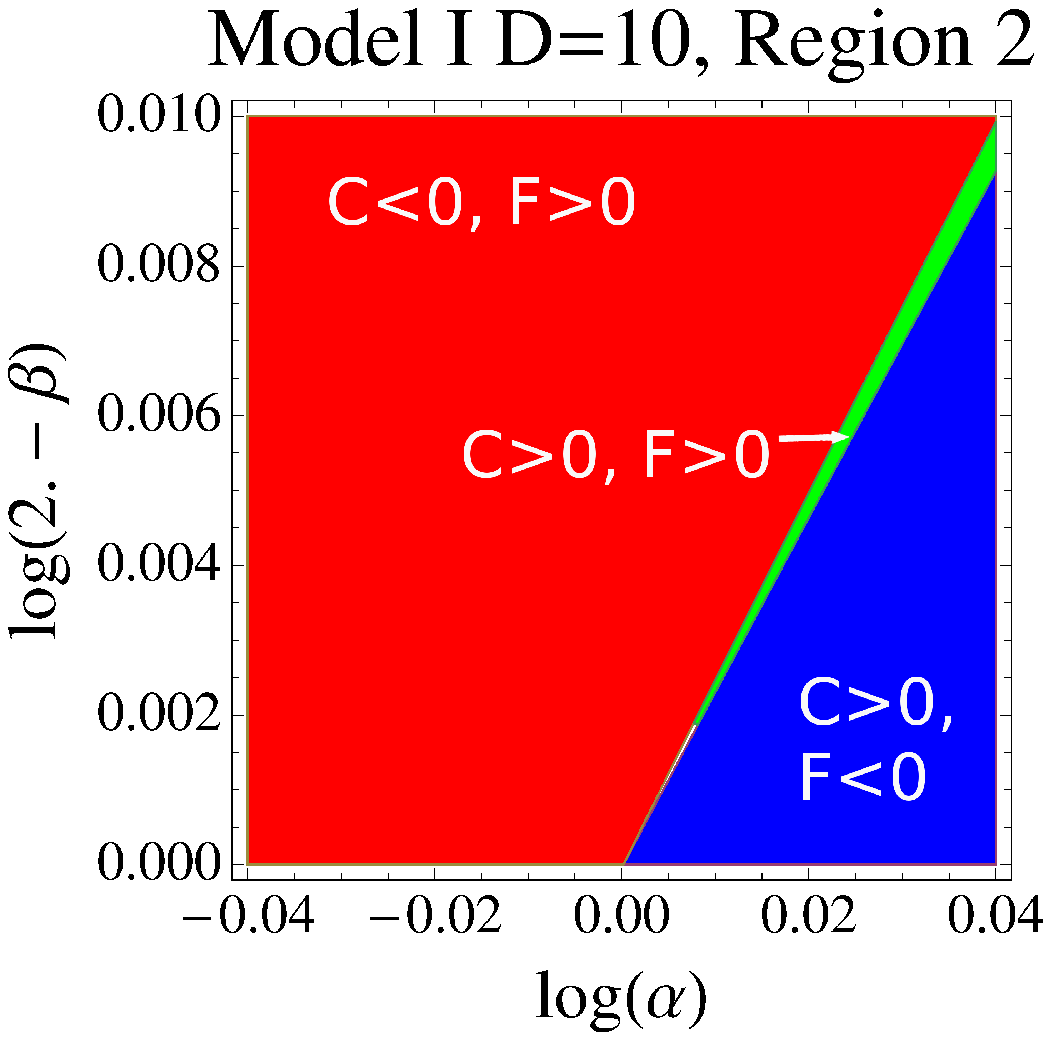}
\end{overpic}
}
\caption{Thermodynamical regions in the $(\alpha,\beta)$ plane for Model I in $D=10$.
Region 1(left),
Region 2 (right).}
\label{M1_figure3}
\end{figure}

In Figures \ref{M1_figure1}, \ref{M1_figure2} and \ref{M1_figure3} we plot the physical regions in the parameter space $(\alpha,\beta)$
corresponding to the different signs of $(C,F)$.
\\

{\bf Model II: $f(R)\,=\, -(-R)^{\alpha}\,\text{exp}(q/R)-R$}

This model may also belong to Class II of $f(R)$ models presented in Section
\ref{sect:fR:Cosmological viability} if $\alpha=1$ as seen from Table 2.1 and could be therefore cosmologically viable.

In this case, a vanishing curvature solution appears
provided $\alpha>1$. In addition, we also have:
\begin{eqnarray}
R_{0}\,=\,\frac{2 q}{2\alpha-D}.
\label{R0_model_II}
\end{eqnarray}
To get $R_{0}<0$ the condition $q(2\alpha-D)<0$ must hold and two separated
regions will be studied: Region $1$ $\{q>0,\, \alpha<D/2\}$ and Region $2$
$\{q<0,\,\alpha>D/2\}$.

\begin{figure}[!hbp]
\subfigure[ \hspace{1ex} Model II, $D=4$, Region 1, $\alpha<2$, $q>0$.]{
\begin{overpic}[width=7.3cm]{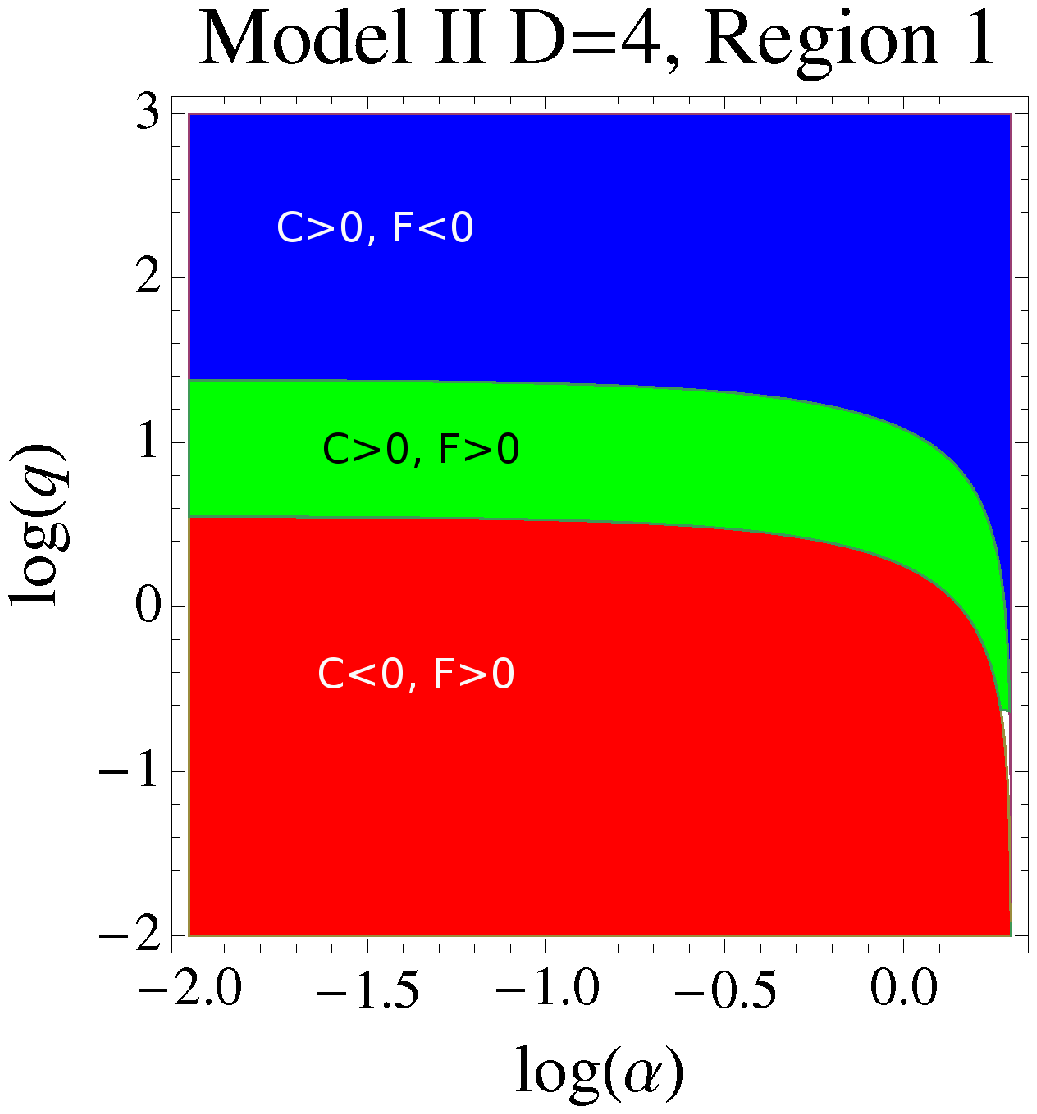}
\label{M2D4_R1}
\end{overpic}
}
\subfigure[ \hspace{1ex} Model II, $D=4$, Region 2, $\alpha>2$, $q<0$.]{
\begin{overpic}[width=7.30cm]{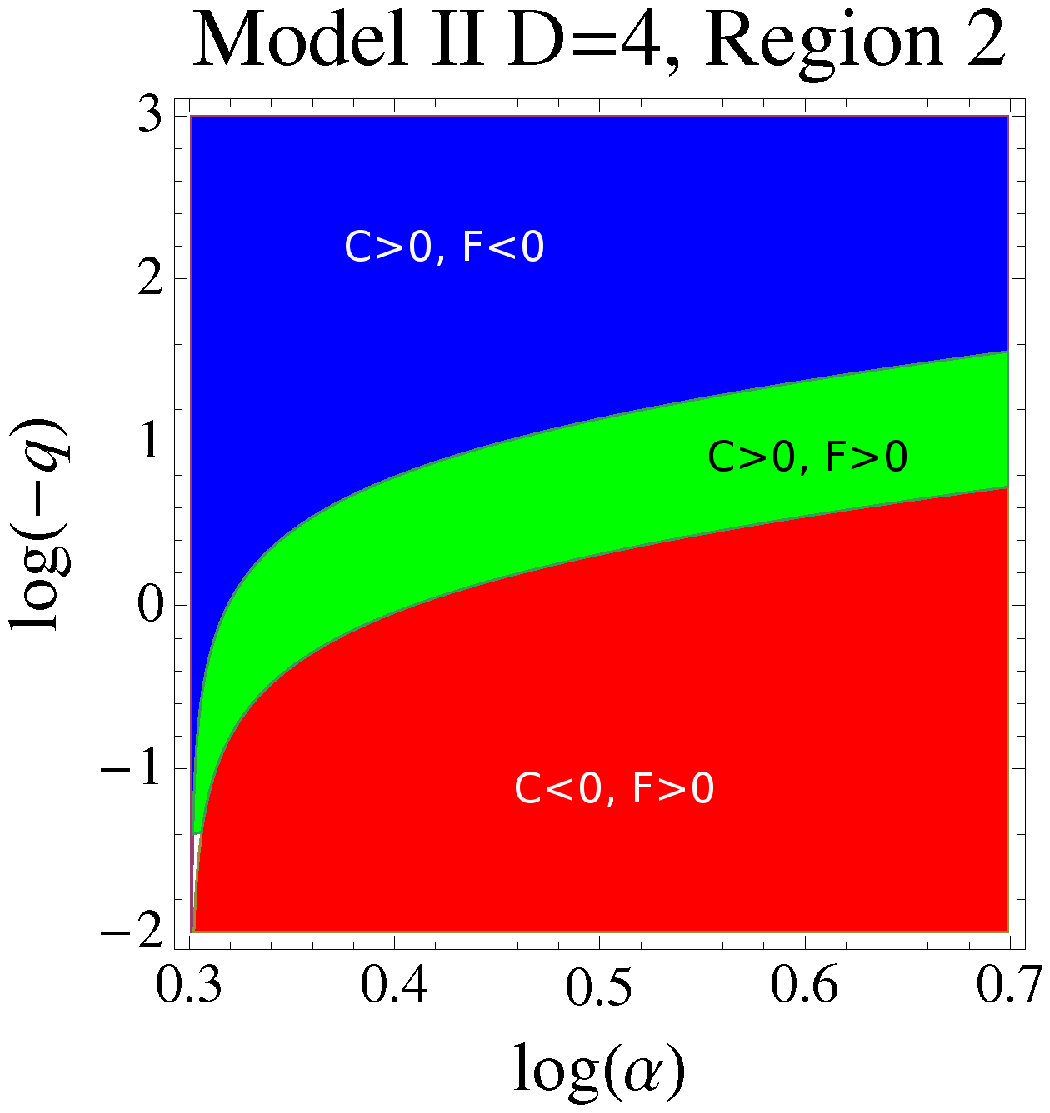}
\label{M2D4_R2}
\end{overpic}
}
\caption{Thermodynamical regions in the $(\alpha,q)$ plane for Model II in $D=4$.
Region 1 (left), Region 2 (right).}
\label{M2_figure1}
\end{figure}

\begin{figure}[!hbp]
\centering
\subfigure[ \hspace{1ex} Model II, $D=5$, Region 1, $\alpha<2.5$, $q>0$.]{
\begin{overpic}[width=7.4cm]{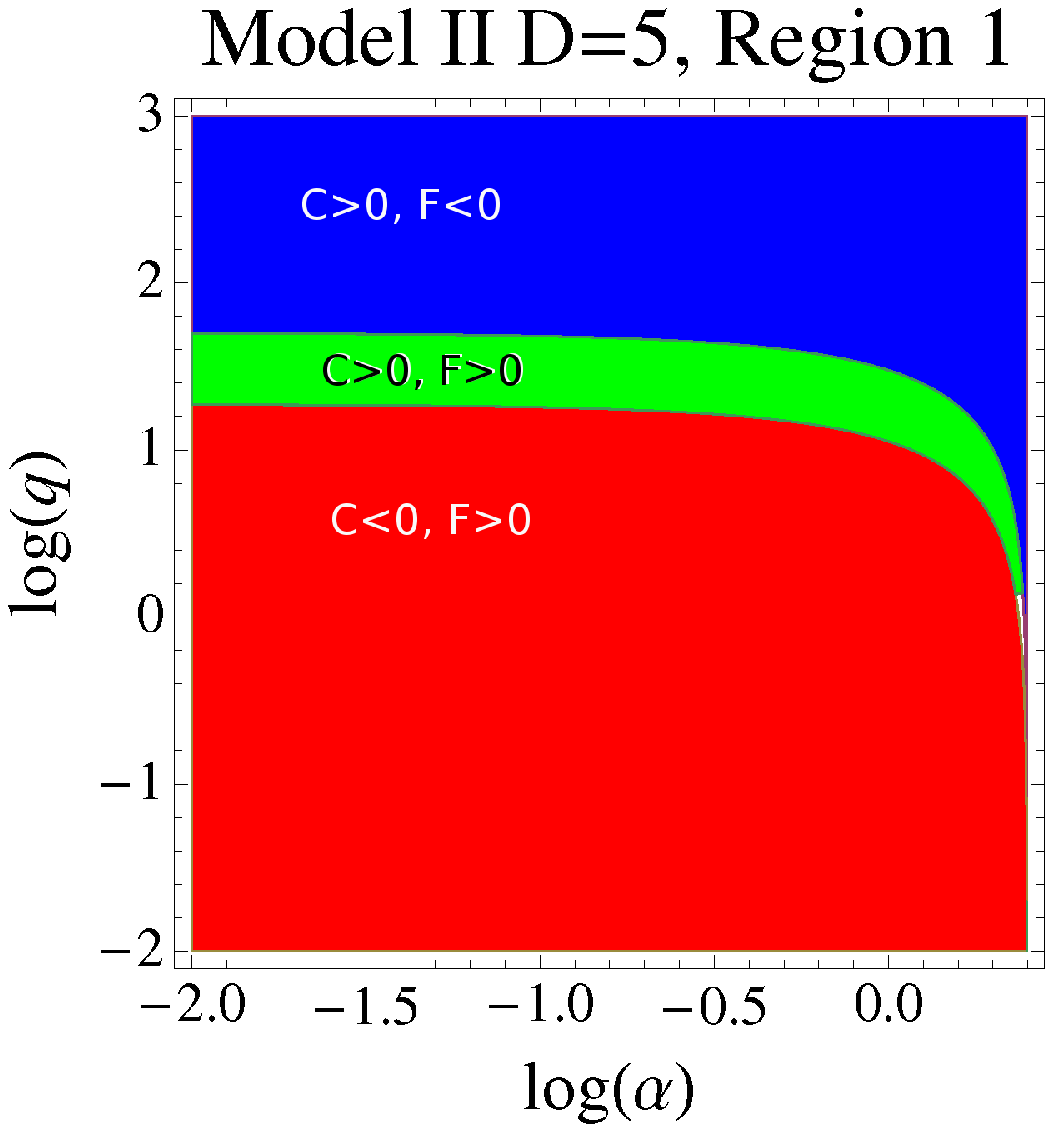}
\label{M2D5_R1}
\end{overpic}
}
\subfigure[ \hspace{1ex} Model II, $D=5$, Region 2, $\alpha>2.5$, $q<0$.]{
\begin{overpic}[width=7.4cm]{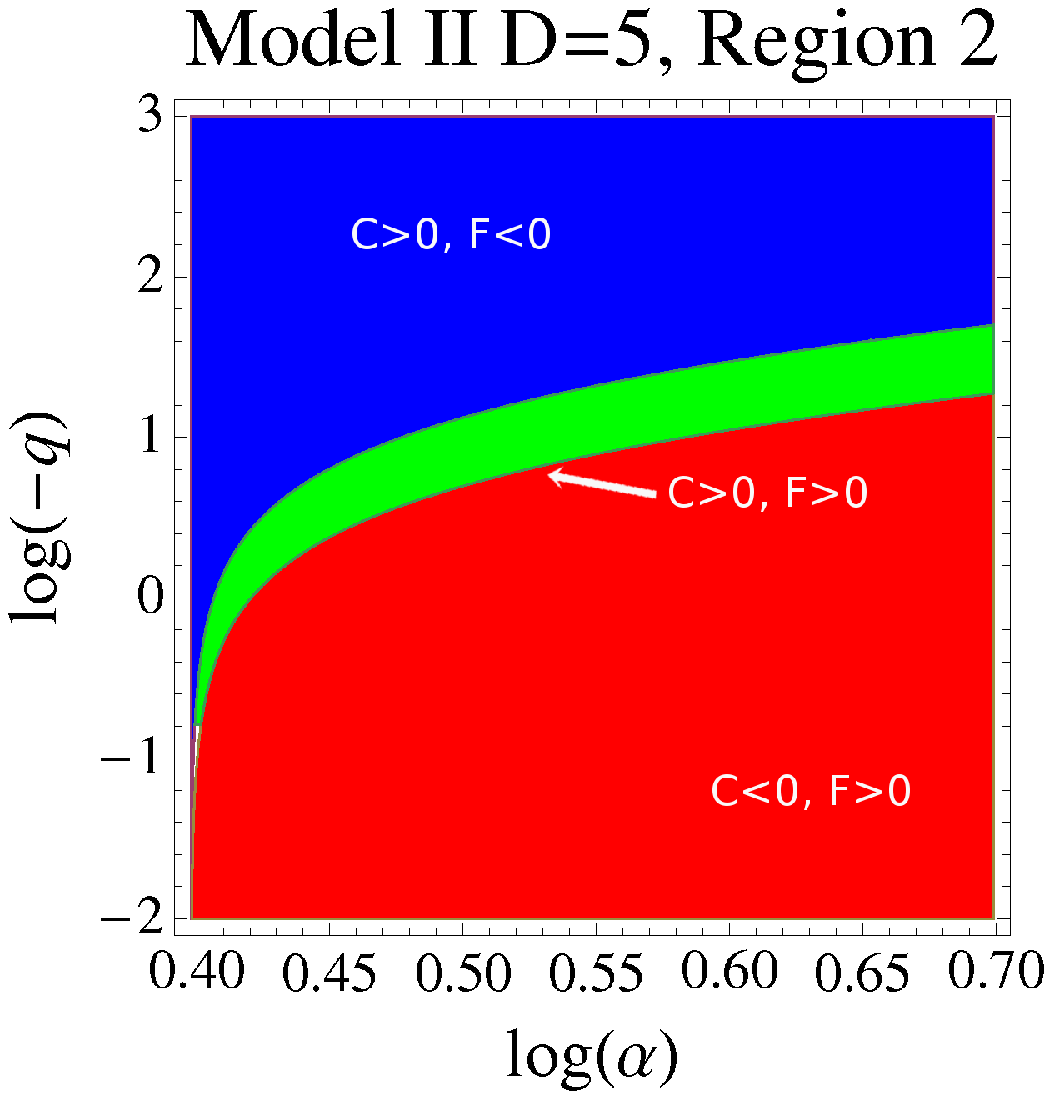}
\label{M2D5_R2}
\end{overpic}
}
\caption{Thermodynamical regions in the $(\alpha,q)$ plane for Model II in $D=5$.
Region 1(left), Region 2 (right).}
\label{M2_figure2}
\end{figure}

\begin{figure}[!hbp]
\centering
\subfigure[ \hspace{1ex} Model II, $D=10$, Region 1, $\alpha<5$, $q>0$.]{
\begin{overpic}[width=7.3cm]{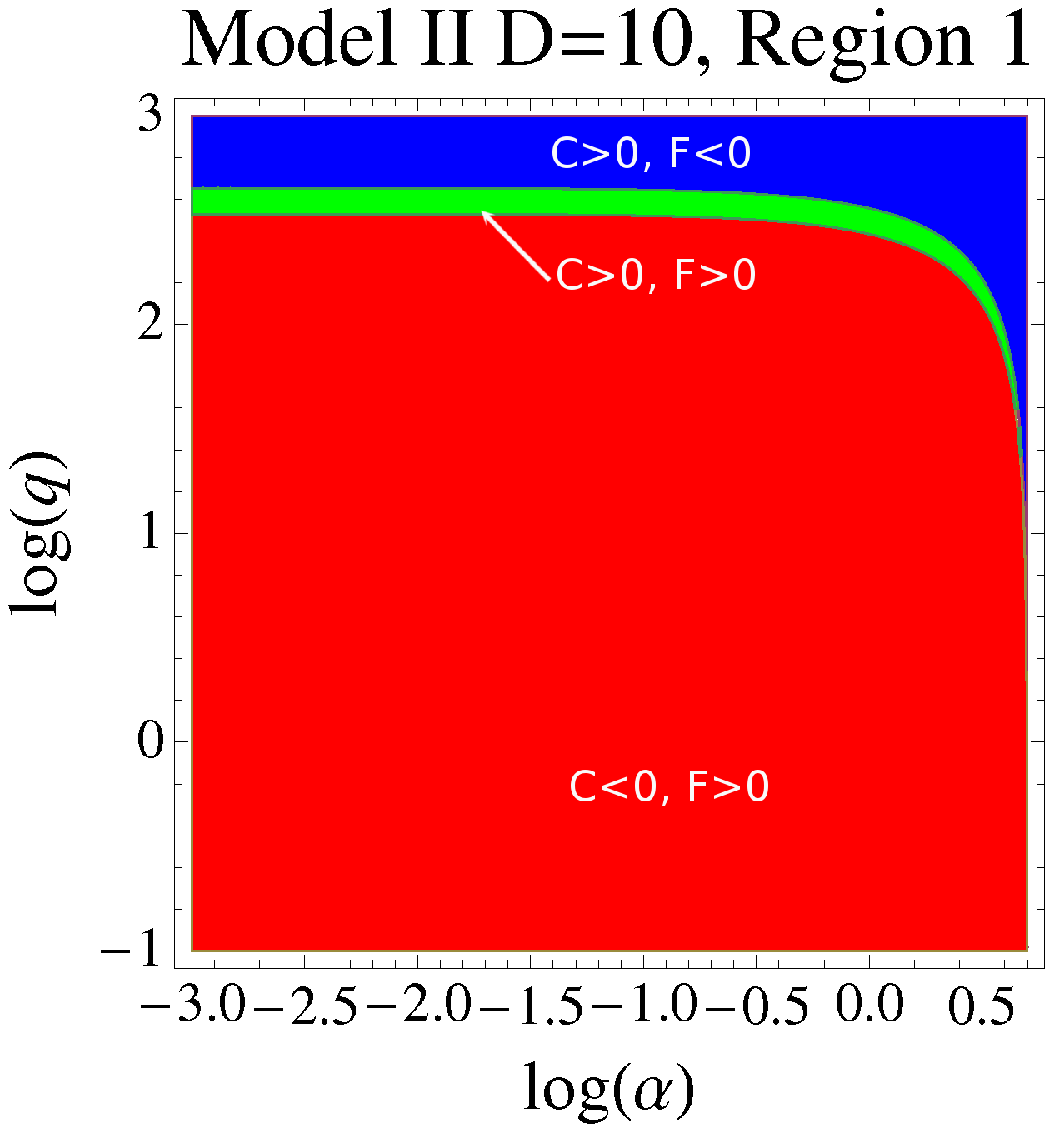}
\label{M2D10_R1}
\end{overpic}
}
\subfigure[ \hspace{1ex} Model II, $D=10$, Region 2, $\alpha>5$, $q<0$.]{
\begin{overpic}[width=7.3cm]{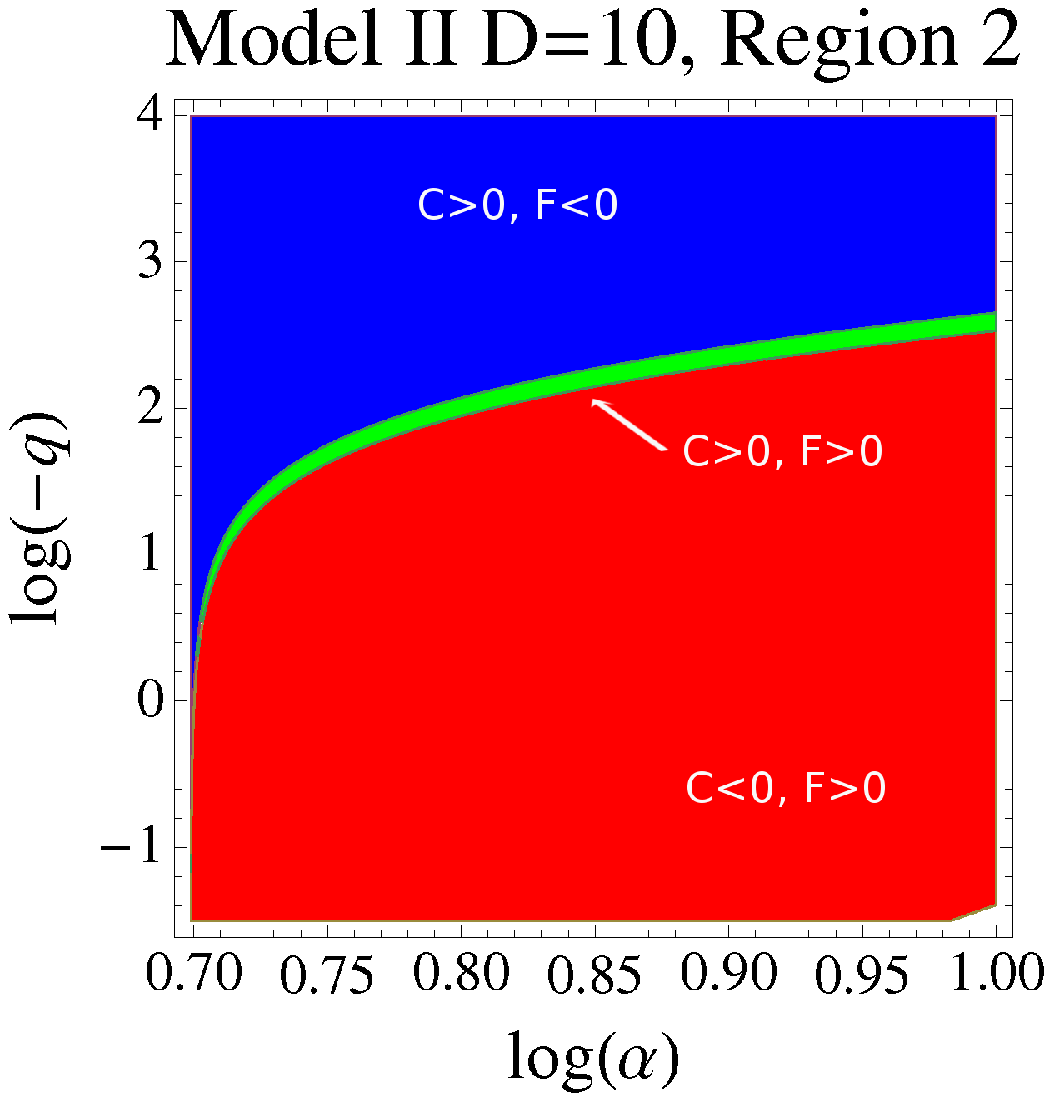}
\label{M2D10_R2}
\end{overpic}
}
\caption{Thermodynamical regions in the $(\alpha,q)$ plane for Model II in $D=10$.
Region 1(left), Region 2 (right).}
\label{M2_figure3}
\end{figure}

In Figures \ref{M2_figure1}, \ref{M2_figure2} and \ref{M2_figure3} we plot the regions in the parameter space $(\alpha,q)$
corresponding to the different signs of $(C,F)$.
\\

{\bf Model III: $f(R)\,=\, R\,\left[\text{log}\left(\alpha R\right)\right]^{q}-R$}

This model may also belong to Class {\bf II} of $f(R)$ models presented in Section
\ref{sect:fR:Cosmological viability} if $q>0$ condition is satisfied and could be therefore cosmologically viable.

A vanishing curvature solution also appears in this model. The nontrivial one
is given by
\begin{eqnarray}
R_{0}\,=\,\frac{1}{\alpha}\text{exp}\left(\frac{2 q}{D-2}\right).
\label{R0_model_III}
\end{eqnarray}
Since $R_{0}$ has to be
negative, $\alpha$ must be negative as well, accomplishing $\alpha R_{0}>0$.
If $q<0$ is considered then, for $D>2$, the expression \eqref{R0_model_III} would imply $\alpha R_0<1$ and then, from $f(R)$
expression for this model, an inconsistency would appear since a negative number would be powered to a negative $q$ value.
Then $q>0$ is the only allowed interval for this parameter and
therefore there exists a unique accessible
region for parameters in this model: $\alpha<0$ and $q>0$.

\begin{figure}[!hbp]
\centering
\subfigure[ \hspace{1ex} Model III, $D=4$, $\alpha<0$, $q>0$.]{
\begin{overpic}[width=7.30cm]{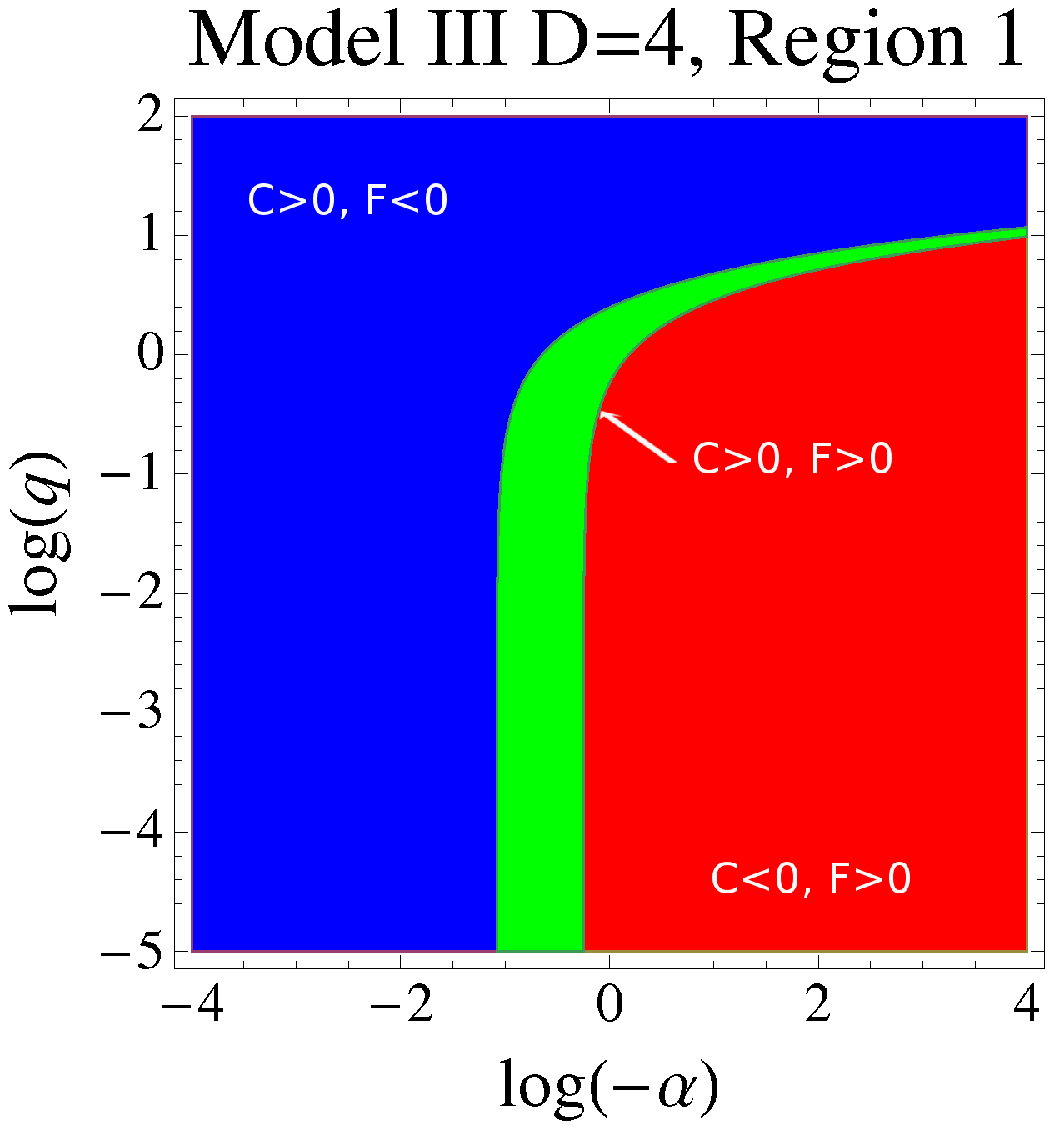}
\label{M3D4}
\end{overpic}
}
\subfigure[ \hspace{1ex} Model III, $D=5$, $\alpha<0$, $q>0$.]{
\begin{overpic}[width=7.30cm]{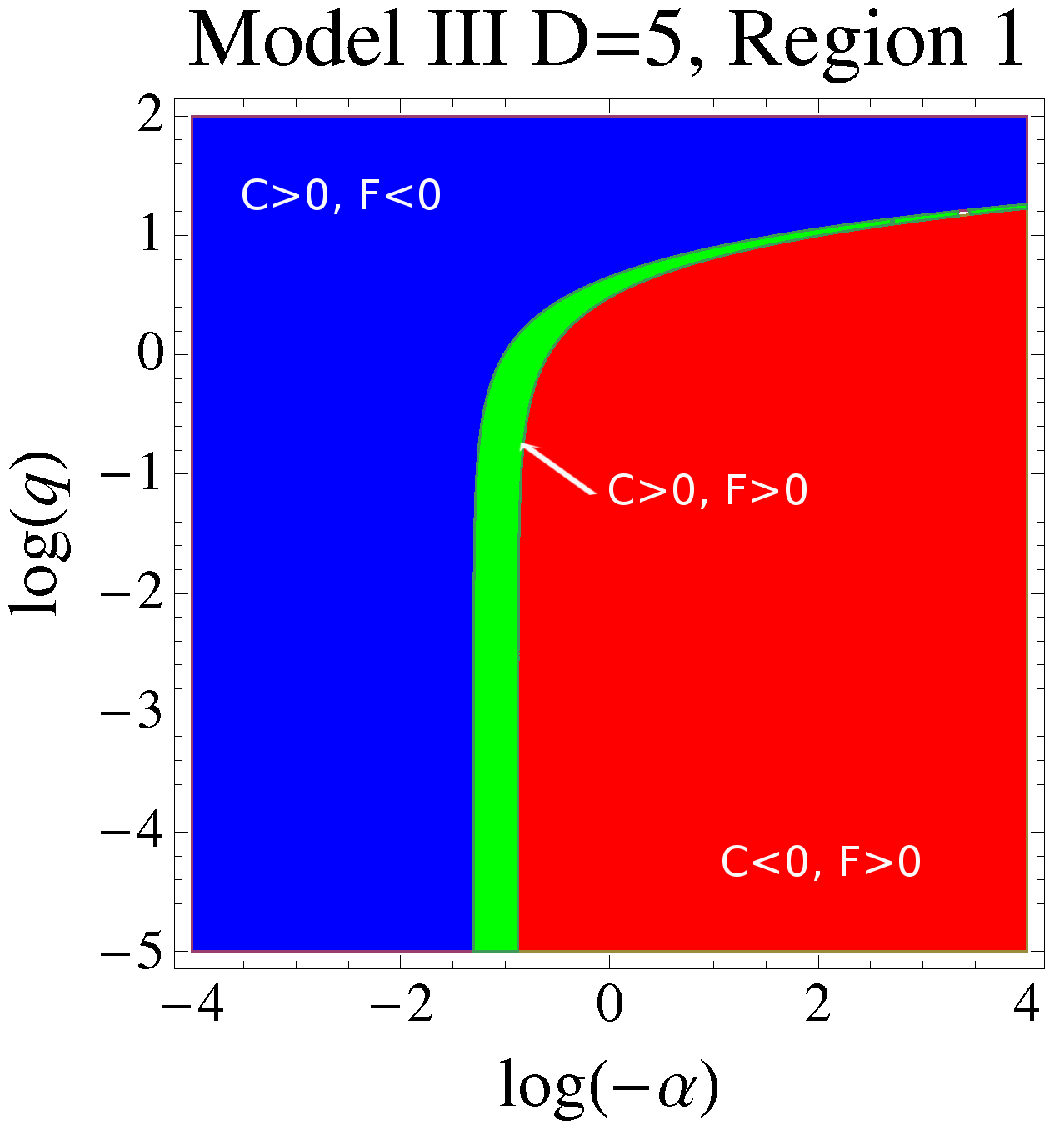}
\label{M3D5}
\end{overpic}
}
\caption{Thermodynamical regions in the $(\alpha,q)$ plane for Model III in $D=4$ (left)
and $D=5$ (right).}
\label{M3_figure1}
\end{figure}

\begin{figure}[!hbp]
\centering
\subfigure[ \hspace{1ex} Model III, $D=10$, $\alpha<0$, $q>0$.]{
\begin{overpic}[width=7.30cm]{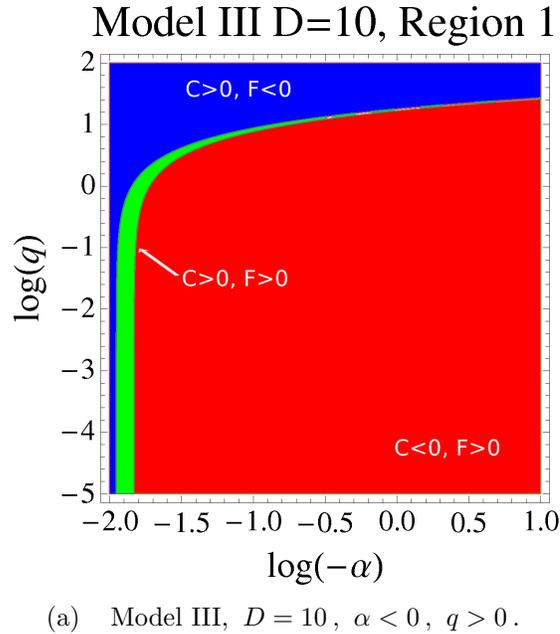}
\label{M3D10}
\end{overpic}
}
\caption{Thermodynamical regions in the $(\alpha,q)$ plane for Model III in $D=10$.}
\label{M3_figure2}
\end{figure}

In Figures \ref{M3_figure1} and \ref{M3_figure2} we plot the regions in the parameter space $(\alpha,q)$
corresponding to the different signs of $(C,F)$.
\\

{\bf Model IV: $f(R)\,=\,-\alpha m_1\left(\frac{R}{\alpha}\right)^{n}\left[1+\beta\left(\frac{R}{\alpha}\right)^n\right]^{-1}$}

As was mentioned in Section \ref{sect:fR:Some viable f(R) models}, this model was originally proposed in \cite{Hu&Sawicki_May_2007} where it was considered to satisfy both cosmological and solar system tests
without a cosmological constant. For this model, $n=1$ was considered for simplicity. Hence denoting $f_{R}(R_0)\equiv\epsilon$ we get
\begin{eqnarray}
m_{1}\,=\,-\frac{(D - 2 (1 + \epsilon))^2}{D^2 \epsilon}
\label{epsilon}
\end{eqnarray}
and a relation between $m_1$, $D$ and $\epsilon$ can be imposed. Therefore this model would only depend on two parameters $\alpha$ and $\beta$.
A vanishing curvature solution also appears in this model and two nontrivial
curvature solutions are given by:
\begin{eqnarray}
R_{0}^{\pm}\,=\,\frac{\alpha}{2\beta(D-2)} \left[D(m_1-2)+4\pm\sqrt{m_1}\sqrt{m_1 D^2-8 D+16}\right].
\label{R0_model_IV}
\end{eqnarray}
The corresponding $1+f_{R}(R_0)$ values for \eqref{R0_model_IV} are
\begin{eqnarray}
1+f_{R}(R_{0}^{\pm})\,=\,1-\frac{4(D-2)^2}{\left(\sqrt{m_1 D^2-8D+16}\pm\,D\sqrt{m_1}\right)^2}
\label{1+f'(R0)_model_IV}
\end{eqnarray}
where $m_1>0$ and $m_1>(8D-16)/D^2$ are required for $R_{0}$ solutions to be real.  Since $1+f_{R}(R_{0})>0$ is also a condition to be fulfilled, that means that $\text{sign}(R_{0}^{\pm})=\text{sign}(\alpha\beta)$.

On the one hand, it can be shown that $1+f_{R}(R_{0}^{-})$ is not positive for any allowed value of $m_1$ and therefore this curvature solution $R_{0}^{-}$ is excluded of our research. On the other hand, $1+f_{R}(R_{0}^+)>0$ only requires $m_1>0$ for dimension
$D\geq4$ and therefore $\epsilon<0$ is required according to
\eqref{epsilon}. Therefore only two accessible regions need to be
studied:  Region $1$ $\{\alpha>0,\, \beta<0\}$ and Region $2$,
$\{\alpha<0,\,\beta>0\}$.

\begin{figure*}[!hbp]
\subfigure{
\begin{overpic}[width=7.30cm]{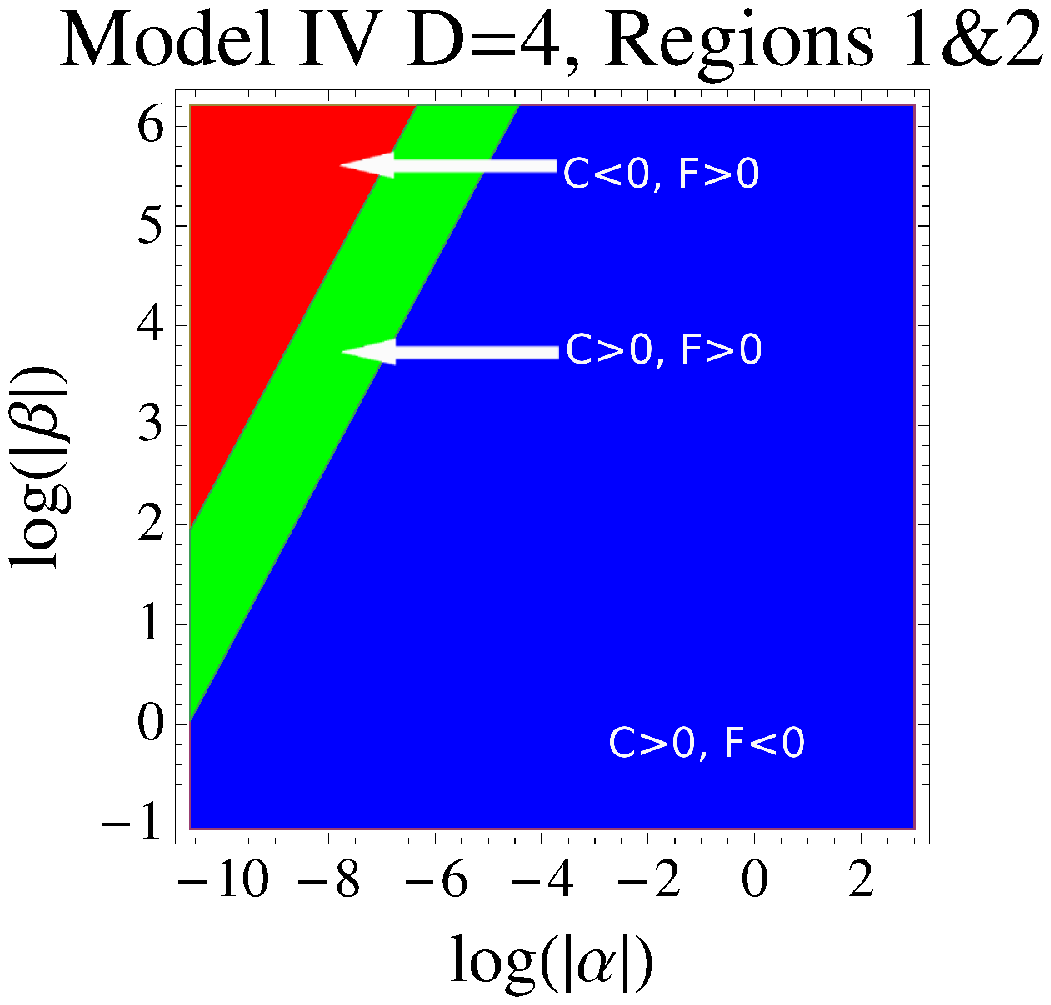}
\label{M4D4}
\end{overpic}
}
\subfigure{
\begin{overpic}[width=7.30cm]{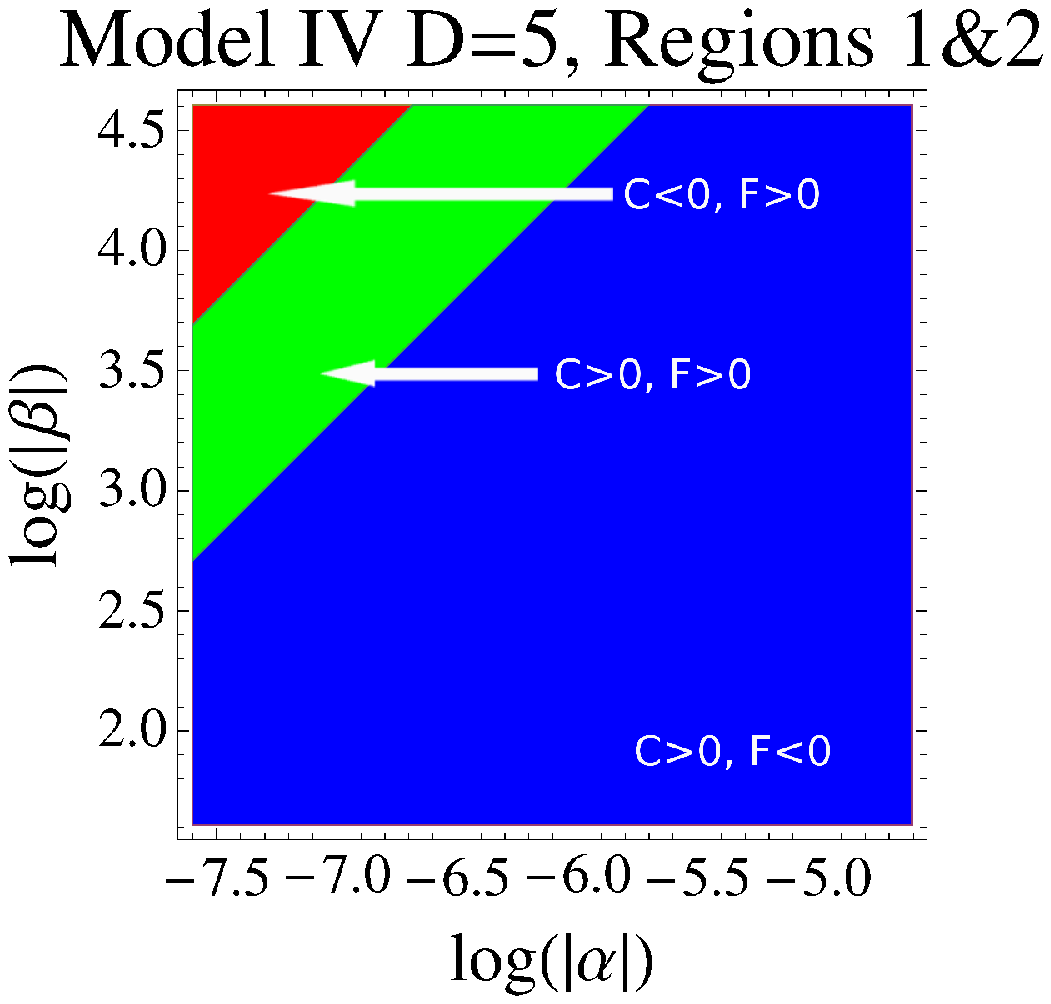}
\label{M4D5}
\end{overpic}
}
\caption{Thermodynamical regions in the $(|\alpha|,|\beta|)$ plane
 for Model IV in $D=4$  (left) and  $D=5$ (right).}
\label{M4_figure1}
\end{figure*}

\begin{figure*}[!hbp]
\subfigure{
\begin{overpic}[width=7.30cm]{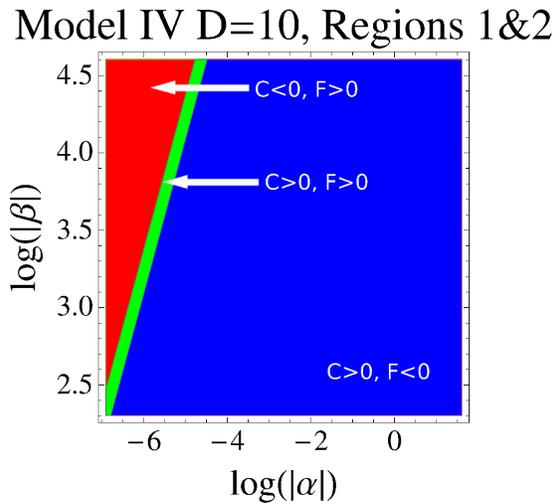}
\label{M4D10}
\end{overpic}
}
\caption{Thermodynamical regions in the $(|\alpha|,|\beta|)$ plane
 for Model IV in $D=10$.}
\label{M4_figure2}
\end{figure*}

In Figures \ref{M4_figure1} and \ref{M4_figure2} we plot the thermodynamical regions in the parameter space $(\alpha,\beta)$ for a chosen $\epsilon=-10^{-6}$. Note that $1+f_{R}(R_{0}^{+})$ does not depend either on $\alpha$ nor on $\beta$ and
that $R_{0}^{+}$ only depends on the quotient $\alpha/\beta$ for a fixed $m_1$.
\section{Figures for thermodynamical regions}
In the following pages we have plotted accessible thermodynamical regions for previously proposed $f(R)$ models I-IV. Thermodynamical regions have been plotted using different colors: red:\,$\{C<0$, $F>0\}$, green:\, $\{C>0$, $F>0\}$ and blue:\,$\{C>0$, $F<0\}$.

Parameter spaces have been chosen in order to show possible
thermodynamical transitions between regions.
\newpage

%

\section{Conclusions}
\label{sec:BH:Conclusions}
In this chapter we have considered static spherically symmetric solutions
in $f(R)$ theories of gravity in arbitrary dimensions. Firstly, we have discussed
the constant curvature case (including charged BH in 4 dimensions). Then,
the general case, without imposing a priori the condition
of constant curvature, has also been studied.

Another important result of this chapter was obtained by performing a perturbative
analysis around the EH case, assuming regular $f(R)$ functions. We have found explicit
expressions up to second order for the metric coefficients. These
coefficients only gave rise to constant curvature (Schwarzschild-$AdS$)
solutions as in the EH case.

On the other hand, we have also calculated thermodynamical quantities for
the $AdS$ BHs and considered the issue of the stability of this kind of
solutions. We have found that the condition for a $f(R)$ theory of gravity to
support this kind of BHs is given by $R_0+f(R_0)<0$ where $R_0$ is the constant curvature
of the $AdS$ space-time. This condition
also implies that the
effective Newton's constant is positive
and that the graviton does not become a ghost. Consequently thermodynamical and cosmological
viabilities of $f(R)$ theories turned out to be related as we have shown.

Finally we have considered several explicit examples of $f(R)$ functions
and studied the parameter regions in which BHs
in such theories are locally stable
and globally preferred. It was found that the qualitative behaviour is the same as in standard EH gravity
but the thermodynamical regions are modified depending on the parameters values in each case.
\newpage
\chapter[Brane-skyrmions and the CMB cold spot]{Brane-skyrmions and the CMB cold spot}
\label{chap:Cold spot}
\section{Introduction}
\label{sec:CS:Introduction}
Although the properties of BH
solutions in BW models have been exhaustively
studied in the literature, this is not the case of a different type of (topological)
spherically symmetric solutions, the so-called brane-skyrmions.
These configurations may appear in a natural way in a broad class of
BW theories.

On the one hand, this type of textures can be understood as holes in the brane
which make possible to pass through them along the extra-dimensional space.
For such objects, cosmological involved scales and consequences have not been already studied in the literature. Therefore, the {\it leitmotiv} of this chapter will be precisely to study some astrophysical and cosmological effects that those configurations may have in the CMB data. 

On the other hand, some striking features have recently been discovered \cite{Vielva} in the CMB data such as the presence of an anomalous cold spot (CS) in the WMAP temperature maps.
This CS has been interpreted in different manners by making use of different physical mechanisms. Among them, one of the most exciting from the physical point of view could justify the appearance of cold spots as the result of a collapsing texture, perhaps coming from some early universe Grand Unified Theory (GUT) phase transition.

In this chapter we shall propose an alternative explanation to those in the existing literature: it will be shown that the brane-skyrmions provide a natural scenario to reproduce the CS features and that typical involved scales, needed for the proposed brane-skyrmions to describe correctly the observed CS, can be as low as the electroweak scale.

The present chapter is thus organized as follows:
%
brane-skyrmions in brane-worlds are presented in Section \ref{sec:CS:CS as skyrmions} and
then we shall determine the possible distortion in the fractional
profile of temperatures generated by the presence of these objects.
Next, in Section \ref{sec:CS:Cold spot in WMAP} we shall present
the most recent results that claim the existence of a CS in the
CMB data. Different possible explanations that have been previously
proposed in the literature will be summarized in Section \ref{sec:CS:Cosmic
texture}.

The subsequent physical interpretation of the calculations
performed in Section \ref{sec:CS:CS as skyrmions} will be presented
in Section \ref{sec:CS:Physical interpretation}. In this section, we
will also show that those results are in agreement with the CMB data
and with other theoretical proposals.
To conclude the chapter, possible future detection
of these brane-skyrmions and some conclusions will be studied in
Section \ref{sec:CS:Future prospects and conclusions}.

The results of this chapter were originally published in \cite{Dombriz&Cembranos:CS}.
\section{Spherically symmetric brane-skyrmions}
\label{sec:CS:CS as skyrmions}
%
%
%
%
In this section we are going to generalize the results presented in Section 1.8 for static brane-skyrmions. Here we shall allow  time dependence for these objects but spherical symmetry will be preserved. Starting from the physical branon fields
equations provided in equation \eqref{eqns_pi}, one
may define the brane-skyrmion spherical coordinates with winding number $n_W$  by the nontrivial mapping
$\pi^\alpha:S^3\longrightarrow S^3$ as follows:
\begin{eqnarray}
\phi_K=\phi\,,\,\,\,\,\,\,\,
\theta_K=\theta\,,\,\,\,\,\,\,\,
\chi_K=F(t,r)\,
\label{skyan}
\end{eqnarray}
with boundary conditions satisfying the requirement
\begin{eqnarray}
F(t,\infty)-F(t,0)=n_W\pi.
\end{eqnarray}
This map is usually referred to as the {\it hedgehog} ansatz.
%
With these coordinates \eqref{skyan} introduced in the metric \eqref{induced_metric_2}, the Nambu-Goto action given in expression \eqref{action_NG} may be rewritten in terms of the previous coordinates as follows:
\begin{eqnarray}
S_{NG}\,=\,-f^4\int \mbox{d}^4x\,\text{sin}\theta\left(2\frac{v^2}{f^4}\text{sin}\left(F\right)\,\text{cos}\left(F\right)\right)\left[1-\frac{v^2}{f^4}\left(\dot{F}^2-F'^2\right)\right]^{1/2}\,.
\label{action_NG_rewritten}
\end{eqnarray}
Varying this action with respect to the function $F(t,r)$ the equation of motion for the
skyrmion profile is obtained and it becomes:
\begin{eqnarray}
\mbox{sin}(2F)-2rF'+\Big(r^2+\frac{v^2}{f^4}\mbox{sin}^2 F\Big)\frac{\ddot{F}-F''+\frac{v^2}{f^4}\left(\ddot{F}F'^2
+F''\dot{F}^2-2F'\dot{F}\dot{F}'\right)}{1-\frac{v^2}{f^4}\left(\dot{F}^2-F'^2\right)}\,=\,0
\label{eqn_motion_BS}
\end{eqnarray}
where dot and prime denote -- throughout this section -- derivatives with respect to $t$ and $r$ respectively.

In this chapter we are interested in the potential cosmological
effects due to the presence of a brane-skyrmion within our Hubble
radius. For that purpose, gravitational field perturbations will
be computed at large distances compared to the size of the extra
dimensions, i.e., we are interested in the region
\begin{eqnarray}
r^2\gg R_B^2\,=\,\frac{v^2}{f^4}.
\label{assumption}
\end{eqnarray}
Notice that in this region, gravity behaves  essentially as in four dimensional space-time and
standard GR can be used in the
calculations. Notice also that in order to simplify these calculations,
the effects due to the universe expansion will be ignored or at least assumed negligible.
This assumption is fully justified provided $r\ll H_0^{-1}$. In such a case
the unperturbed (ignoring the presence of the defect) background metric can be
taken as Minkowski, i.e.,
$\tilde g_{\mu\nu}=\eta_{\mu\nu}$ in expression \eqref{induced_metric_2}. From assumptions
in \eqref{assumption}, the equation of motion \eqref{eqn_motion_BS} reduces to:
\begin{eqnarray}
r^2\left(\ddot{F_0}-F_0''\right)+\mbox{sin}(2F_0)-2rF_0'\,=\,0
\label{EqFzeroth}
\end{eqnarray}
which is equivalent to expression $(3)$ in reference \cite{turok2}.
Notice that this is an expected result since, as shown in \cite{turok1,Press},
at large distances, i.e. except in the microscopic unwinding regions,
the dynamical evolution of the fields is completely
independent from the symmetry breaking mechanism, it simply depends on the
geometry of the coset manifold $K$. On small scales \cite{turok1,Press} it is
possible that higher-derivative terms could affect the dynamics and even
stabilize the textures, this is also the case of
brane-skyrmions \cite{CDM}, although generically they could unwind by means of
quantum-mechanical effects.

The equation \eqref{EqFzeroth} admits an exact solution with
winding number $n_{W}$ equals to unity of the  following form:
\begin{eqnarray}
F_{0}(t,r)\,=\,2\,\mbox{arctan}\left(-\frac{r}{t}\right)
\label{F_0_exacta}
\end{eqnarray}
with $t<0$ since as explained below, we will be interested
in photons passing the texture before it collapses.

Our approximated equation \eqref{EqFzeroth}
is consistent with the complete equation \eqref{eqn_motion_BS} for the
above solution \eqref{F_0_exacta} since the second term in the numerator in equation \eqref{eqn_motion_BS}
vanishes for  $F\equiv F_{0}$ and
\begin{eqnarray}
\dot{F}_{0}^2-F_{0}'^2\,=\,4\frac{r^2-t^2}{(r^2+t^2)^2}\,\,\,\,;\,\,\,\,
\sin^2 F_{0}\,=\,4\frac{r^2\,t^2}{(r^2+t^2)^2}
\end{eqnarray}
so that in the considered regime, the neglected terms in equation \eqref{eqn_motion_BS} are indeed
irrelevant for all $r$  and $t$ values.

Now that $F_{0}(t,r)$ has been determined,
we may calculate the energy-momentum tensor components also in this
region from the Nambu-Goto action \eqref{action_NG} as:
\begin{eqnarray}
T^{\mu\nu}\,=\,  2\vert \tilde{g} \vert^{-1/2}\frac{\delta S_{NG}}{\delta \tilde{g}_{\mu\nu}}
\end{eqnarray}
where note that $\tilde{g}$ is the involved metric to determine this tensor.
In spherical coordinates they become
\begin{eqnarray}
T_{00}\,&=&\,-\frac{2v^2(r^2+3t^2)}{(t^2+r^2)^2}\,\,\,\,;\,\,\,\,
T_{rr}\,=\,\frac{2v^2(t^2-r^2)}{(t^2+r^2)^2}\nonumber\\
T_{0r}\,&=&\,\frac{4v^2 r t}{(t^2+r^2)^2}\,\,\,\,;
\,\,\,\,T_{\theta\theta}\,=\,\frac{2v^2 r^2(t^2-r^2)}{(t^2+r^2)^2}\nonumber\\
T_{\phi\phi}\,&=&\,\mbox{sin}^2 \theta \, T_{\theta\theta}
\label{T_spherical}
\end{eqnarray}
and note that $\nabla_{\mu} T^{\mu}_{\,\nu}$  identically vanishes.

We shall now determine the background metric $\tilde g_{\mu\nu}$ in the
$r\gg R_B$ region and in the presence of the  brane-skyrmion
as a small perturbation on the Minkowski metric, i.e.
$\tilde g_{\mu\nu}=\eta_{\mu\nu}+h_{\mu\nu}$.

Thus, for the scalar perturbation of the  Minkowski space-time
in the longitudinal gauge, the line element will adopt the form, straightforwardly obtained from equation \eqref{perturbed_metric} in Chapter 3 if $a(\t)\equiv1$, that follows:
\begin{equation}
\text{d}s^2\,=\,(1+2\Phi)\text{d}t^2-(1-2\Psi)(\text{d}r^2+r^2\text{d}\Omega_{2}^2)
\label{perturbed_metric_Minkowski}
\end{equation}
and thus
the perturbed Einstein tensor components in cartesian coordinates are the following
(see \cite{Giovannini}):
 \begin{eqnarray}
\delta G^{0}_{0}\,&=&\,2\nabla^2\Psi\nonumber\\
\delta G^{j}_{i}\,&=&\,-[2\ddot{\Psi}+
\nabla^2(\Phi-\Psi)]\delta_{i}^{j}
+\partial_{i}\partial^{j}(\Phi-\Psi)\nonumber\\
\delta G^{0}_{i}\,&=&\,2\partial_{i}\dot{\Psi}
\end{eqnarray}
with $i,j=1,2,3$ and $\nabla^2\equiv\sum_{i=1}^{3}
\partial_{i}\partial^{i}$. Using Einstein's equations
$\delta G^{\mu}_{\,\nu}\,=\,-8\pi G\, T^{\mu}_{\,\nu}$
we determine that the potentials $\Phi$ and $\Psi$ are
\begin{eqnarray}
\Psi\,\equiv\,\Phi\,=\,4\pi G_{}\,v^2\,\mbox{log}\left(\frac{r^2+t^2}{t^2}\right)\,.
\label{Phi}
\end{eqnarray}
The physical metric on which photons propagate is not
the $\tilde g_{\mu\nu}$ we have just calculated, but the
induced metric \eqref{induced_metric_2}. However, using the solution in
(\ref{F_0_exacta}), we  find that the contribution from branons
fields is $\Od( R_B^2/r^2)$, i.e. negligible in the region we are interested in, so that
 $g_{\mu\nu}\simeq \tilde g_{\mu\nu}$.

Photons propagating on the perturbed metric will suffer red (blue)-shift
due to the Sachs-Wolfe (SW) effect \cite{Giovannini}. The full expression for the
temperature fluctuation is given by:
\begin{eqnarray}
\left(\frac{\Delta T}{T}\right)_{\text{SW}}\,
=\,- [\Phi]^{\tau_{f}}_{\tau_{i}}\,
+\,\int^{\tau_{f}}_{\tau_{i}}\left(\dot{\Psi}
+\dot{\Phi}\right)\mbox{d}\tau\,
=\,
- [\Phi]^{\tau_{f}}_{\tau_{i}}\,
+\,\int^{\tau_{f}}_{\tau_{i}}2\dot{\Phi}\mbox{d}\tau
\label{SW_effect}
\end{eqnarray}
where we have considered local and integrated SW effects and
neglected the Doppler contribution. $\tau_{i}$
and $\tau_{f}$
are the initial and final times respectively required
to study temperature fluctuation in that interval.
Substituting expression (\ref{Phi}) in the previous one and using
that
\begin{eqnarray}
r^2\,=\,z^2+R^2\,\,\,;\,\,\, z\,=\,t-t_{0}
\end{eqnarray}
as may be seen at the schematic representation in Figure \ref{figura_turok},
the fractional distortion we obtain is:

\begin{figure}
\begin{center}
{\epsfxsize=10.0cm\epsfbox{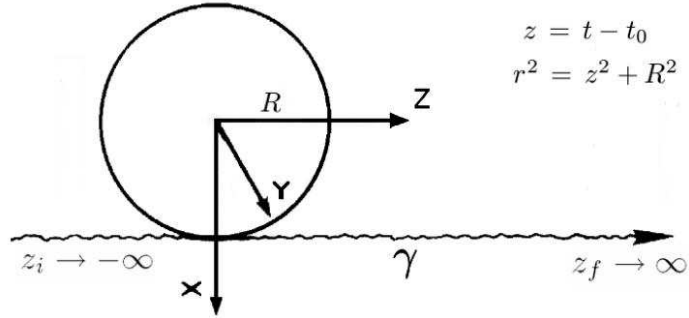}}
\end{center}
\caption{Schematic representation of photon trajectory passing near the brane-skyrmion as originally presented in \cite{turok2}. $R$ is the impact parameter (the radius in the plotted circle) and the photon trajectory through the bottom horizontal line is chosen along the Z axis.}
\label{figura_turok}
\end{figure}
\begin{eqnarray}
\left(\frac{\Delta T}{T}\right)_{\text{SW}}
\,=\,8\pi G_{} v^2\left(\frac{t_0}{\sqrt{2R^2+t_0^2}}\,
\mbox{arctan}\left(\frac{t_0+2z}{\sqrt{2R^2+t_0^2}}\right)-\mbox{log}|t_0+z|\right)_{z_i}^{z_f}
\end{eqnarray}
where $R$ is the impact parameter and $t_0$ is the time at which the
photon passes the texture position at $z=0$.
In the limit where $z_{i}\rightarrow -\infty$ and
$z_f\rightarrow\infty$ the result is
\begin{eqnarray}
\left(\frac{\Delta T}{T}\right)_{SW} \,=\,\epsilon\frac{t_0}{\sqrt{2 R^2+t_{0}^2}}
\label{distorsion_BW}
\end{eqnarray}
with $\epsilon\,\equiv\,8\pi^{2}G v^2$.
As shall be seen in the next section, this temperature profile is the one which fits the apparently observed anomaly in the CMB data.
\section{Cold spot in WMAP data}
\label{sec:CS:Cold spot in WMAP}
One of the most important pieces of information about the
history and nature of our universe comes from the Cosmic Microwave
Background (CMB). Measurements of the CMB temperature anisotropies
obtained by WMAP \cite{WMAP,Hinshaw} have been thoroughly studied in
recent years. Such anisotropies have been found to be Gaussian as
expected in many standard cosmological scenarios corresponding to
density fluctuations of one part in a hundred thousand in the early
universe. However, by means of a wavelet analysis, an anomalous CS,
 apparently inconsistent with homogeneous
Gaussian fluctuations, was found in \cite{Vielva,Cruz1} in the
southern hemisphere centered at the position $b=-57$$^\circ$,
$l=209$$^\circ$ in galactic coordinates.
\footnote{where $b$ is the galactic
latitude measured from the plane of the galaxy to the object using
the Sun as vertex. The galactic longitude is referred to as $l$ and
it is measured in the plane of the galaxy using an axis pointing
from the Sun to the galactic center.}
The characteristic scale in the sky of the CS is about 5$^\circ$.

\begin{figure}
\begin{center}
{\epsfxsize=8.0cm\epsfbox{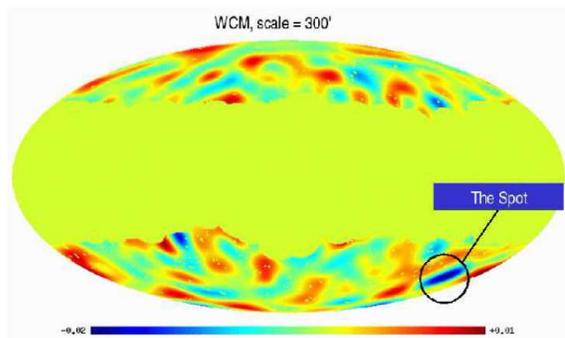}}
\end{center}
\caption{The combined and foreground cleaned Q-V-W WMAP map after convolution with the SMHW at scale $R_{9}$
as shown in \cite{Cruz:2006}. The CS position is marked and seen at the bottom left part of the plot.}
\label{Figure1_CS}
\end{figure}

The CS position in usual WMAP plots is shown in Figure \ref{Figure1_CS}. The existence of this CS has been
claimed to be confirmed more recently in reference \cite{Nas}.
However, in recent references \cite{WMAP_2010} it has been argued
that there is no compelling evidence for deviations from the $\Lambda\text{CDM}$ model in the
WMAP data. In particular, it is claimed that the evidence that the CS is statistically anomalous is not
robust.

%
%
%
%
%
%
Concerning the origin of the CS, different explanations have been proposed in the literature:  in references
\cite{Inoue:2006fn,Inoue:2006rd,Rudnick:2007kw} the CS was tried to be explained by voids in the matter distribution
whereas the reference \cite{Chang:2008} dealt with possible string theory bubble origin. Other authors
proposed in \cite{Cruz:2006} the Sunyaev-Zeldovich (SZ) effect as possible explanation
showing that the flat frequency dependence
of the CS is incompatible with being caused by a SZ signal alone. However, a combination of CMB plus SZ effect may explain the spot and could have a sufficiently flat frequency spectrum.

\section{Cold spot as a cosmic texture}
\label{sec:CS:Cosmic texture}
A very interesting possibility about the CS origin has recently been proposed in \cite{cruzturok}. According to it, some
theories of high energy physics predict the formation of various
types of topological defects, including cosmic textures
\cite{turok1} which would generate hot and cold spots in the CMB
\cite{turok2}. These textures would be the remnants of a symmetry breaking
phase transition that took place in the early universe.

In order to produce textures, the cosmic phase transition must be related to a
global symmetry breaking pattern from one group $G$ to a subgroup
$H$ so that the coset space  $K=G/H$ has a nontrivial third homotopy group.
A  typical example is $K=SU(N)$, which has associated
$\pi_3(K)= \Bbb{Z}$ for $N\geq 2$.
Notice that, as usual, in order to get a texture formed
in the transition, the symmetry breaking must correspond to a global
symmetry since if it were local it could be gauged away.

Textures can be understood as localized wrapped field
configurations which collapse and unwind on progressively larger
scales. These textures can produce a concentration of energy which gives
rise to a time dependent gravitational potential. CMB photons
traversing the texture region will suffer a red (blue) shift
producing a cold (hot) spot in the CMB maps.
In reference \cite{cruzturok} the authors
consider a $SU(2)$ non-linear sigma model
to build up a model of texture that could explain the observed CS.
They simulate the unwinding texture by using a
spherically symmetric scaling solution and they find a fractional
temperature distortion given by:
\begin{equation}
\frac{\Delta T}{T}(\theta)=\pm\epsilon\,\frac{1}{\sqrt{1+4\left(\frac{\theta}{\theta_c}\right)^2}}
\label{distorsion_CS}
\end{equation}
where $\theta$ is the angle from the center, $\epsilon$ is a measure
of the amplitude and $\theta_c$ is the scale parameter that depends
on the time at which the texture unwinds. The best fit of the CS is found
for $\epsilon=7.7\times10^{-5}$ and $\theta_c=5.1^\circ$.
Furthermore, the parameter $\epsilon$
is given by $\epsilon\equiv 8\pi^2 G_{} \Phi_0$,
where $\Phi_0$ is the fundamental symmetry breaking scale which is then
set to be $\Phi_0\simeq 8.7 \times 10^{15}$ GeV. This scale is
nicely close to the GUT scale thus making the results given in
\cite{cruzturok}  extremely interesting.

Nonetheless, it is important to stress that textures require having
a global symmetry breaking but what one finds typically in GUTs is a
local symmetry breaking producing the Higgs mechanism and then
destroying the topological meaning of texture or any other possible
defect appearing in the cosmic transition.
\section{Physical interpretation of the results and involved scales}
\label{sec:CS:Physical interpretation}
As has been shown in Sections \ref{sec:CS:CS as skyrmions} and
\ref{sec:CS:Cosmic texture}, the results for the fractional
temperature distortion provoked by the CS given by expression
\eqref{distorsion_CS} coincides with the fractional distortion
caused by a brane-skyrmion given by expression
\eqref{distorsion_BW} provided that:
\begin{eqnarray}
\Phi_0\,=\,v\,\,\,\,;\,\,\,\, 2\left(\frac{\theta}{\theta_c}\right)\,=\,\frac{R}{t_0}
\end{eqnarray}
where $\theta_c$ was defined in reference \cite{cruzturok}.

Let us now estimate the required scales in order for this kind of brane-skyrmion to be able
to explain the observed CS. For the minimal model supporting brane-skyrmions with three extra
dimensions $\delta=3$, the relation \eqref{M_F} approximately becomes
\begin{eqnarray}
M_{P}^2\simeq R_{B}^{3}M_{D}^5
\end{eqnarray}
since gravitation is embedded in  $D=7$  dimensions and therefore $R_{B}^3$ is the characteristic volume
for the compactified extra dimensions. Therefore
\begin{eqnarray}
R_{B}\,\simeq\,\left(\frac{M_{P}^2}{M_{D}^5}\right)^{1/3}
\end{eqnarray}
and consequently
\begin{eqnarray}
v^2\,\equiv\,f^4R_{B}^2\,\simeq\,f^4\left(\frac{M_{P}^2}{M_{D}^5}\right)^{2/3}\,.
\end{eqnarray}
If the parameter $\epsilon$ needs to be fixed around $7.7 \cdot 10^{-5}$, then the required $v$ would be
\begin{eqnarray}
v\simeq 1.2 \cdot 10^{16}\,\text{GeV}
\end{eqnarray}
which in fact can be achieved with $M_D\sim f\sim$ TeV. Notice that for this parameter
range, the radius of the extra dimension is around
\begin{eqnarray}
R_B\sim 10^{-8}\,\text{m},
\end{eqnarray}
 i.e., previous approximations in expression \eqref{assumption} are totally justified.

In addition to the previous estimations, we have also checked
that the possible effects coming from a nonvanishing branon mass
are suppressed by $\Od(M^2/v^2)$. Therefore for mass values also
around $M\sim $ TeV, which are typical of branon as dark
matter candidate, see \cite{DM2}, such effects are negligible.

In other words, brane-skyrmions provide an accurate description for the CS without
any need of introducing very high energy (GUTs) scales since
the correct temperature fluctuation amplitude can be
obtained with natural values around the electroweak scale.


\subsection{Brane-skyrmions abundance}
The abundance of brane-skyrmions in this model is given by its
low-energy (large distance) dynamics  as described by a non-linear sigma model as that in \cite{cruzturok}.
%
This is nothing but the well-known fact that, except in the
microscopic unwinding region, the field evolution only depends on
the geometry of the coset space $K$, but not on the details of the
symmetry breaking mechanism. As was shown in \cite{Press} in a
simple model with a potential term, the final abundance of defects
and other properties of the pattern of density perturbations are
expected to be not very sensitive to the short distance physics,
once the texture unwinds, making this kind of theories highly
predictive. For that reason, we expect that provided the same kind
of initial conditions are imposed in both models, the predicted
abundance of hot and cold spots agree with that obtained from
simulations in \cite{cruzturok}. Such simulations show that the
number of unwinding textures per comoving volume and conformal
time $\tau$ can be estimated as
\begin{eqnarray}
\frac{\text{d}n}{\text{d}\tau}\,=\,\frac{\nu}{\tau^{4}}
\end{eqnarray}
with $\nu\simeq 2$. This allows to estimate the number of hot and cold spots in
a given angular radius interval. As commented above, this is
a quite robust result at late times with little effects for short
distance dynamics.
In the case of brane-skyrmions, the short distance
effects will be embodied in the higher-derivative terms appearing
in the expansion of the Nambu-Goto action or even in possible induced
curvature terms generated by quantum effects \cite{CDM}.
Provided that such terms do not stabilize the brane-skyrmions,  we expect
that such an abundance  could be directly applied
in our case.
\newpage
\section{Future prospects and conclusions}
\label{sec:CS:Future prospects and conclusions}
In this section we dedicate some lines to mention how both present and future experiments may confirm the
validity of the BW models. Then we shall sketch the most important conclusions and consequences of the
brane-skyrmion model presented in this chapter as a viable explanation for the observed CS.

On the one hand, the fact that the fundamental scales of the theory are of the order of the TeV opens the possibility
to test this explanation with collider experiments through the production of real or virtual branons and KK-gravitons.
The expected signatures of the model \cite{Mirabelli:1998rt}
from the production of KK-gravitons come fundamentally from the
single photon channel studied by LEP, which restrict ${M}_D> 1.2$ TeV
at 95\% of confidence level.

On the other hand, the LHC accelerator will be able to test the model up to ${M}_D=3.7$ TeV, analysing
single photon and monojet production to establish the number of extra dimensions as was explained in Section
\ref{sec:Int:BW:BH}.

Finally, there also exists the possibility to find
signatures of the model at low energies associated with branon
phenomenology. This case is more interesting from the cosmological point of
view, since branons can constitute the nonbaryonic dark matter abundance as
typical WIMPs \cite{DM2,Cembranos:2004jp}. Present
constraints coming from the single photon analysis realized by L3 (LEP)
imply $f > 122$ GeV (at 95 \% C.L.) \cite{Alcaraz:2002iu,Achard:2004uu} and
the LHC will be able to check this model up to $f=1080$ GeV through
monojet production \cite{Cembranos:2004jp}. The idea to test the physics associated with the
CS with the next generation of colliders at the TeV scale is a very
intriguing and distinctive property of this texture.

\chapter{Conclusions and prospects}
\label{chap:Conclusions}
%
%
%
%
In this thesis we have studied some
cosmological and astrophysical consequences
in two different types of modified gravity
theories: $f(R)$ theories and brane-world extra dimensions theories.

First, we have shown some relevant results for $f(R)$ modified gravities. Thus, we have studied how
some features of general relativity can be mimicked by $f(R)$ models if certain
conditions are imposed on those models.
We have also proved that there exists a class of $f(R)$ models which can indeed
reproduce the present cosmological evolution --from matter-radiation equality till today -- as
described by general relativity within the standard cosmological concordance model. These $f(R)$ models
allow to remove the cosmological constant term and therefore the
observed acceleration is understood to have a purely geometrical
interpretation with no dark energy origin. Furthermore, initial
conditions may be imposed on such a class of theories in order
to recover a null scalar curvature solution for a vacuum scenario.
Unfortunately such functions are not cosmologically viable and
should be considered as effective models to reproduce
$\Lambda\text{CDM}$ evolution.

%
%
%
Also, it has been proved that any perfect fluid parameterized by a
constant equation of state can be reproduced by the presence of
$f(R)$ terms. Thus, these functions
could provide a mechanism to reproduce the
cosmological behaviour of dark energy type fluids.
It therefore seems
worthwhile to try to search for a unique phenomenological $f(R)$
theory able to explain the main features in the cosmological
evolution, from inflation till late time observed acceleration.

Next, we have considered the modification introduced by the new $f(R)$ terms
in the evolution of cosmological perturbations. Here,
%
an analysis concerning first order cosmological scalar perturbations has been performed.
We have presented a general method in order to obtain the evolution equation
of the matter density contrast for arbitrary $f(R)$ theories. That procedure was proven to be valid
regardless of the chosen $f(R)$ model or
the size of the involved scales.
An important by-product of the calculations is the conclusion that the
usual approximations made in the literature for sub-Hubble modes are not
always valid. In fact, the required hypotheses to get the so-called quasi-static
approximation have been established here. 
For those sub-Hubble modes, it was found
that only for $f(R)$ models satisfying the local gravity constraints, the evolution
of perturbations is indistinguishable from that obtained using the quasi-static approximation but
perfectly distinguishable from the evolution obtained by the $\Lambda\text{CDM}$ model.

Now that those theoretical calculations were made,
the correct evolution for matter perturbations in sub-Hubble modes for
$f(R)$ models is obtained. Therefore
some robustness tests may be used to compare
experimental data with expected theoretical results. Thus
both the validity and certain constraints on those $f(R)$ functions
may be established. This procedure shows how
our theoretical results may be used
in order to confront the predictions made by a given $f(R)$ theory with large scale structure observations.

%
%
%
%

As a final subject within our $f(R)$ models research, we dealt
with some features of black holes in arbitrary dimensions. First, the scalar
constant curvature case for static and spherically symmetric solutions
in vacuum was studied. It was determined that the only possible
solution for the modified Einstein equations in this case was the
Schwarzschild-(anti-)de Sitter solution.
As a complementary result to the previous research, we studied
the static and spherically symmetric case without imposing
the constant curvature condition. To do so, a perturbative approach was
performed around the standard
Schwarzschild-anti-de Sitter solution
of general relativity. It was found that, up to second order in perturbations,
the only solution was the generalized Schwarzschild-anti-de Sitter solution with modified
coefficients in terms of $f(R)$ and its derivatives evaluated at the background curvature.

This part of the thesis was finished by studying the thermodynamic properties of
Schwarzschild-anti-de Sitter black holes in $f(R)$ theories. Thus
accessible thermodynamical regions for several $f(R)$ models were
studied in detail.
In that realm, a very
interesting property, which encourages further
investigation, was found: it turned out that thermodynamical viability of $f(R)$
theories in constant curvature solutions was embodied by the condition
$1+\text{d}f(R)/\text{d}R > 0$. This condition was also required
to ensure gravitational viability for $f(R)$ models. Consequently two
different aspects of the viability of $f(R)$ theories
have been brought together.
%
%
%

The last part of the thesis was devoted to a study of some
consequences of the existence of topologically nontrivial
brane configurations, the so-called brane-skyrmions, in brane-world theories.
In particular we have considered the effects of
brane-skyrmions on the temperature
fluctuations of the CMB and their potential
connection with the so-called CMB cold spot.
We have shown that brane-world theories can naturally accommodate
this striking cosmological feature
by invoking these nontrivial topological configurations. A different
type of textures related to Grand Unified
Theories has also been considered in
the literature, but the model presented
here shows that this type of defect could
be explained from the electroweak scale physics.

%
%
To conclude, the results included in this thesis have shown that
modified gravity theories remain compelling candidates to
describe the properties of the gravitational interaction on very large scales.
Both the validity and viability of these theories have still to be
subjected to many theoretical and experimental tests. To do so, the
possibility of reproducing standard cosmological results, first order
perturbations, black holes and features on the CMB temperature
maps were
considered as interesting aspects which could shed some
light on the fundamental properties of the
gravitational interaction.

\newpage
$ $
\newpage
\makeatletter
\renewcommand{\chapter}{
                    \thispagestyle{plain}%
                    \global\@topnum\z@
                    \@afterindentfalse
                    \secdef\@chapter\@schapter}
\def\@chapter[#1]#2{
                         \refstepcounter{chapter}%
                         \typeout{\@chapapp\space\thechapter.}%
                    \chaptermark{#1}
                    \@makechapterhead{#2}%
                     \@afterheading
\addcontentsline{toc}{chapter}{\protect\numberline{\thechapter}#1}
}
\renewcommand{\@makechapterhead}[1]{%
  {\ifnum \c@secnumdepth >\m@ne
        \Large\bfseries \@chapapp\space \thechapter
        \par\nobreak
    \fi
    \interlinepenalty\@M
    \Large\bfseries #1\par\nobreak
    \vskip 0\p@
  }}
 \renewcommand{\section}{\@startsection {section}{1}{\z@}%
                                   {-3.5ex \@plus -1ex \@minus -.2ex}%
                                   {2.3ex \@plus.2ex}%
                                   {\normalfont\large\bfseries}}

\makeatother
\newpage\appendix
\newpage
\chapter[Coefficients in $f(R)$ cosmological perturbations]{Coefficients in $f(R)$ cosmological perturbations}
\label{app:perturbation_coefficients}
In this appendix we include the coefficients for the series expansions given in equation \eqref{delta_equation_separated} in Chapter 3.
All the $\alpha$'s terms (coming from EH-part) and
the first four $\beta$'s terms (coming from $f$-part) for each
$\delta$ derivative, i.e. $\delta^{iv},...,\delta,$, in equation \eqref{delta_equation_separated} have been included once
the condition $\vert f_{R}\vert\ll 1$ has been imposed to simplify
their expressions. For sub-Hubble scales the excluded $\beta$'s
terms are negligible with respect to the written ones.
%
%
%
%
\section{Appendix $\bf{I}$: $\alpha's$ and $\beta's$ coefficients}
\label{Appendix I}
Coefficients for $\delta^{iv}$ term:
\begin{eqnarray}
\beta^{(1)}_{4,f}\,&\simeq&\,8f_{R}^{4}(1+f_{R})^{6}f_{1}^{4}\epsilon^{2}\nonumber\\
\beta^{(2)}_{4,f}\,&\simeq&\,72f_{R}^{3}f_{1}^{3}\epsilon^{4}(-2+\kappa_2)\nonumber\\
\beta^{(3)}_{4,f}\,&\simeq&\,216f_{R}^{2}f_{1}^{2}\epsilon^{6}(-2+\kappa_{2})^{2}\nonumber\\
\beta^{(4)}_{4,f}\,&\simeq&\,216f_{R}f_{1}\epsilon^{8}(-2+\kappa_{2})^{3}.
\label{delta_4_coefficients}
\end{eqnarray}
Coefficients for $\delta'''$ term:
\begin{eqnarray}
\beta^{(1)}_{3,f}\,&\simeq&\,8f_{R}^{4}(1+f_{R})^{5}f_{1}^{4}
\mathcal{H}\epsilon^{2}[3+f_{R}(3+f_{1})]\nonumber\\
\beta^{(2)}_{3,f}\,&\simeq&\,6f_{R}^{3}f_{1}^{2}\mathcal{H}\epsilon^{4}
\{8f_{2}(-2+\kappa_{2})+4f_{1}[12\kappa_{1}+9\kappa_{2}-2(9+\kappa_{3})]\}\nonumber\\
\beta^{(3)}_{3,f}\,&\simeq&\,-72f_{R}^{2}f_{1}\mathcal{H}\epsilon^{6}
(-2+\kappa_{2})[-4f_{2}(-2+\kappa_{2})+f_{1}(19-23\kappa_{1}-10\kappa_{2}
+4\kappa_{3})]\nonumber\\
\beta^{(4)}_{3,f}\,&\simeq&\,-216f_{R}\mathcal{H}\epsilon^{8}(-2+\kappa_{2})^{2}[-2f_{2}(-2+\kappa_{2})+f_{1}(7-11\kappa_{1}-4\kappa_{2}+2\kappa_{3})].
\label{delta_3_coefficients}
\end{eqnarray}
Coefficients for $\delta''$ term:
\begin{eqnarray}
\alpha^{(1)}_{2,\text{EH}}\,&=&\,432(1+f_{R})^{10} \mathcal{H}^{2} \epsilon^{8} (-1+\kappa_1)(-2+\kappa_2)^{3}\nonumber\\
\alpha^{(2)}_{2,\text{EH}}\,&=&\,1296(1+f_{R})^{10} \mathcal{H}^{2} \epsilon^{10} (-1+\kappa_1)^{2}(-2+\kappa_2)^{3}\nonumber\\
\alpha^{(3)}_{2,\text{EH}}\,&=&\,3888(1+f_{R})^{10} \mathcal{H}^{2}\epsilon^{12}(-1+\kappa_{1})^{2}(-2+\kappa_2)^{3}\nonumber\\
\beta^{(1)}_{2,f}\,&\simeq&\,8f_{R}^{4}(1+f_{R})^{6}f_{1}^{4}\mathcal{H}^{2}\nonumber\\
\beta^{(2)}_{2,f}\,&\simeq&\,88f_{R}^{3}f_{1}^{3}\mathcal{H}^{2}\epsilon^{2}(-2+\kappa_2)\nonumber\\
\beta^{(3)}_{2,f}\,&\simeq&\,24f_{R}^{2}f_{1}^{2}\mathcal{H}^{2}\epsilon^{4}(-2+\kappa_2)(-28+2\kappa_{1}+13\kappa_{2})\nonumber\\
\beta^{(4)}_{2,f}\,&\simeq&\,72f_{R}f_{1}\mathcal{H}^{2}\epsilon^{6}(-2+\kappa_2)^{2}(-14+4\kappa_{1}+5\kappa_{2}).
\label{delta_2_coefficients}
\end{eqnarray}

Coefficients for $\delta'$ term:
\begin{eqnarray}
\alpha^{(1)}_{1,\text{EH}}\,&=&\,432(1+f_{R})^{10} \mathcal{H}^{3} \epsilon^{8} (-1+\kappa_1)(-2+\kappa_2)^{3}\nonumber\\
\alpha^{(2)}_{1,\text{EH}}\,&=&\,2592(1+f_{R})^{10} \mathcal{H}^{3} \epsilon^{10}(-1+\kappa_1)^{2}(-2+\kappa_{2})^{3}\nonumber\\
\alpha^{(3)}_{1,\text{EH}}\,&=&\,-7776(1+f_{R})^{10} \mathcal{H}^{3}\epsilon^{12}(-1+\kappa_{1})^{3}(-2+\kappa_{2})^{3}\nonumber\\
\beta^{(1)}_{1,f}\,&\simeq&\,8f_{R}^{4}(1+f_{R})^{6}f_{1}^{4}\mathcal{H}^{3}\nonumber\\
\beta^{(2)}_{1,f}\,&\simeq&\,88f_{R}^{3}f_{1}^{3}\mathcal{H}^{3}\epsilon^{2}(-2+\kappa_2)\nonumber\\
\beta^{(3)}_{1,f}\,&\simeq&\,24f_{R}^{2}f_{1}^{2}\mathcal{H}^{3}\epsilon^{4}(-2+\kappa_2)(-28+2\kappa_{1}+13\kappa_{2})\nonumber\\
\beta^{(4)}_{1,f}\,&\simeq&\,72f_{R}f_{1}\mathcal{H}^{3}\epsilon^{6}(-2+\kappa_2)^{2}(-14+4\kappa_{1}+5\kappa_{2}).
\label{delta_1_coefficients}
\end{eqnarray}

Coefficients for $\delta$ term:
\begin{eqnarray}
\alpha^{(1)}_{0,\text{EH}}\,&=&\,432(1+f_{R})^{10} \mathcal{H}^{4} \epsilon^{8} (-1+\kappa_1)(2\kappa_1-\kappa_2)(-2+\kappa_2)^{3}\nonumber\\
\alpha^{(2)}_{0,\text{EH}}\,&=&\,1296(1+f_{R})^{10}\mathcal{H}^{4} \epsilon^{10}(-1+\kappa_1)^{2}(-1+4\kappa_1-\kappa_2)(-2+\kappa_{2})^{3}\nonumber\\
\alpha^{(3)}_{0,\text{EH}}\,&=&\,3888(1+f_{R})^{10} \mathcal{H}^{4}\epsilon^{12}(-1+\kappa_{1})^{2}(2\kappa_{1}^{2}-\kappa_2)(-2+\kappa_{2})^{3}\nonumber\\
\beta^{(1)}_{0,f}\,&\simeq&\,-\frac{16}{3}f_{R}^{4}(1+f_{R})^{5}f_{1}^{4}\mathcal{H}^{4}
[2+f_{R}(2+2f_{1}-f_{2}-2\kappa_1)-2\kappa_1]\nonumber\\
\beta^{(2)}_{0,f}\,&\simeq&\,112f_{R}^{3}f_{1}^{3}\mathcal{H}^{4}\epsilon^{2}(-1+\kappa_{1})(-2+\kappa_{2})\nonumber\\
\beta^{(3)}_{0,f}\,&\simeq&\,48f_{R}^{2}f_{1}^{2}\mathcal{H}^{4}\epsilon^{4}(-1+\kappa_{1})(-2+\kappa_{2})(-16+2\kappa_{1}+7\kappa_{2})\nonumber\\
\beta^{(4)}_{0,f}\,&\simeq&\,144f_{R}f_{1}\mathcal{H}^{4}\epsilon^{6}(-1+\kappa_{1})(-2+\kappa_{2})^{2}(-6+4\kappa_{1}+\kappa_{2}).
\label{delta_0_coefficients}
\end{eqnarray}

\newpage
\section{Appendix $\bf{II}$: $c's$ coefficients}
\label{Appendix II}
The coefficients appearing in equation \eqref{eq_delta_cs} are explicitly provided here once the condition
$\vert f_{R}\vert\ll 1$ has been imposed.
%
\begin{eqnarray}
c_{4}\,&=&\,-f_{R}f_{1}[-f_{R}f_1k^{2}-3\mathcal{H}^2(-2+\kappa_2)]^{3}\nonumber\\
&&\nonumber\\
c_{3}\,&=&\,-3f_{R}\mathcal{H}[-f_{R}f_1k^2-3\mathcal{H}^2(-2+\kappa_2)]\{
f_{R}^{2}f_{1}^{3}k^4+6f_2\mathcal{H}^4(-2+\kappa_2)^2+f_1\mathcal{H}^2(-2+\kappa_2)\nonumber\\ &&\times[2f_{R}f_2k^2+
3\mathcal{H}^2(-7+11\kappa_1+4\kappa_2-2\kappa_3)]+2f_{R}f_1\mathcal{H}^2k^2(-6+6\kappa_1+3\kappa_2-\kappa_3)\}\nonumber\\
&&\nonumber\\
c_{2}\,&=&\,[-f_{R}f_1k^2-3\mathcal{H}^2(-2+\kappa_2)]^{2}\nonumber\\
&&\times[f_{R}^{2}f_{1}^{2}k^4+5f_{R}f_1\mathcal{H}^2k^2(-2+\kappa_2)+6\mathcal{H}^4(-1+\kappa_1)(-2+\kappa_2)]\nonumber\\
&&\nonumber\\
c_{1}\,&=&\,[-f_{R}f_1k^2-3\mathcal{H}^2(-2+\kappa_2)]^{2}\nonumber\\
&&\times[f_{R}^{2}f_{1}^{2}k^4\mathcal{H}+5f_{R}f_1\mathcal{H}^3k^2(-2+\kappa_2)+6\mathcal{H}^5(-1+\kappa_1)(-2+\kappa_2)]\nonumber\\
&&\nonumber\\
c_{0}\,&=&\,\frac{2}{3}\mathcal{H}^2(-1+\kappa_1)
[-f_{R}f_1k^2-3\mathcal{H}^2(-2+\kappa_2)]^{2}
[2f_{R}^{2}f_{1}^{2}k^4+9f_{R}f_1
\mathcal{H}^2k^2(-2+\kappa_2)\nonumber\\
&+&9\mathcal{H}^4(2\kappa_1-\kappa_2)(-2+\kappa_2)].
\label{coefficients_cs}
\end{eqnarray}
\newpage
$ $
\newpage

\makeatletter
      \renewcommand{\@evenhead}{\thepage\hfil}%
      \renewcommand{\@oddhead}{\hfil\thepage}%
\makeatother
\newpage
\centering
\
\vskip 5.5 cm
\begin{minipage}{\textwidth}
This thesis has given rise to the following publications in regular journals:\vskip 1 cm

A.~de la Cruz-Dombriz, A.~Dobado and A.~L.~Maroto,\\
\emph{Comment on 'Viable singularity-free $f(R)$ gravity without a cosmological  constant' [arXiv:0905.1941]},\\
  Phys.\ Rev.\ Lett.\  {\bf 103}, 179001 (2009)
  [arXiv:0910.1441 [astro-ph.CO]].
\par\ \par

A.~de la Cruz-Dombriz, A.~Dobado and A.~L.~Maroto,\\
\emph{Black Holes in $f(R)$ theories},\\
  Phys.\ Rev.\  D {\bf 80}, 124011 (2009)
  [arXiv:0907.3872 [gr-qc]].
\par\ \par

J.~A.~R.~Cembranos, A.~de la Cruz-Dombriz, A.~Dobado and A.~L.~Maroto,\\
\emph{Is the CMB Cold Spot a gate to extra dimensions?},\\
  JCAP {\bf 0810}, 039 (2008)
   [arXiv:0803.0694 [astro-ph]].
\par\ \par

A.~de la Cruz-Dombriz, A.~Dobado and A.~L.~Maroto,\\
\emph{On the evolution of density perturbations in $f(R)$ theories of gravity},\\
  Phys.\ Rev.\  D {\bf 77}, 123515 (2008) [arXiv:0802.2999 [astro-ph]].
\par\ \par

A.~de la Cruz-Dombriz and A.~Dobado,\\
 \emph{A $f(R)$ gravity without cosmological constant},\\
   Phys.\ Rev.\  D {\bf 74}, 087501 (2006) [arXiv:0607118 [gr-qc]].
\par \par

\vskip 2 cm

The work developed in this thesis has been presented in a
series of scientific meetings with the following contributions to
the corresponding proceedings: \vskip 1 cm

A.~de la Cruz-Dombriz, A.~Dobado and A.~L.~Maroto,\\
\emph{Black holes in modified gravity theories},\\
Proceedings of the Spanish Relativity Meeting ERE 2009,\\
Journal of Physics: Conference Series (JPCS) [arXiv:1001.2454 [gr-qc]].
\par\ \par

A.~de la Cruz-Dombriz, A.~Dobado and A.~L.~Maroto,\\
\emph{Cosmological density perturbations in modified gravity theories},\\
AIP Conf.\ Proc.\  {\bf 1122} 252 (2009) [arXiv:0812.3048 [astro-ph]].
\end{minipage}

\end{document}